# Instability, Intermixing and Electronic Structure at the Epitaxial $LaAlO_3/SrTiO_3(001)$ Heterojunction

S.A. Chambers[a], M.H. Engelhard[a], V. Shutthanandan[a], Z. Zhu[a], T.C. Droubay[a], L. Qiao[a], P.V. Sushko[b], T. Feng[c], H. D. Lee[c], T. Gustafsson[c], E. Garfunkel[c], A.B. Shah[d], J.-M. Zuo[d] and Q.M. Ramasse[e]
[a]Pacific Northwest National Laboratory, Richland, WA
[b]University College, London, United Kingdom
[c]Rutgers University, Piscataway, NJ
[d]University of Illinois at Urbana-Champaign
[e]SuperSTEM Laboratory, STFC Daresbury, UK

## Outline

## Abstract

The question of stability against diffusional mixing at the prototypical $LaAlO_3/SrTiO_3(001)$ interface is explored using a multi-faceted experimental and theoretical approach. We combine analytical methods with a range of sensitivities to elemental concentrations and spatial separations to investigate interfaces grown using on-axis pulsed laser deposition. We also employ computational modeling based on the density function theory as well as classical force fields to explore the energetic stability of a wide variety of intermixed atomic configurations relative to the idealized, atomically abrupt model. Statistical analysis of the calculated energies for the various configurations is used to elucidate the relative thermodynamic stability of intermixed and abrupt configurations. We find that on both experimental and theoretical fronts, the tendency toward intermixing is very strong. We have also measured and calculated key electronic properties such as the presence of electric fields and the value of the valence band discontinuity at the interface. We find no measurable electric field in either the $LaAlO_3$ or $SrTiO_3$, and that the valence band offset is near zero, partitioning the band discontinuity almost entirely to the conduction band edge. Moreover, we find that it is not possible to account for these electronic properties theoretically without including extensive intermixing in our physical model of the interface. The atomic configurations which give the greatest electrostatic stability are those that eliminate the interface dipole by intermixing, calling into question the conventional explanation for conductivity at this interface – electronic reconstruction. Rather, evidence is presented for La indiffusion and doping of the $SrTiO_3$ below the interface as being the cause of the observed conductivity.

## 1. Introduction

As a class of materials, complex oxides exhibit an exceedingly wide range of structural, compositional, and functional properties. The working definition of a complex oxide is an inorganic solid consisting of more than one metal cation and oxygen anions. The simplest of the complex oxides contain (only) two metal cations in distinct, well-defined sublattices. For example, the perovskites have the formula $ABO_3$, where twelve-coordinate A-site cations are at the corners of the unit cell (u.c.), six-coordinate B-site cations are at body center positions, and six-coordinate O anions are at face-center positions. The structural and compositional diversity of complex oxides is realized by the ease with which different metal cations that can be placed within the A- and B-site sublattices. For instance, the A sites can be populated by mixtures of alkaline earth and rare earth cations, and the B sites can be occupied by first and second row transition metal cations, in addition to several of the Group IIIA, IVA and VA cations, giving rise to a wide range of compositions, many of which are achieved via solid solution formation. The formal charge degree of freedom on the B-site when transition metal cations are used allows mixing and matching of A-site cations to achieve a range of charge configurations. As the chemical identities of A and B sites are varied, the resulting structures change in response to variable cation radii. Complex oxides can be characterized by useful metrics such as the tolerance factor ($f$) for perovskites, defined as $f = (r_A + r_O)/\sqrt{2}(r_B + r_O)$ [1]. Cubic perovskite structures result when $f$ is near unity. Distorted perovskites with rhombohedral and eventually orthorhombic structures occur as the size of the A-site cation drops, and the B-O-B bond angle deviates from $180^o$. B-O-B bond angle distortion in turn results in a decrease in the one-electron bandwidth due to a drop in the *d*-electron transfer amplitude between adjacent B sites associated with changes in B-O 3*d*-2*p* hybridization. The functional impact of this kind of structural distortion is a change in the metal-insulator transition temperature.

The wide range of functional properties exhibited by complex oxides is then a direct consequence of the chemical identity of the constituent cations and the associated structural distortions. Perhaps the best know example of this phenomenon is the occurrence of high-$T_c$ superconductivity in the layered cuprates ($La_{2-x}Sr_xCuO_4$), brought about by Sr-induced hole doping of the formally charge-neutral (Cu–O) state to $Cu^{(2+x)+}$( Cu–O)$^{x+}$. A less well known example is that of doped $SrTiO_3$ (STO), a cubic, diamagnetic band insulator with an optical gap of 3.2 eV. Doping the A site with La(III) or the B site with Nb(V) at the ~1 at. % level transforms STO into an *n*-type oxide semiconductor [2]. Conversely, doping $LaTiO_3$, an antiferromagnetic Mott insulator, with Sr(II) at the ~5 at. % level results in a phase transition to a paramagnetic metal ground state [3].

The range of functional properties that can be achieved in bulk complex oxides by mixing and matching cations generates tantalizing possibilities when considering single layers and superlattices prepared by epitaxial thin-film growth techniques. By combining a high degree of stoichiometric control with the reduced dimensionality in the growth direction achievable by ultrathin film and small-period superlattice growth, it is in principle possible to create artificially structured materials with novel properties not realized in the bulk. An excellent example of this phenomenon is recent work by Santos et al. [4] in which ozone-assisted molecular beam epitaxy (MBE) was used to grow small-period superlattices consisting of alternating unit cells (u.c.) of $SrMnO_3$ and $LaMnO_3$ on $SrTiO_3$(001), and then comparing the resulting electronic and magnetic properties with those of random alloys of MBE-grown $La_{\sim 0.5}Sr_{\sim 0.5}MnO_3$ . Differences were observed in the saturation magnetization and resistivity between superlattices and the solid solution of the same overall composition, particularly at low temperatures, and novel tuning of the magnetic properties could be achieved by injected an extra u.c. of $SrMnO_3$ and $LaMnO_3$. However, the same forces of nature that allow us to generate a wide range of compositions in complex oxides in the first place can also promote solid solution formation at interfaces that we intend to be abrupt. These forces include mutual solubility and structural similarity, both of which can readily homogenize an otherwise abrupt interface at the junction of materials with different compositions. The problem is exacerbated by the thermal energy available to the system from substrate heating which is in turn required to achieve epitaxial growth, as well as the some times high ion energies of species within the laser plume generated during pulsed laser deposition (PLD), an exceedingly popular method for complex oxide film growth. It is thus of considerable interest to engineer the deposition conditions so that intermixing is kinetically constrained at the interface, giving rise to metastable structures that retain a high degree of abruptness.

Against this backdrop, we consider the $LaAlO_3/SrTiO_3(001)$ (LAO/STO) interface. This material system has attracted wide-spread interest over the past several years because of the observation of conductivity near the interface under certain deposition conditions, despite the fact that both constituent materials are band insulators. Starting with seminal work by Ohtomo and Hwang [5], soon thereafter reproduced and built upon by Thiel et al. [6], several experimental groups worldwide have synthesized this interface in various forms and have made a common set of observations. These include the following: (1) when using on-axis PLD (the most commonly used growth method for LAO/STO interface preparation), layer-by-layer growth and conductive interfaces occur only when the oxygen partial pressure is between $\sim 10^{-4}$ and $\sim 10^{-6}$ Torr, (2) interface conductivity is unambiguously observed only when LAO is grown on $TiO_2$-terminated STO(001), which can be realized by a buffered HF etch followed by a tube furnace anneal in oxygen, (3) in the absence of external perturbations, conductivity occurs at and above a threshold LAO thickness of 4 u.c., (4) the measured carrier concentration is less than that expected based on electronic reconstruction arguments involving transfer of half an electron per unit cell from LAO into the STO, unless the film is not grown and/or annealed in sufficient oxygen to prevent reduction of the STO, and, (5) the interface appears to be abrupt when examined using high-angle annular dark field scanning transmission electron microscopy (HAADF-STEM), as reviewed recently by Muller [7]. In HAADF-STEM images, the high degree of atomic number (Z) contrast realized by collecting scattered electrons at high angles, in conjunction with a highly focused beam and aberration correction in the lenses, results in La cations "lighting up" relative to Sr cations, giving the appearance of a high degree of interfacial abruptness. Imaging alone is not sufficient to rule out cation disorder at the LAO/STO interface, as high resolution electron energy loss measurements in conjunction with HAADF-STEM have shown [8]. However, a majority of experimentalists and theorists who have published on this system to date tend to model the interface as if it were abrupt. If cation disorder (intermixing) is considered at all, it is thought to be limited approximately at most one unit cell on either side of the interface.

The presumed cause of LAO/STO interface conductivity is alleviation of the so-called "polar catastrophe", which results from forming a junction between a polar material (LAO) and a nonpolar material (STO). STO consists of alternating layers of $(Sr^{2+}O^{2-})^0$ and $(Ti^{4+}O^{2-}{}_2)^0$ along [001]. Both constituent layers are formally charge neutral. In contrast, LAO consists of alternating $(La^{3+}O^{2-})^+$ and $(Al^{3+}O^{2-}{}_2)^-$ layers, and thus exhibits a polarity along [001]. Atomically abrupt interface formation between LAO and STO thus gives rise to a polar discontinuity which extends to the LAO surface. Elementary electrostatic considerations suggest that layer-by-layer growth of LAO on STO and abrupt interface formation will lead to a diverging electric potential as a result of the accumulation of dipoles within the LAO film – the so-called "polar catastrophe". This unstable situation can in principle be mitigated by transfer of half an electron per unit cell from $(La^{3+}O^{2-})^+$ to $(Ti^{4+}O^{2-}{}_2)^0$ at the interface for $TiO_2$-terminated STO, or half a hole per unit cell from $(Al^{3+}O^{2-}{}_2)^-$ to $(Sr^{2+}O^{2-})^0$ for SrO-terminated STO. Electronically, these two interfaces are formally n-type and p-type, respectively, and are routinely referred to as such in the literature. In principle, both interfaces should exhibit some degree of conductivity, albeit with opposite majority carriers. Moreover, the carriers should be confined to the interface to form a two-dimensional electron or hole gas (2DEG or 2DHG) if there is sharp band bending there. However, conductivity has been observed only at the n-type interface, and it is routinely ascribed to the presence of a 2DEG.

Nakagawa et al. [8] carried out cross-sectional HAADF-STEM measurements on PLD-grown LAO/STO(001) heterojunctions of both polarities (n- and p-type). These authors found that cation disorder in the form of La and Ti cross diffusion exists at both kinds of interfaces, as seen in Fig. 1a & b. However, the p-type interface was found to be somewhat more abrupt than the n-type interface, with the interface width being slightly less than 2 nm for the p-type interface and slightly greater than 2 nm for the n-type interface. Additionally, the presence of Ti(III) within a few nm of the n-type interface was deduced by fitting Ti L-edge spectra to linear combinations of appropriate reference spectra (Fig. 1a). Likewise, O vacancies ($V_O$) were deduced at both interfaces by fitting O K-shell spectra to those of bulk LAO, bulk STO and O-deficient STO ($SrTiO_{3-\delta}$, where $\delta = 0.25$), as seen in Figs. 1c&d. Ironically, more $V_O$ were found at the p-type interface, despite the virtual absence of Ti(III). The electrical asymmetry between n- and p-type interfaces was rationalized as follows. Extra electrons from the LaO interfacial layer result in the partial reduction of Ti ions near the interface where $V_O$ would normally be found. The absence of free holes at the p-type interface was ascribed to the presence of compensating $V_O$ with the attendant pair of electrons per vacancy. Intrinsic to this argument is the claim that oxygen vacancies at this interface are qualitatively different in origin and effect than those in bulk oxides. Moreover, these authors argue that the electrical and roughness asymmetries are related in the following way. The increase in interface dipole energy resulting from the spread of electrons from the LaO interfacial layer across several $TiO_2$ layers near the n-type interface is compensated by enhanced Sr-La exchange, which reduces the dipole energy and roughens the interface. They suggest that the absence of itinerant charge at the p-type interface eliminates the need for cation exchange, resulting in a sharper interface.

Jia et al.[9] also used STEM to investigate the LAO/STO interface prepared by high-oxygen-pressure radiofrequency sputtering, but did not carry out EELS measurements. These authors observed an increase (decrease) in HAADF intensity as the interface was approached from the STO (LAO) side and interpreted this finding as being due to Sr-La intermixing over a few u.c.

Willmott et al. [10] used surface x-ray diffraction (SXRD) to deduced atom profiles and displacements across the PLD-grown n-type LAO/STO interface by fitting the experimental data to simulations. The results are summarized in Fig. 2. These authors concluded that intermixing occurs and extends over greater distances for Sr and La than for Ti and Al. Their model of the interface includes a few u.c. of predominantly $LaTiO_3$ and a region of $La_{1-x}Sr_xTiO_3$ on the STO side of the interface. Such a structure would contain a significant quantity of Ti(III) with the resulting effect of lattice dilation in the $Z$ direction by virtue of the larger ionic radius of Ti(III) compared to Ti(IV), which was consistent with the diffraction data. Density functional theory (DFT) calculations based on the interface composition extracted from SXRD led to the prediction of a significant enhancement in band bending in both the LAO and STO relative to those in the abrupt n- and p-type interfaces. These authors note that $La_{1-x}Sr_xTiO_3$ is conductive for a wide range of x, and at least imply that the La doping of the underlying STO may be at least partially responsible for electron conduction at the n-type interface.

Kalabukhov et al. [11] have used medium energy ion scattering (MEIS) along with atomic force and scanning Kelvin probe (SKP) microscopies to probe atom profiles, morphology and electrostatic potential at PLD-grown n-type LAO/STO(001) interfaces of various thicknesses. The MEIS spectra and associated modeling in the channeling and random directions yield clear evidence for La indiffusion and both Sr and Ti outdiffusion to the surface for LAO thicknesses of up to 4 u.c. Fig. 3 shows the results for 1 u.c. LAO/STO(001) in the random geometry in which the incident beam is not aligned along a low-index direction. Simulations reveal that only ~ 50% of the A sites within the top u.c. are populated with La and that significant La atomic fractions must be included in the A sites of the first three u.c. of STO to account for the measured yield. Moreover, SKP measurements reveal inhomogeneities in the surface potential with characteristic sizes of 100 – 1000 nm which are suggestive of compositional inhomogeneities and, thus, "filamentary" interdiffusion. These authors suggest that fully stoichiometric LAO nucleates on an intermixed phase that forms during nucleation of the first 3 u.c., giving rise to an insulator-to-metal transition at 4 u.c., perhaps as a result of the formation of a $La_xSr_{1-x}TiO_3$ conductive layer. It is not clear why the threshold for conductivity should be 4 u.c. unless this thickness of LAO is required to drive enough La diffusion into the STO to form a *continuous*, conductive doped layer.

While the experiments discussed above were well conceived and well executed, the associated results are often ignored, particularly by those who carry out first principles calculations of electronic structure at the LAO/STO interface. For example, a recent topical review by Pentcheva and Pickett [12] states, "Layer-by-layer growth allows synthesis of phases that are not thermodynamically stable. Recent development of growth techniques like PLD and MBE have enabled the synthesis of oxide superlattices with atomic precision." The theoretical work reviewed thereafter and, indeed, the vast majority of all theoretical modeling of this interface, starts with a completely abrupt and structurally perfect interface. Similarly, while the possibility of intermixing is sometimes admitted, an abrupt interface paradigm tends to dominate the thinking of experimentalists [13]. However, what references [8-11] unambiguously show is that at the very least, interfacial intermixing *can* occur when LAO/STO is prepared by PLD and reactive sputtering, and perhaps also by MBE as well, although very little MBE growth of this system has been reported [14]. The fundamental unanswered question is whether intermixing at the LAO/STO interface is the exception or the rule. Is intermixing an unfortunate consequence of non-optimized growth conditions or ion induced diffusion, or is it a natural result of the thermodynamics of interface formation?

In light of the potential importance of intermixing at the LAO/STO interface in determining electronic structure, and the need to find reliable ways to characterize intermixing at complex oxide interfaces in general, we have undertaken a multi-technique, multi-institutional investigation of interface composition using samples prepared in the pioneering laboratories of Professors Jochen Mannhart and Harold Hwang at the Universities of Augsburg and Tokyo, respectively. This report summarizes our investigation. The analytical techniques we have brought to bear on the problem include Rutherford backscattering spectrometry (RBS), time-of-flight secondary ion mass spectrometry (ToF-SIMS), high-angle annular dark field scanning transmission electron microscopy and electron energy loss spectroscopy (HAADF STM/EELS), angle-resolved x-ray photoelectron spectroscopy (ARXPS) and medium energy ion scattering (MEIS). We have also undertaken first principles DFT as well as classical force field calculations in which interfacial intermixing is explicitly included in our structural models in order to determine the energetics of intermixing, as well as the effect of intermixing on key electronic properties such as valence band offsets and internal electric fields, or band bending. The Report is organized as follows. Section 2 discusses details of film growth. Sections 3 & 4 cover interface composition in thicker (25 u.c.) and thinner (4 u.c.) films, respectively. For each thickness, we utilize techniques well suited to the thickness – RBS, ToF-SIMS and HAADF-STEM/EELS for 25 u.c. films, and ARXPS, MEIS and theory for 4 u.c. films. Moreover, we used these different techniques on the same samples from the two laboratories in order to directly compare data from the different methods on a given sample. Section 5 presents results on the determination of band offsets and band bending at the interface of 4 u.c. films based on high-energy resolution XPS along with theoretical prediction of these quantities, and Section 6 summarizes the Report. Our purpose is not to write a comprehensive review of the LAO/STO field; other recent reviews can be found [12-13, 15]. Rather, we focus specifically on interfacial composition and its relationship to electronic structure, a topic that is inadequately dealt with in the LAO/STO literature in our view.

## 2. Film growth

All LAO films were grown by on-axis PLD using a KrF excimer laser and $TiO_2$-terminated STO substrates prepared by buffered HF etching and oxygen tube furnace annealing [16-17]. The laser ablation targets were LAO single crystals. The substrate size was either 10 mm x 10 mm x 1mm or 5mm x 5mm x 0.25mm. The growth parameters are summarized in Table 1. The Tokyo films were grown with one of three values of laser energy density on the target – 0.7, 1.1 or 1.6 J/cm$^2$. Ion scattering and x-ray photoelectron spectroscopy measurements described in detail below did not reveal significant differences between films grown under these three laser focusing conditions. Following growth, the Tokyo films were cooled to room temperature in 1 x $10^{-5}$ Torr oxygen at a rate was 30 °C/min. In contrast, the Augsburg were cooled to room temperature over a 2.5 hour period in 400 mbar of oxygen with a one hour anneal at 600°C.

All growths apparently proceeded in an essentially layer-by-layer fashion, as judged by the persistence of reflection high-energy electron diffraction (RHEED) intensity oscillations throughout the growth period. Atomic force microscopy (AFM) images reveal a well-defined terrace-step structure with a minimum step height of ~0.4 nm, as expected for heteroepitaxial growth on STO substrates of a single termination. Due to the lack of sharpness of most AFM tips, these results do not rule out the possibility of nanoscale roughness on portions of the terraces, or incomplete nucleation of a given layer on one portion of the surface before the next layer begins to nucleate elsewhere. However, the AFM results in concert with the RHEED intensity oscillations show that the growth mode is predominantly Frank-van de Merwe.

The groups at the Universities of Augsburg and Tokyo who grew these films have published extensively on LAO/STO and used growth conditions that consistently lead to interface conductivity. The presence of interface conductivity via transport measurements could not be verified on films grown on 5mm x 5mm substrates because contact metallization would have obliterated too much of the pristine surface to allow us to make interface characterization measurements. However, Au metallization pads were deposited in the corners of some films grown on 10mm x 10 mm substrates, and conductivity was verified using a four-point probe apparatus. For these and other 4 u.c. films grown on 10mm x 10mm substrates without Au pads, x-ray photoemission spectroscopy could be done without charge compensation by making electrical contact with these Au pads, or by using a stiff spring clip to press through the 1.5 nm thick film to the interface. Interface conductivity was verified on at least one 4 u.c. film from each lab for which an extensive set of interface characterization measurements was also made. Thus, this investigation is of direct relevance in addressing the relationship between interface composition and electronic structure for samples which exhibit conductivity.

## 3. Film and interface composition – 25 unit cell LAO films

### 3.1 Rutherford backscattering

Rutherford backscattering spectrometry (RBS) has been extensively used to investigate stoichiometry, structure and thickness of epitaxial films of a variety of materials [18-22]. In RBS, the probe particle is typically a $He^+$ ion beam of energy between ~0.5 and ~2.0 MeV. In this energy regime, Coulomb scattering of the incident ion by nuclei in the solid can be treated classically and reasonably accurate numerical simulations can be performed at this level of approximation. RBS is element specific since the recoil energy of the backscattered $He^+$ ion is dependent on the mass of the scatterer. Since the ion loses energy by means of inelastic scattering as it travels through the target material, an energy spectrum of the backscattered ions also yields information about the depth at which a given backscattering event occurs. Because high-energy ion beams penetrate deeply into materials, RBS can be used to study the buried interfaces and diffusion profiles.

The probability of Rutherford backscattering is quantified by the RBS differential cross section, $d\sigma/d\Omega$, which was derived by Rutherford in 1911 based entirely on classical electrodynamics. The operative interaction is the Coulomb repulsion between the incoming probe ion (kinetic energy = $E_0$, charge = $Z_1$) and the target nucleus (kinetic energy = 0, charge = $Z_2$). Solving the equations of motion and utilizing the spherically symmetric nature of the Coulomb force yields,

$$\frac{d\sigma}{d\Omega} = \left(\frac{Z_1 Z_2 e^2}{4E_0}\right)^2 \frac{1}{\sin^4(\theta/2)} \tag{1}$$

where $\theta$ is the scattering angle for the incident ion. The differential cross section is very strongly forward peaked, reflecting the fact that most incident particles pass through the considerable volume of the solid not consisting of nuclei, and only a very small fraction will pass close enough to a nucleus to have a Coulombic interaction leading to a change in trajectory of any significant magnitude. Moreover, the differential cross section falls off by a factor of four in going from $\theta = 90^{o}$ to $\theta = 180^{o}$. Thus, a significant gain in sensitivity can be realized by operating at scattering angles closer to $90^{o}$.

Conservation of energy and momentum within the two-particle system consisting of the incident ion (mass = $M_1$) and the target nucleus (mass = $M_2$) allows the energetics of RBS to be elucidated to a high degree of accuracy. The resulting kinematic factor ($K$) is the ratio of the energies of the incident ion after ($E_1$) and before ($E_0$) the backscattering event, and is given by

$$K = \frac{E_1}{E_0} = \frac{\frac{1}{2}M_1 v_1^2}{\frac{1}{2}M_1 v_0^2} = \left[\frac{(M_2^2 - M_1^2 \sin^2\theta)^{1/2} + M_1 \cos\theta}{M_1 + M_2}\right]^2 \tag{2}$$

Here, $\theta$ and $\phi$ are the scattering angles for the incident ion and stationary nucleus, respectively, as seen in Fig. 4a. We show in Fig. 4b&c plots of $K$ vs. $\theta$ for a fixed incident ion mass and three target nuclei masses (b), as well as three incident masses and a fixed target mass (c). Several physical insights are gleaned from inspection of these plots. First, the larger the mass of the target nucleus, the less target recoil energy and, thus, the closer $K$ is to unity. As a result, backscattering peaks for less massive nuclei fall at lower incident ion energies than do those of more massive nuclei. Second, for a given target mass, there is more energy transfer at higher scattering angle, as expected for more "head-on" collisions. Third, for any two target masses, the ratio of the associated kinematic factors increases with increasing scattering angle. Thus, mass resolution is enhanced at higher scattering angles. Fourth, as the mass of the incident particle increases relative to that of the stationary nucleus, the backscattering peak energy drops more rapidly with scattering angle. These features of the RBS process are useful in designing experiments on real materials, as we shall see below. In addition to the primary elastic backscattering, there are also small-angle inelastic scattering events which occur as the incident particle traverses the solid. These events result in a broadening of the peak which increases as the angle of the outgoing relative to the surface plane particle is lowered, resulting in a loss of mass resolution for closely spaced peaks. A detailed description of the RBS and accelerator facility we have used is given elsewhere [23].

In the present work, RBS measurements were carried out using 1 - 3 MeV $He^+$ ions and a current of 5 nA or less. There was no evidence of specimen damage at this current. Damage would appear in the form of increased minimum yield with beam exposure time. The minimum yield ($\chi$) is the ratio of the RBS signal for a given element with the incident beam aligned in a random direction to that measured with the incident beam aligned along a low-index direction. Beam damage results in an increase in local crystallographic disorder, which in turn causes $\chi$ to go up. However, no such increase in $\chi$ was observed. Typical random spectra are shown using a 2 MeV incident ion energy for 25 u.c. films from Augsburg and Tokyo in Figs. 5 & 6, respectively, along with the optimized simulated spectra. Two different geometries were used. In one, the incident beam was $60^o$ off the surface normal and the scattering angle was $150^o$ (Figs. 5a & 6a). This geometry produced superior mass resolution because the high scattering angle and because the backscattered beam traversed close to the surface normal, resulting in a relatively short path length and thus minimal broadening by inelastic scattering. In the other, the incident beam was $7^o$ off the surface normal and the scattering angle was $96^o$ (Figs. RBS 5b & 6b). This geometry produced slightly better depth resolution and higher sensitivity to the weak Al backscattering peak than does the $150^o$ scattering angle. (In point of fact, sensitivity to all elements is enhanced by using the lower scattering angle, as discussed above. However, our primary interest in going to $96^o$ is to enhance our sensitivity to Al, since the Al feature is particularly weak.) The SIMNRA simulation program was used to model experimental RBS spectra, yielding the stoichiometries and thicknesses of the films [24]. In these simulations, the sample is divided into several thin layers of variable thickness and composition. Experimental parameters such as incident ion energy, atomic number of the ion, incident and scattering angles, energy calibration values for the detectors, solid angle of detection and the total charge deposited by the incident beam are inputs. From this information and a trial sample structure, the program calculates the backscattered spectra. The composition profiles and film areal densities (in units of atoms/cm$^2$) are then systematically varied until the best fit to experiment is obtained. The areal density is the product of the atom number density of atoms and the film thickness.

Quantification of the La concentration is most accurate since the La peak is isolated from all other peaks. The La atom percents are slightly different at the two scattering angles, but are within experimental error, which is the same for the two scattering angles. Likewise, the Al concentrations extracted from the spectra at the two angles are nominally within the experimental error. However, the error bar is smaller for the Al spectrum taken at 96$^o$ due to the better counting statistics.

Both spectra in Fig. 5 reveal that this film is slightly Al rich and La poor, and the same is true for all other films from Augsburg. Due to the low scattering cross section for O, the O stoichiometric coefficient is not independently varied in the SIMNRA simulator. Rather, it is taken to be $5.0 - x_{Al} - x_{La}$, where $x_{Al}$ and $x_{La}$ are the stoichiometric coefficients of Al and La, respectively. Therefore, any deviations in the values of the O stoichiometric coefficients from 3.0 resulting from SIMNRA simulations do not represent the presence of either O vacancies or excess O, but rather are artifacts of the simulation algorithm. In the absence of any actual insight into the O stoichiometry in these films, we have set the O coefficient to 3.0.

In contrast to the Augsburg films, the Tokyo film is typically slightly La rich and Al poor (Fig. 6). The counting statistics on the Al peak and, therefore, the uncertainty on the Al concentration, improve by going to the 96$^o$ scattering angle. Nevertheless, the Tokyo film departs from perfect stoichiometry in the opposite direction from the Augsburg film. A summary of the film composition and film thickness obtained from these two samples is shown in Table 2.

Interestingly, the structural properties of 25 u.c. films from Augsburg and Tokyo are nominally the same; the films are coherently strained to the substrate and the *c* lattice parameter is reduced relative to the bulk value, in keeping with the expected tetragonal distortion accompanying tensile in-plane strain. We show in Fig. 7a&b channeling rocking curves for 25 u.c. films from both labs with the incident beam varied about [101] 45° off normal. As expected, the Sr backscattering peak goes through a deep minimum along [101], corresponding to backscattering within the unstrained substrate. In contrast, the La backscattering peaks are shifted 1.4° in the more off normal direction as a result of film tetragonality. From these data, the *c*/*a* ratio is determined to be tan(43.6°) = 0.95. In light of the fact that based on high-resolution x-ray diffraction (not shown) the LAO is coherently strained to the substrate (*a* = 3.91Å), *c* is thus 3.73Å, in agreement with the XRD value. These results thus show that LAO is able to accommodate nonstoichiometries in the La and Al atom counts in a way that allows pseudomorphic growth on STO for La:Al atom ratios slightly above and slightly below unity.

A subtle feature in these RBS spectra is the presence of counts between the energies of 1700 and 1760 keV and, in particular, the small shoulder appeared at the low-energy side of the La peak, as seen in Figs. 5 & 6. There should not be any counts here if the interface is atomically abrupt. However, counts are expected to appear in this energy range if some interdiffusion occurs. Indeed, the simulations shown in Figs. 5 & 6 for both scattering angles include some indiffused La, the details of which are shown in Table 3. Without indiffusion included in the model, the fit in the valley would be quite poor. For example, we show in the inset to Fig. 6b fits for the abrupt and optimized intermixed configurations. We also show in Fig. 8 breakdown of how the different layers in Table 3 contribute to this simulation. In order to adequately account for all counts in the valley, a long La diffusion tail that extends to ~1000Å into the STO must be included. However, it is in actuality not possible to determine how deeply the La diffuses from these data because of overlap of this valley region with the Sr peak. We need to know the channel at which the tail on the low energy side of the La peaks goes to zero. This can only be done by increasing the separation between the La and Sr peaks, and doing so requires an increase in ion energy while maintaining the 150° scattering angle.

To this end, we have taken data at 3 MeV using the Tokyo 25 u.c. film, and the results are shown in Fig. 9. The counts do indeed go to zero near the center of the valley, indicating the maximum depth to which La diffuses, which is not detectable at 2 MeV (Figs. 5&6). The intermixed interface simulation seen in Fig. 9 includes a La diffusion profile that has been varied to optimize agreement with experiment, paying particular attention to the low-energy tail on the La peak (Fig. 9b); an abrupt-interface simulation is also shown. The composition of the film proper is within experimental error of that obtained by fitting the 2 MeV spectra. However, with the 3 MeV spectrum, we can obtain a more reliable estimate of the extent of La diffusion into the STO and determine how the concentration varies with depth. As with the data taken at 2 MeV, we have modeled the La atom profile as five distinct La-containing regions. The concentrations in each layer are shown in Table 4. In contrast to the profile extracted from the 2 MeV data for the same sample, this profile goes to zero La concentration at a depth of ~500Å. This diffusion profile is more accurate because the La tail does not interfere with the Sr peak.

There are also counts on the high-energy side of the Sr peak in Fig. 9 which might be due to Sr outdiffusion. In order to investigate the possibility, we included a Sr atom profile which includes 5 at. % St in the first ~7 u.c. of the LAO out from the interface and 1 at. % in the next ~6 u.c. This profile approximates Sr outdiffusion in a reasonable way. However, as seen in Fig. 9c, inclusion of this outdiffused species does not account for these counts. The origin of these counts is thus not known at this time.

The counts in the valley between the La and Sr peaks which we interpret as being due to interdiffusion may also be of spurious origin. There are three artifacts in RBS that can generate counts between otherwise distinct peaks – pulse pile up, straggling, and multiple and/or plural scattering. In what follows we explore the possibility that one or more of these are giving rise to the counts we observe in the valley, rather than interdiffusion.

Pulse pile-up is caused by electronic detector delays in the data collection system. When a particle hits the Si detector, an electron-hole pair is created and the two charges are drawn to opposite ends of the Si by an applied voltage. The resulting charge pulse is then amplified and processed by an analog-to-digital converter (ADC). If during the time required to process one pulse, a second particle of the same energy strikes the detector, the ADC processes the two pulses simultaneously as one pulse with twice the charge. The higher charge will appear to the algorithm as a particle of higher energy and cause some unwanted backgrounds in the RBS spectrum. Pulse pile up typically occurs at higher primary beam currents and is accompanied by measurable detector dead time. In order to check for pulse pile up in the present experiments, a series of spectra were measured at beam currents above and below the rather low value we used for our normal data acquisition (5 nA, for which the detector dead time is less than 1%), with special attention paid to behavior in the valley between the La and Sr peaks. The results are shown in Fig. 10a. For beam currents of 2.5, 5, 10, and 15 nA, the backscattering yield in the valley is approximately the same. However, the yield in the valley grows considerably when the beam current is increased to 30 nA. Increased counts also appear above the La peak when 30 nA is used. These results reveal that pulse pile up does not occur to any significant effect until the beam current exceeds 15 nA. Even when the effect is occurring at a minimal level, the SIMNRA simulator can detect and correct for it in two methods [24-25]. In one method, the program calculates the pile up contribution from the experimental spectrum and subtracts this contribution for the raw spectrum. In the other method, SIMNRA calculates the pile up contribution from a simulated spectrum and adds this contribution to the calculated spectrum. We typically use the first of these methods, and we illustrate its effectiveness in Fig. 10b, which shows the data in Fig. 10a after correction. After correction, the counts on the higher energy side of the La peak are near zero for all beam currents, indicating the effective removal of pile up. However, counts in the valley are still present, indicating an origin other than pulse pile up. Moreover, the amplifier in the detector circuit also has built-in pulse-pile-up rejecter circuit which rejects at least ~40 % of the dual pulses when they occur due to high primary beam current. The remainder can be dealt with as described above

The second candidate artifact is straggling. Straggling is energy broadening due to inelastic scattering of the beam as it traverses the solid, and the effect increases with increasing sample thickness. In order to determine if straggling results in any appreciable counts in the valley, we have measured RBS spectra on specimens with different LAO film thicknesses, paying particular attention to the low-energy tail on the La peak. This tail should increase in magnitude with increasing film thickness if it is due to straggling. In contrast, this tail will not change with thickness if it is due to La indiffusion. We show in Fig. 11a the La peak for films with areal densities of $68 \times 10^{15}$, $97 \times 10^{15}$, $105 \times 10^{15}$, and $160 \times 10^{15}$ atoms/cm$^2$. The corresponding thicknesses for this series of films are approximately 80, 110, 120 and 180Å. The La peak positions have been shifted so the low-energy tails directly overlap. There is no detectable change in the magnitude of the tail with thickness, as seen more clearly in the expanded region in Fig. 11b. Therefore, the low energy tail on the La peak is not due to straggling.

Multiple and plural scattering can result in trajectory changes that can also produce spurious counts in the valley between La and Sr peaks. SIMNRA uses straight-line trajectories for the ingoing and outgoing particles, with a single backscattering event resulting in the change in trajectory. In reality, the particles undergo many small-angle deflections due to multiple elastic scattering. The particles may also undergo more than one large-angle scattering event, and this phenomenon is known as plural scattering. Plural scattering is responsible for the background on the low energy side of peaks associated with high-Z elements when they sit atop low Z elements, as well as a steeper increase on the low-energy side of peaks than is calculated with a single scattering model. Furthermore, the lower the energy of the incident beam, the higher the background yield from multiple scattering. Thus, varying the energy of the incident beam is another way to assess the effect of multiple scattering. To this end, RBS data were collected from samples with incident beam energies of 1, 2, and 3 MeV. The spectra are shown in Fig. 12. The peaks have been aligned so the low-energy tails overlap. As can be seen, the counts in the valley are almost the same for incident energies of 2 MeV and 3 MeV, and are only slightly higher for 1 MeV. The lack of an effect between 3 and 2 MeV suggests that multiple scattering is not a significant contributor to these counts.

### 3.2 Time-of-flight secondary ion mass spectrometry

ToF-SIMS depth profiling provides information on elemental abundances as a function of depth by physically sputtering the specimen and then measuring mass spectra after each sputtering cycle. An advantage of the method is a very high degree of sensitivity which in turn varies with the mass and charge of the analyzing ion. Absolute elemental concentrations cannot be obtained without the use of suitable standards. The drawbacks are that the method is destructive in the region where the analysis is done, and that knock on, in which lattice ions at the analysis front are implanted rather than sputtered away by the analyzing ion, can occur. Knock on can spuriously enhance the concentration of atoms from the film material in the substrate. However, knock on cannot artificially increase the concentration of substrate atoms within the film.

A dual beam approach is normally used in ToF-SIMS depth profiling experiments [26], and we have done so here as well. We employed a TOF.SIMS5 spectrometer, manufactured by IONTOF GmbH in Germany. The pulsed analysis beam is of high energy (10-30 keV) and low current ($10^{-2}$-1 pA), as is normally used in static SIMS surface analysis. To achieve high sputtering rates and maintain a relatively smooth surface in the analysis region, a low-energy (0.2-2 keV) high-current ($10^{1}$-$10^{3}$ nA) sputtering beam was used. Cesium ($Cs^+$) and oxygen ($O_2^+$) beams have been the standard sputtering beam materials in SIMS analysis for more than thirty years. Cesium atoms are strong electron donors. Thus, $Cs^+$ sputtering enhances the yield of negative ions. Molecular oxygen enhances positive ion signal through surface oxidation.

Here we have used 1000 eV $O_2^+$ ions with a current of ~100 nA for physical sputtering with a scanning area of 300 μm × 300 μm. A pulsed 25 keV $Bi_5^+$ beam with a current of ~0.05 pA was used for analysis. The $Bi_5^+$ beam was focused at the center of the $O_2^+$ sputter crater and scanned over an area of 100 μm × 100 μm. For quantification purposes, we use the near-surface region of the LAO film, as well as LAO single crystals, as standards for La and Al, and the substrate well beyond the interface as a standard for Sr and Ti. Fig. 13 show depth profiles for 25 u.c. films from Augsburg and Tokyo. We have normalized the La and Al (Sr and Ti) counts to unity near the surface (deep in the bulk) for easy comparison of concentrations. We estimate the interface to be at the depth corresponding to a sputter time of ~90 seconds. It is clear that there is outdiffused Sr and Ti in both LAO films. For both specimens, the Sr and Ti signals reach ~1% of their bulk STO values at a sputter time of ~70 seconds, which corresponds to a distance of ~ 2 nm out from the interface. Likewise, for both films, the Al signal is not down to the 1%-of-bulk level until ~150 seconds, corresponding to a distance of ~6 nm in from the interface, and the La signal does not reach 1% of bulk until ~200 seconds (~11 nm in from the interface). Some of the persistence of the La and Al signals with depth may be due to knock-on effects. However, the early appearance of the Sr and Ti signals cannot be due to anything other than outdiffusion into the LAO. For both films, the La signal decays more slowly with depth than the Al signal, which may be indicative of preferential indiffusion of La.

### 3.3 Scanning transmission electron microscopy/electron energy loss spectroscopy

High-resolution transmission electron microscopy (HRTEM) offers the substantial advantage of ultra-high spatial resolution when probing buried interfaces. High-angle annular dark field scanning transmission electron microscopy (HAADF STEM) offers the additional advantage of improved atomic number (Z) contrast by virtue of the strong dependence of the electron scattering cross section on Z at higher scattering angles. When combined with electron energy loss spectroscopy (EELS) obtained with a highly focused beam and an electron monochromator, HAADF STEM can in principle produce the most detailed local chemical and electronic data at the highest levels of spatial and energy resolution [7]. As we shall see below, EELS is essential to obtain an accurate determination of compositional variations with distance from a buried interface when using HAADF STEM because the strong Z contrast inherent in the method can make an interface appear to be more abrupt than it actually is when there are marked differences in Z across the interface. Accordingly, we have used state-of-the-art HAADF STEM/EELS to examine 25 u.c. LAO/STO(001) interfaces. We compare these results with those obtained using the volume averaging methods discussed above.

25 u.c. LAO/STO(001) samples for electron microscopy were cut into small pieces and glued to thin glass slides with an epoxy. The specimens were wedge polished at a 1° angle and ion milled in a Gatan model 691 Precision Ion Polishing System for several hours at 4.0 kV, followed by fine polishing at 2.0 kV for 20 minutes. Care was taken to reduce the ion milling beam current to 22 μA or less at 4.0 kV to reduce the ion milling rate and specimen heating. Immediately prior to observation in the electron microscope, the specimens were plasma cleaned for 1 minute.

HAADF STEM imaging was carried out using a JEOL JEM2200FS microscope at the University of Illinois and the TEAM 0.5 Titan microscope at the National Center for Electron Microscopy at Lawrence Berkeley National Laboratory. In each case, the aberration correctors were tuned to optimum probe sizes of ~1.0 Å and 1.5 Å at 200 kV and 80 kV, respectively. Atomically resolved EELS studies were carried out using the TEAM 0.5 Titan microscope coupled with a Gatan Image Filter (Tridiem ER). The acceleration voltage was 80 kV to reduce electron beam knock-on damage to the specimen. The monochromator was used to select an appropriate balance between beam current and the energy spread from the electron source, resulting in a well controlled electron dose at a slightly improved energy spread from the 0.8 eV delivered by the high brightness Scottky emitter (XFEG). The spectrometer semi-collection angle was set at ~60 mrad and the dispersion was set to 0.1 eV per channel for a maximum energy loss range of 204.8 eV. At 0.1 eV per channel, subtle changes in EELS fine structure can be observed at the EELS edges, which aids in the determination of interface sharpness and electronic structure. The Ti $L_{2,3}$ and O K edges could be collected simultaneously, as well with the Al $L_{2,3}$ and La $N_{4,5}$ edges, while the La $M_{4,5}$, Al K, and Sr $L_{2,3}$ were recorded separately. The Ti $L_{2,3}$, O K, Al $L_{2,3}$ and La $N_{4,5}$ spectral data presented here were treated with a power law background subtraction. Principal component analysis was performed on the Al K and Sr $L_{2,3}$ edges to reduce noise levels in the raw EELS spectra and a power law was used for background subtraction [27]. The EELS spectra were measured at discrete points along lines perpendicular to the interface in the STEM mode. Images acquired during STEM/EELS had an inner cut-off angle of 120 mrad for the annular dark field detector. Consequently, the images acquired with the EELS line profiles have weaker HAADF intensity than what is ideal for STEM imaging at the highest level of spatial resolution. Stage drift was checked for by comparing HAADF images recorded immediately before and after the EELS line scans. At least two EELS line scans were measured for each core level and we have based our analysis on scans that exhibited a drift of less than 1 u.c.

We used the JEOL JEM2200FS microscope to obtain high-resolution HAADF STEM images at 200 kV. An inner cut-off angle of 100 mrad was used for high-resolution HAADF STEM imaging. For both imaging and EELS, several thin areas of the specimen were investigated for a representative study in order to avoid the experimental bias that may arise from limited sampling. Large field-of-view atomic resolution STEM images were acquired to survey characteristic features at the interface. Large scans required for atomic resolution in large field of view images were made possible by the aberration-corrected STEM, due to the aberration corrected optics and increased stability in a state-of-the-art microscope room. Nanoarea electron diffraction patterns were acquired over nanometer scale areas in the JEOL JEM2010F transmission electron microscopy (TEM) operating at 197 kV. The probe size was ~55 nm and the divergence angle was less than 0.05 mrad.

### 3.3.1 LAO film structure

Fig. 14a shows a representative HAADF image from a 25 u.c. Augsburg film oriented along the [001] zone axis taken with a HAADF inner collection angle of ~100 mrad. The top surface is protected by a thin layer of epoxy. At the large cut-off angle used for imaging, the weaker scattering within the epoxy generates very little intensity in the HAADF images in comparison with the much heavier crystal and is therefore not easily observed within the dynamic range of the printed grayscale images. Both the LAO film and the first ~5 nm of the underlying STO substrate are slightly misoriented relative to the rest of the substrate. This misorientation results in a smearing of the HAADF contrast for the B-site Al and Ti atomic columns. There is a narrow band of reduced contrast just below the LAO-STO interface marked by arrows containing wave-like regions. This reduced contrast is suggestive of Sr, Ti, and/or O, vacancies, or substitution of lighter elements in these atomic columns. Moreover, contrast in the LAO film is reduced in several patchy areas in the lattice image. These lighter regions point to possible La deficiencies. The inset from a selected area of the LAO film shows that the top layer of LaO is partially formed and not perfectly sharp. Additional STEM images also show valleys in the top layer of LaO. At the film-substrate interface, the misorientation in the film and the first 5 nm of the underlying STO are a strong indication of strain. A representative electron diffraction pattern in Figure 14b shows a clear splitting between reflections in the out-of-plane direction. The film is strained in the out-of-plane direction with a lattice parameter of 3.797 Å and is fully coherent in the in-plane directions with the STO substrate with a lattice parameter of 3.905 Å. This *c* value is somewhat larger than that from RBS channeling and XRD. In bulk form, Howard et al. [28] reported $LaAlO_3$ is cubic at temperatures above ~ 830 K and a rhombohedrally distorted perovskite at lower temperatures. The electron diffraction results suggest a tetragonal $LaAlO_3$ film constrained by the STO substrate. A survey of STEM images in different regions revealed no misfit dislocations in agreement with electron diffraction patterns recorded from both LAO and a part of the STO substrate, which showed no splitting of diffraction spots in the direction normal to in-plane lattice planes. In contrast, analogous STEM images for the Tokyo films, which we do not have permission to publish, exhibit a large density of misfit dislocations at the interface. Based on HAADF images, which are dominated by the Z contrast associated with the La, the extent of interdiffusion appears to be limited to at most ~1 nm.

### 3.3.2 Diffusion lengths by electron energy loss spectroscopy line profiles

EELS line profiles must be measured with care and properly interpreted in order to extract meaningful information about diffusion lengths near interfaces. In general, minimizing: (i) the incident probe diameter, (ii) the cross sectional specimen thickness, and, (iii) the extent of crystallographic misalignment results in minimizing the extent of artificial atom profile broadening. In addition, using lower incident beam voltages help reduce electron beam knock-on damage to the sample. While these are experimental parameters that can be optimized, there are fundamental physical broadening effects that must be considered as well. There are two broadening mechanisms one caused by elastic scattering and the other by inelastic scattering. Elastic scattering causes a spread of the electron probe and degrades the spatial resolution of the inelastic scattering signal, especially in thick samples. Its effect can be reduced by using thin samples. The extent of electron probe spreading can also be checked using the contrast in HAADF STEM images; the electron probe spread increases the background intensity and reduces the image contrast. The extent of beam delocalization increases with: (i) increasing primary beam energy, (ii) decreasing energy loss, and (iii) increasing atomic number at the scattering site [29]. Thus, shallow core-level energy loss processes at high-Z elements will result in more beam delocalization than deep-core level losses at low-Z sites in the same material. For the materials of interest here, the effective EELS delocalization diameter is of the order of 0.4 – 0.6 nm (~1 – 1.5 u.c.) for the La $N_{4,5}$, Al $L_{2,3}$, Ti $L_{2,3}$ and O K edges, and 0.2 – 0.3 nm (~0.5 – 0.7 u.c.) for the Sr $L_3$ and Al K edges using the 50% intensity criterion.

Fig. 15 shows EELS line scans for the Ti $L_{2,3}$ and O K edges from a 25 u.c. Augsburg film. A specimen area close to the zone axis orientation in both the LAO and STO regions was chosen so that both film and substrate were in focus. The line scan is constructed from spectra acquired at intervals of ~1 Å to allow regions on and off atomic columns to be sampled. Atomic scale EELS resolution is evidenced by the alternating contrast in the Ti $L_{2,3}$ spectrum image within the substrate. The small electron probe is also evidenced by the image contrast in HAADF STEM images recorded before and after the EELS acquisition. A line scan across the top of LAO film shows that the HAADF STEM image intensity drops to 10% within a distance of 1 u.c.. Over the course of the 3.5 minute line scan acquisition time, the sample stage drifted downward by ~2 Å, as determined by pre-EELS and post-EELS STEM images and the simultaneously acquired HAADF signals. The horizontal drift is of the same order. This amount of drift is at the low end of the typical range of stage drift (0.5 - 2 Å per minute) for aberration corrected microscopes with a side-entry stage and stable room environment. The long acquisition times we used allowed for good counting statistics on energy-loss near-edge structure (ELNES) with a large collection angle. The Ti $L_{2,3}$ loss feature shows the characteristic splitting of the $t_{2g}$ - $e_g$ orbitals into four peaks, indicating a formal charge state of +4 for Ti throughout the interfacial region.

To measure the extent of Ti outdiffusion into the LAO film, we integrated the intensity under the EELS edges for each spectrum along the line scan. We define the diffusion length to be the distance over which the loss signal drops to ~10% of its maximum value. The Ti signal extends up into the first three unit cells of the LAO film, indicating an upper limit of detectable diffusion over this distance. Some Ti signal is expected in the LAO film from electron probe spread and delocalization by inelastic scattering. But the extents of these two effects are smaller than what is observed. In addition, the spectral image of the Ti $L_{2,3}$ edge shows atomic column centered contrast, which is not expected from delocalization alone. Across the EELS line scan, the simultaneously measured HAADF signal, which is most sensitive to La and Sr atomic columns, was displaced by ~2 Å relative to a HAADF signal from a STEM image line profile taken before the EELS scan, as seen at the bottom of Fig. 15.

Detailed analysis of Ti $L_{2,3}$ and O K EELS fine structure yields an understanding of the interfacial electronic states between the LAO film and STO substrate, as well as a determination of the Ti valence in this region. Fig. 16 shows individual background subtracted Ti $L_{2,3}$ and O K edge spectra taken along a line perpendicular to and cutting through the interface. Consecutive spectra were acquired every 1 Å. The O K edge spectra were smoothed with a 2.0 eV low pass filter in the energy axis to reduce the noise from gain variations of the detector pixels, as well as to show the significant differences between spectra. Within the STO substrate, the Ti $L_{2,3}$ edge shows the characteristic $t_{2g}$-$e_g$ splitting, and indicates a 4+ charge state for Ti. As the probe scans beyond the LAO-STO interface, the Ti signal is still relatively strong two unit cells into the LAO film, and becomes weaker, but is still clearly visible, three unit cells into the LAO film. This result was consistently found in multiple EELS line scans taken in different regions of the sample. Based on measurements made under similar conditions for MBE-grown complex oxide films on STO(001) substrates for which the interfaces are substantially more abrupt than in the present case, we judge that the extent of interface broadening, which we interpret as being due to interdiffusion, exceeds what is expected from signal delocalization within the microscope.

Additionally, there is no shift in the Ti $L_{2,3}$ peak energies, which should be clearly visible if the Ti valence is 3+ or a mixture of 3+ and 4+ near the interface [30]. This result differs from that of Nakagawa et al. [8], who observed a small amount of $Ti^{3+}$ on the STO substrate side of the interface and suggested that charge transfer from the polar LaO layer into the interfacial $TiO_2$ layer had occurred [5]. Recent work by Verbeeck et al. [31] also confirmed the absence of $Ti^{3+}$ at the interface, and estimated the Ti valence near the interface to be 3.8+ or greater. Verbeeck *et al.* did not detect Ti in the LAO. However, these authors used a significantly larger probe size that we have used (~5 Å versus ~1.5 Å) for EELS acquisition and coarser sampling than we did (one spectrum every 3.0 - 7.5 Å versus one spectrum per 1.0 Å). We also used a larger spectrometer collection angle (60 mrad versus 25 mrad), which yields better signal-noise ratio and thus is more sensitive to smaller concentrations of cations. Our finer grained approach and smaller probe size allows us to isolate the electron probe to the interface with improved spatial resolution and more spatially precise sampling.

The O K edge can be used to observe the electronic states at the interface [8, 32]. The first three peaks in the O K EELS are attributed to excitation to Ti 3d (first or pre-edge peak), 4d states of Sr and 5d states of La (second peak), and 4sp states of Ti (third peak) [33-35]. The reference spectra from bulk materials in Nakagawa et al. [8] show that the pre-edge peak is strong in STO films and absent in LAO films [8]. In the present work, the pre-edge peak intensity can be used to correlate diffusion lengths as it is present in STO but absent in LAO. The O K-edge line scan in Fig. 16c shows that the pre-edge feature at ~529 eV is still strong two unit cells into the LAO film, indicating an STO-like electronic perturbation within the LAO. The deepest position within the LAO at which the perturbation occurs is marked by a circle in Fig. 16a and an arrow in Fig. 16c. Fig. 16d and 16e show the corresponding Ti and O spectra at the positions marked by arrows in Fig. 16b & c, and these spectra clearly show four strong peaks in the Ti $L_{2,3}$ edge and a pre-edge peak in the O K edge, consistent with both Sr and Ti being in the first three u.c. of the LAO film. (A reference spectrum of the O K edge deeper into LAO is provided in Figure 16e). If Ti alone diffused to this distance, the Ti valence would be expected to be 3+, and the four white lines in the Ti $L_{2,3}$ edge should merge into two lines. Moreover, the white line positions should exhibit a shift if $Ti^{3+}$ is present. However, the presence of $Sr_xLa_{1-x}Ti_yAl_{1-y}O_3$ on the LAO side of the interface can account for the observed spectral lineshapes.

The second peak can also be used to corroborate Sr diffusion into the LAO. Since the Sr $L_{2,3}$ edge has a low cross section, fingerprinting with the O K edge is more useful in determining the extent of intermixing. There are no fingerprints of intermixed $SrTiO_3$ and $LaAlO_3$ in the literature at the time this Report was written, so we used closely matched compounds. X-ray absorption spectroscopy (XAS) studies of Sr-doped $LaTiO_3$ [36] as well as Sr-doped $LaFeO_3$ and $LaMnO_3$ [37] show the second O K-edge loss peak shifts toward higher energy loss as the Sr doping level is increased. This shift is attributed to changes in the conduction band as Sr 4d states replace La 5d states [37]. The valley between the second and third peaks also shifts to higher loss energy as the Sr concentration increases. The dotted lines in Figs. 16a & c show this shift from the last STO layer to the first two unit cells of the LAO film. The fact that the shifts in the second peak and valley are gradual across the interface indicates the Sr concentration decreases with distance into the first ~2 unit cells of LAO, similar to the Ti.

Fig. 17 shows EELS line scans for the Al $L_{2,3}$ and La $N_{4,5}$ edges at ~90 eV and ~120 eV, respectively. The La signal starts to decrease within the LAO film at position 3.0 nm as the interface is approached along the line and we judge the LAO-STO interface to be at position of 7.2 nm from the HAADF image contrast. Thus, the La concentration starts to decrease 4.2 nm, or approximately 11 unit cells, from the interface. The integrated signal drops to 10% of its maximum value at position 9.0 nm, which is ~4 u.c into the STO. A portion of the signal drop is expected from electron probe delocalization. To compare the apparent La movement from diffusion and delocalization distances, it is helpful to take a derivative of the EELS line profile. The La $N_{4,5}$ integrated intensity was first smoothed with a 10-pass Gaussian filter in order to reduce the effective changes in La concentration due to on-column versus off-column sampling. Figure 17 shows the resultant derivative of the smoothed integrated intensity overlapped with the unprocessed integrated intensity. The derivative shows that the rate of decay of the La signal is asymmetric about the interface position at 7.2 nm. A symmetric derivative would be expected for a step function for La concentration convoluted with a delocalized electron probe. This result clearly shows that the La concentration is not atomically abrupt, that La diffusions well into the substrate, and that several u.c. of LAO immediately adjacent to the interface are partially depleted of La. The Al $L_{2,3}$ edge signal also begins to drop within the LAO film a few u.c. before the interface, and does not appear to drop to zero until a distance of more than 6 u.c. within the STO. However, there is likely interference from the nearby broad edges of Sr $M_{4,5}$ and possibly the Ti $M_1$ edges within the EELS spectrum in STO. Visual inspection of the white line intensity in the Al spectral image shows that the intensity appears to approach a minimum value very near the LAO-STO interface.

In order to avoid overlap with substrate EELS features, the Al K edge at ~1575 eV was also acquired, and these data are shown in Fig. 18. Although this loss feature is relatively weak, it is more localized than the Al $L_{2,3}$ edge because Al is a light element and the K-edge loss energy is high, resulting in reduced broadening at the K edge relative to that at the Al L edge, as discussed above. This line scan shows that Al diffuses into the STO substrate over a distance of 2 u.c., or ~8 Å. Fig. 18 also shows an EELS line scan for the Sr $L_{2,3}$ edge at 1945 eV. As with the Al K edge, this loss feature is more localized due to the high energy loss. Sr diffuses into the LAO film over a distance of ~4 u.c. The sensitivity to concentration for Al and Sr is low at these higher loss energies because the loss features are weak relative to the background from multiple scattering. Additionally, noise signals from the detector are moderate in this energy range. The combination of these factors makes background subtraction and integrated intensities somewhat problematic. Therefore, the diffusion lengths determined from the Al K and Sr $L_3$ edge intensities may be underestimated.

In summary, the three techniques brought to bear on the 25 u.c. films paint a consistent picture of rather more extensive intermixing at the interface than has been previously thought. All three techniques have inherent strengths and limitations. RBS is exquisitely sensitive to La and Sr, but is much less sensitive to Ti and Al. Moreover, RBS lacks depth and spatial resolution. ToF-SIMS depth profiling is highly sensitive to all elements, but is destructive to the sample, resulting in potential complications due to knock-on effects when looking at elements found in the film. Complications due to knock on do not occur for substrate elements that have diffused out into the film. ToF-SIMS also lacks spatial resolution in plane, but has reasonable depth resolution due to the low penetration depth of the analyzing ion. HAADF STEM/EELS has the best spatial resolution by far, but lacks elemental sensitivity for Al and Sr. In light of the fact that each method has at least one substantial weakness, it is risky to draw firm conclusions about interface composition from any one method. However, taken together, data from these three methods support the conclusion that all four elements (La, Al, Sr and Ti) diffuse at least a few u.c. into the oxide on the opposite side of the interface, giving rise to what amounts to a complex quaternary interface with non-negligible concentration gradients normal to the interface. Moreover, there is clear evidence that La diffuses deeply into the substrate, and this result has important ramifications for electronic structure, as discussed in Section 5.

## 4. Film and interface composition – 4 unit cell LAO films

We now consider 4 u.c. films of LAO on STO(001). We employ experimental tools which have the required depth sensitivity to be able to probe ultra-thin films, as well as classical and quantum mechanical theory, applied to slabs of dimensions consisting of a few u.c.of both materials.

### 4.1 Angle-resolved x-ray photoelectron spectroscopy

Another way to obtain elemental abundances as a function of depth, but without damaging the sample, is to measure core-level photoelectron yields as a function of polar or take-off angle, $\theta_t$, relative to the surface plane [38-39]. This method works for the intended purpose by virtue of the fact that the probe depth goes as the photoelectron attenuation length, $\lambda$, which can be defined as the depth of an emitter at which the photoelectron yield is reduced to 1/e of the value for the same emitter at the surface, multiplied by $\sin\theta_t$. Thus, for a given value of $\theta_t$, ~95% of the signal originates within a depth of $\sim 3\lambda \sin\theta_t$, and this quantity is a convenient way to quantify the probe depth. As long as the probe depth exceeds the film thickness ($d$) over at least some range of $\theta_t$, the photoelectron yields will be sensitive to the distribution of atoms in the interfacial region. The physical origin of photoelectron attenuation is inelastic scattering, whereby photoelectrons lose energy due to electronic excitations during propagation from the emitter site to the surface. This process is analogous to EELS associated with STEM (Section 3.3). These excitations can be associated with either inter-band or core-to-conduction band transitions. The former can occur throughout the solid whereas the latter are localized at atomic positions. Inelastic attenuation can be treated to first approximation in terms of an isotropic damping of the outgoing photoelectron wave amplitude in which the distance over which the damping occurs is characterized by the attenuation length $\lambda$. Photoelectron intensities are also sensitive to the positions of atoms close to the photoelectron emitter via elastic scattering, which can be thought of classically as a change in electron trajectory brought about by Coulomb interaction with ion cores, again analogous elastic scattering in STEM. If neighboring atoms are arranged in a periodic lattice, there are marked intensity variations with angle due to elastic scattering and interference, otherwise known as photoelectron diffraction (PED) [38-40]. A fully quantitative treatment of elastic scattering effects requires a full quantum mechanical multiple scattering formalism [41]. Here outgoing spherical photoelectron waves, modulated by the differential photoelectric cross section appropriate for the core state in question and attenuated by a damping factor of the form $\exp(-r/\lambda)$, are emitted from each lattice site containing the atom in question and are allowed to scatter some number of times (up to 10 for full convergence) from every atom within a few $\lambda$ of the emitter. For any given atomic arrangement, this calculation is quite CPU intensive, but yields useful information on atomic positions with accuracies on the same order as those available from low-energy electron diffraction (LEED) I-V analysis and surface extended x-ray absorption fine structure (SEXAFS). However, when the distribution of elements across the interface itself is also a variable, the calculations become prohibitively expensive as the number of configurations that must be modeled can number in the tens of thousands for clusters of adequate size to represent multilayer systems.

In the case of LAO/STO, PED effects come into play because LAO/STO is an epitaxial system. As has been shown previously, PED is a very useful way to determine structure in ultrathin epitaxial films [42] as well as indiffusion of ordered monolayers of Group IV and Group VI elements on III-V compound semiconductor surfaces [43-46]. In these papers, single scattering theory, an approximation to the more exact multiple scattering approach, was used to interpret experimental results and determine the extent of adatom indiffusion. However, for more complex interfaces such as LAO/STO in which all four cations can in principle undergo site exchange, even single scattering calculations are prohibitively time consuming due to the sheer number of atomic arrangements that are possible and must be modeled. Alternatively, the diffraction modulation can be dealt with by the use of LAO and STO single crystal standards. In this approach, the polar intensity profile, $I(\theta_t)$, from LAO/STO(001) for a given azimuthal orientation and core-level photoelectron should show similar intensity modulations with angle as do analogous scans for bulk or bulk-like LAO(001) and STO(001). Moreover, bulk-like LAO(001) and STO(001) should also serve as useful standards for atomic concentrations in the different layers below the surface for stoichiometric LAO and STO, provided the surface terminations are the same in the standards as for the heterojunctions under study. This last condition must be met because changing surface termination (i.e. AO → $BO_2$, or vice versa) has a measurable effect on the core-level intensities from A and B; the attenuation length is a few to several times the interlayer spacing, depending on the photoelectron kinetic energy, resulting in nonnegligible changes in photoelectron yields from the A and B sites as the surface termination is changed.

All ARXPS measurements were carried out using a Gamma Data/Scienta SES 200 analyzer and a monochromatic AlKα x-ray source ($h\nu$ = 1487 eV). X-rays resulting from 500W of electrical power were focused into an area of dimensions ~9 mm x ~1 mm on the sample surface. The angular acceptance of the analyzer was ±7$^o$ in both polar and azimuthal directions. Polar scans were measured in the (100) azimuthal plane. A schematic of the experiment is seen in Fig. 19a.

All specimens were cleaned on the bench using a standard UV/ozone treatment as well as in the load lock using activated oxygen from an electron cyclotron resonance oxygen plasma source. The surfaces were found to be free of adventitious carbon. Atmospheric exposure resulted in hydroxylation of the top layer of $AlO_2$ of all films, as judged by the presence of a second O 1s peak ~2 eV to higher binding energy from the lattice O peak characteristic of OH. This feature was greatly enhanced by going to low take-off angles. Fig. 19b shows O1s spectra for a typical 4 u.c. film measured over a range of take-off angles. Significant enhancement (suppression) of the OH (lattice) peak at low angles is clearly seen, as qualitatively expected for surface bound OH. Fig. 19c shows the integrated areas for the two spectral features as a function of take-off angle. Fig. 19d shows the ratio of OH to lattice O intensity vs. take-off angle, along with the results of a simple inelastic attenuation calculation (discussed below) in which it was assumed that each undercoordinated Al on the surface chemisorbs an OH from adventitious $H_2O$, and the associated H binds to an undercoordinated surface O. The good level of agreement between experiment and this simple model, at least for $\theta_t \geq$ ~20$^o$, confirms that the OH is surface bound.

The analytical approach we have taken to determining interface stoichiometry is to compare Sr3d, Ti 2p, La 4d and Al 2p polar scans measured for 4 u.c. LAO/STO(001) samples with those measured for $TiO_2$-terminated STO(001) bulk crystals and 25 u.c. LAO/STO(001) samples grown in the same laboratory as the 4 u.c. specimen. These four core levels were chosen because they exhibit good intensities and have fairly similar kinetic energies, leading to similar attenuation lengths. As noted in section 3.1, 25 u.c. films from Augsburg and Tokyo are somewhat off stoichiometry in terms of their La/Al atom ratios, and they are off stoichiometry in opposite directions. Therefore, we can use the thicker films from each lab as standards in determining the compositions of the thinner films. 4 u.c. and 25 u.c. LAO films from both labs are expected to be $AlO_2$-terminated because all growths were stopped after an integral number of RHEED intensity oscillations, and the films were grown on $TiO_2$-termianted STO substrates, leading to initial nucleation of a LaO layer. Likewise, the STO substrates used as reference samples were chemically etched and $O_2$ tube furnace annealed to result in a $TiO_2$ termination. Thus, our standards have the same surface and interface terminations as the 4 u.c films, allowing direct and meaningful comparison.

We have used simple expressions for the polar intensity profiles based on isotropic inelastic scattering of the various photoelectrons. PED effects have been ignored in our theoretical treatment for simplicity and ease of analysis. For a 4 u.c. film of thickness $d$ with the same stoichiometry as that of the 25 u.c. standard that makes an abrupt interface with STO, the Sr 3d and Ti 2p polar intensity profiles, $I_{4uc}(\theta_t)$, should be related to those for bulk STO, $I_{STO}(\theta_t)$, by the expression $I_{4uc}(\theta_t) = I_{STO}(\theta_t)\exp(-d/\lambda\sin\theta_t)$. If $d$ is accurately known, the quantity $[d\sin\theta_t]/\{\ln[I_{STO}(\theta_t)] - \ln[I_{4uc}(\theta_t)]\}$ should be constant for all take-off angles and equal to $\lambda$ [47-48]. As discussed in section 2, these films grew in an essentially layer-by-layer fashion, as verified by the observation of RHEED intensity oscillations during growth, and precisely 4 u.c. of LAO was grown in each case. The films were quite flat, as judged by AFM images which exhibited a clear terrace-step structure with no obvious roughness on the terraces. Additionally, high-resolution x-ray diffraction (XRD) verified that the 25 u.c. LAO films were coherently strained to the substrate, with $c$ values ranging from 0.372 to 0.374 nm. Assuming the same value of $c$ for the 4 u.c films, $d$ is 1.50 nm. Plots of $[d\sin\theta_t]/\{\ln[I_{STO}(\theta_t)] - \ln[I_{4uc}(\theta_t)]\}$ vs. $\theta_t$ are shown in Fig. 20 for Sr 3d and Ti 2p core-level intensities for both specimens. At higher angles ($\theta_t \geq \sim40^o$), the apparent values of $\lambda$ are reasonable for the associated photoelectron kinetic energies, ~2.0 nm for both Ti 2p and Sr 3d [49]. However, there is a marked divergence at low angles in which $\lambda$ reaches ~11 nm and ~9 nm for Sr 3*d* and Ti 2*p*, respectively, at $\theta_t = 5^o$ for the Augsburg film, and ~25 nm and ~17 nm, respectively, for the Tokyo film. These values of $\lambda$ are physically unreasonable, and their extraction from the formula above suggests that the interface is *not* abrupt, but rather that Sr and Ti diffuse out into the film, resulting in closer proximity to the surface. Moreover, based on these plots, the Tokyo film exhibits more Sr and Ti outdiffusion than the Augsburg film.

The actual attenuation lengths, which are needed for further analysis, are most likely less than 2 nm for both Ti 2p and Sr 3d because the "interface" is closer to the surface than expected for an atomically abrupt system due to Sr and Ti outdiffusion. Moreover, based on their respective kinetic energies ($E_k$), $\lambda$ for Ti 2p should be somewhat lower than $\lambda$ for Sr 3d, La 4d and Al 2p. Sr 3d, La 4d and Al 2p all have $E_k$ values of ~1400 eV when excited by AlKα x-rays whereas Ti 2p has an $E_k$ of ~1000 eV for the same x-ray excitation source. Thus, Ti 2p should have a $\lambda$ that is ~0.2 nm less than those for Sr 3d, La 4d and Al 2p, based on the $E^{1/2}$ dependence of $\lambda$ above ~150 eV [50].

That Sr and Ti outdiffusion into the 4 u.c. films occurs is also evident from the Sr 3d and Ti 2p polar scans which are shown in the top panels of Figs. 21&22. Here we compare the experimental data with scans measured using bulk STO(001) scaled by a factor of $\exp(-d/\lambda\sin\theta_t)$. A range of $\lambda$ values has been used in each case (1.5 – 2.0 nm for Sr 3d and 1.3 – 1.8 nm for Ti 2p). We know from the data summarized in Fig. 20 that $\lambda$ must be less than ~2.0 nm for all core levels, as discussed above. Therefore, the $\lambda$ ranges we have chosen are reasonable. If the 4 u.c. interfaces were atomically sharp, the polar scans (marked as diamonds) should overlap the band of $I_{STO}(\theta_t)\exp(-d/\lambda\sin\theta_t)$ plots (solid curves connected by hatch) for all angles. While there is some overlap at higher angles, at least for the Augsburg film, the actual 4 u.c. intensities are clearly greater than those of $I_{STO}(\theta_t)\exp(-d/\lambda\sin\theta_t)$ at lower angles. Indeed, $I_{STO}(\theta_t)\exp(-d/\lambda\sin\theta_t)$ goes to zero at $\theta_t \leq$ ~10$^{o}$ for both Sr 3d and Ti 2p because the path length through the film exceeds the probe depth at low angles (e.g. $\exp(-d/\lambda\sin\theta_t) \rightarrow 0$). Yet, there are clearly Sr 3d and Ti 2p yields at very low angles, as seen by the spectra at $\theta_t = 5^{o}$ which are shown as insets. This result is consistent with Sr and Ti outdiffusion. In fact, the presence of Sr 3d and Ti 2p intensity at $\theta_t = 5^{o}$ suggests that these elements are resident within the top u.c. of LAO as minority species. Pinholes (if present) in the 4 u.c. films cannot account for the persistent Ti 2p and Sr 3d yields at low angles because photoelectrons from the substrate would be blocked by the steep walls surrounding the pinholes. The extent of Sr and Ti outdiffusion appears to be more extensive for the Tokyo sample, as judged by the fact that the 4 u.c. experimental scans exceed $I_{STO}(\theta_t)\exp(-d/\lambda\sin\theta_t)$ to a greater extent than those for the Augsburg sample.

The diffraction modulation seen in the 4 u.c. films is similar, but not identical, to that seen in $I_{STO}(\theta_t)\exp(-d/\lambda\sin\theta_t)$ for both Sr 3d and Ti 2p. Most notable is the higher intensity in $I_{STO}(\theta_t)\exp(-d/\lambda\sin\theta_t)$ in the Sr 3d polar scans near $\theta_t = 45^o$. In the (100) azimuth, this take-off angle corresponds to the [011] low-index direction, which consists of chains of alternating Sr and O atoms with a Sr – O spacing of $a/\sqrt{2} = 0.276$ nm. This relatively low value of inter-atomic spacing results in zeroth-order forward focusing which universally leads to intensity maxima along low-index directions in single-crystal specimens. The same phenomenon has been observed and theoretically accounted in bulk and epitaxial MgO(001) [51-53]. Forward focusing is not as strong along [011] for Ti 2p because this portion of the unit cell consists of chains of Ti atoms separated by $\sqrt{2}\,a = 0.553$ nm, and the forward focusing effect falls off with increasing distance. Strong forward focusing is also seen in $I_{STO}(\theta_t)\exp(-d/\lambda\sin\theta_t)$ for both Ti 2p and Sr 3d along the [001] direction at $\theta_t = 90^o$. This direction consists of chains of Sr atoms of spacing $a =$ 0.391 nm and chains of alternating Ti and O atoms separated by a/2 =0.196 nm. The analogous forward-focusing effects are much weaker in the 4 u.c. films because of multiple-scattering-induced defocusing [54-55]. Here (nonemitting) A- and B-site atoms in the film defocus photoelectrons generated below the buried interface. Photoelectrons emitted in the substrate are first forward focused via small-angle elastic scattering by atoms in the outgoing path immediately adjacent to the emitter, leading to intensity enhancements along low-index directions. These same electrons are then defocused by overlayer atoms in the film material along the exit path, the result being a loss of intensity along the low-index direction which grows with increasing film thickness. A similar phenomenon has been observed for the epitaxial Ge/GaAs(001) system [42, 48].

Also shown in Figs. 21&22 (bottom panels) are polar scans in (100) of La 4d and Al 2p intensities for the 4 u.c. films (diamonds), along with 25 u.c. film polar scans, designated as $I_{LAO}(\theta_t)$, scaled by a factor of $(1 - \exp(-d/\lambda\sin\theta_t))$ (solid curves connected by hatch). An atomically abrupt 4 u.c. film with the same composition as the 25 u.c. film should exhibit nearly the same core-level intensities for all angles as $I_{LAO}(\theta_t)[1 - \exp(-d/\lambda\sin\theta_t)]$. However, the 4 u.c. La 4d and Al 2p peak areas are considerably lower in value than those for $I_{LAO}(\theta_t)[1 - \exp(-d/\lambda\sin\theta_t)]$ for all angles for both films, but more so for the Tokyo film. The same range of attenuation lengths was used for La 4d and Al 2p as for Sr 3d, because these three photoelectrons have nearly the same kinetic energy. These results suggest that La and Al indiffusion into the substrate occurs, and that the extent of diffusion is slightly greater for the Tokyo film. The diffraction modulation is largely the same in the 4 u.c. and scaled 25 u.c. films, as expected because of the epitaxial relationship between LAO and STO. In contrast to the Ti 2p and Sr 3d results, there is no multiple-scattering-induced defocusing in the 4 u.c. film because La 4d and Al 2p photoelectrons are generated all the way to the surface.

The extent of intermixing can be estimated by modeling the ARXPS data using the simple inelastic attenuation model described above. However, the PED intensity modulation must first be removed to the greatest possible extent. We do so by dividing the 4 u.c. polar scans by the appropriately scaled bulk or bulk-like polar scans. We then compare these data to inelastic attenuation simulations in which the elemental compositions of the different A- and B-site layers were systematically varied. In this approach, the normalized intensities from atoms in the $n_{\mathrm{th}}$ A- and B-site layers from the surface are given by,

$$A(\theta_t, n) = \exp\left[\frac{-z_n}{\lambda \sin\theta_t}\right] \quad (3)$$

and

$$B(\theta_t, n) = \exp\left[\frac{-z_n}{\lambda \sin\theta_t}\right] \quad (4)$$

where $z_n$ is the depth of the $n_{\mathrm{th}}$ layer in each sublattice. If the fractional occupancies of the four atoms in the $n_{\mathrm{th}}$ layer are denoted by $\{p_n\}$, the four core-level intensities for a given atomic configuration can be written as,

$$I_{Sr3d}(\theta_t) = \sum_j p_j^{Sr} A(\theta_t, j) \quad (5)$$

$$I_{Ti2p}(\theta_t) = \sum_j p_j^{Ti} B(\theta_t, j) \quad (6)$$

$$I_{La4d}(\theta_t) = \sum_j p_j^{La} A(\theta_t, j) \quad (7)$$

$$I_{Al2p}(\theta_t) = \sum_j p_j^{Al} B(\theta_t, j) \quad (8)$$

The sums are over a sufficient number of layers to reach convergence. The ratio of the intensities given in eqns. 5-8 for various atom profiles with the analogous intensities for the abrupt interface were then compared to their experimental counterparts, $I_{4uc}(\theta_t)/I_{STO}(\theta_t)\exp(-d/\lambda\sin\theta_t)$ for the substrate peaks, and $I_{4uc}(\theta_t)/I_{LAO}(\theta_t)[1 - \exp(-d/\lambda\sin\theta_t)]$ for the film peaks. The results of this analysis are summarized in Figs. 23&24. Inspection of this figure reveals that taking the ratio of $I_{4uc}(\theta_t)$ and $I_{STO}(\theta_t)\exp(-d/\lambda\sin\theta_t)$ does not completely eliminate the diffraction modulation. Although the intensity maxima and minima occur at the same angles for a give core level, the magnitudes are not the same due to differences in thicknesses and substrate defocusing effects, as discussed above. Therefore, we cannot get a highly precise fit between the model and experiment. Additionally, a specific value of $\lambda$ must be chosen to carry out the simulations, and we have used 1.5 nm for La 3d, Al 2p and Sr 3p and 1.3 nm for Ti 2p. However, the uncertainty in $\lambda$ propagates uncertainty into the atom profiles. Finally, because of depth resolution issues (discussed below), a unique fit is not forthcoming from this analysis. As a result, the atom profiles extracted from the ARXPS data by this method are not of highly quantitative value. However, inasmuch as the ratio $I_{4uc}(\theta_t)/I_{LAO}(\theta_t)[1 - \exp(-d/\lambda\sin\theta_t)]$ would be near unity for all $\theta_t$ if the interface was perfectly abrupt, this analysis makes it clear that extensive intermixing occurs.

The solid curves in Figs. 23c & 24c are the atom profiles that give rise to the simulated intensity ratios in Figs. 23ab & 24ab, and the analogous abrupt interface atom profiles are shown as dotted lines. A clear "smearing" of the interface is evident for all four elements. It should be noted that in this model, the ratio was taken as "abrupt to intermixed", rather than the reciprocal, in order to avoid singularities in the Sr 3d and Ti 2p polar scan intensity ratios for abrupt interfaces (either scaled bulk scans or as computed using eqn. 5-6) at low angles. Therefore, some information from the low-angle data is lost, since the abrupt-to-intermixed ratio goes to zero. Nevertheless, it is clear from this analysis that intermixing can account for the ARXPS data in a satisfactory way, and that the data are incompatible with an abrupt interface model, which would yield a near unity ratio for all angles. Depth resolution comes into play in the following way. The probe depth goes as $\lambda\sin\theta_t$, and the sensitivity to individual atomic planes drops exponentially with the depth of the plane. At the lowest angles, the signal arises from the top u.c. at the surface, and thus the depth resolution is ~0.4 nm at $\theta_t = 5^o$. However, at higher angles, more layers are sampled, and the measured signal contains contributions from many layers, precluding our ability to selectively probe any given layer. Thus, we have the equivalent of one equation in several unknowns at most angles, and a unique solution is not achievable.

Finally, we consider the effect of nonidealities other than interfacial intermixing to see if the ARXPS data can be accounted for in some other way. Pinholes in the film associated with incomplete layer formation cannot account for the present results because such pinholes would be perpendicular to the interface and have large depth-to-diameter aspect ratios. Such defects cannot constitute viable exit paths for photoelectron escape except near $\theta_t = 90^o$. Interface roughness may occur if there is a high step density on the STO substrate. In this scenario there may be a significant variation in the layer thickness ($d$) across the area over which photoelectrons are collected. We have simulated this effect, and the results are shown in Fig. 25. Using the Augsburg 4 u.c. film to illustrate the result, we compare the 4 u.c. data (diamonds) with $I_{STO}(\theta_t)\exp(-d/\lambda\sin\theta_t)$ and $I_{LAO}(\theta_t)[1 - \exp(-d/\lambda\sin\theta_t)]$ for simulated interfaces consisting of a uniform 4 u.c. film, as well as equal mixtures of 4 and 3 u.c., and 4 and 2 u.c.. It is clear that if good agreement is reached for a uniform 4 u.c. thickness for Sr 3d and Ti 2p for a given choice in $\lambda$, the agreement is quite poor for La 4d and Al 2p. Conversely, if good agreement is reached for a mixture of, for example, 4 u.c. and 2 u.c. for La 4d and Al 2p for a given choice in $\lambda$, agreement is quite poor for Sr 3d and Ti 2p. In other words, it is not possible to achieve good agreement *simultaneously* for substrate *and* overlayer elements for a given extent of interfacial roughness and a given choice of $\lambda$. Therefore, interfacial roughness cannot account for the complete experimental data set. Likewise, it is not possible to achieve simultaneous agreement on all four core levels assuming a perfectly abrupt and flat interface using the *same* value of $\lambda$ for La 4d, Al 2p and Sr 3d, and a slightly lower value for Ti 2p. Yet, the first three of these core level photoelectrons should exhibit the same $\lambda$ through a given material because they have nearly the same kinetic energy. Using the Augsburg sample again for illustration, while reasonably good agreement can be achieved at higher angles for the Sr 3d scan at $\lambda = 2.0$ nm, $\lambda$ values of 3.0 nm and 2.6 nm are required to achieve good high-angle agreement for the La 4d and Al 2p scans, respectively. Moreover, for any given core level, there is no single value of $\lambda$ that gives good agreement over the entire range of angles, as discussed above, ruling out uncertainty in $\lambda$ as the cause for the poor level of agreement. Therefore, strong intermixing is the best and, indeed, the only way to satisfactorily account for the ARXPS data. Although the atom profiles extracted from the data are not highly accurate, intermixing is sufficiently extensive that the basic conclusion is unavoidable. As we show in section 4.2, the conclusions drawn from ARXPS are strongly corroborated and made more quantitative by MEIS.

### 4.2 Medium energy ion scattering

MEIS [56] is a low-energy, high-resolution version of Rutherford backscattering. The excellent depth resolution of MEIS is due to the use of a high-energy-resolution toroidal electrostatic analyzer [57] and the fact that the electronic stopping power of the probe ions is at its maximum for $H^+$, and is close to its maximum for $He^+$, in the MEIS energy regime (100 – 200 keV) [58]. As a result, the depth resolution in the present work is ~3Å in the near-surface region, which corresponds to ~1 u.c of LAO. Quantitative depth profile information can be extracted from the energy distributions of these backscattered ions. Knowledge of the scattering cross section, ion stopping power, and extent of energy straggling allows one to extract elemental depth profiles from MEIS data assuming the density of the film is known or can be calculated.

Two scattering geometries were utilized - channeling and random. In the channeling geometry, the ion beam was incident along the surface normal, [001], and the detector was aligned along the [111] direction of the STO(001) substrate, resulting in significant suppression of scattering in the substrate [59]. This geometry is very useful for determining composition in the outermost LAO unit cell. In the random geometry, the incident ion beam is not aligned with any major crystallographic axis, resulting in much less shadowing and, therefore, more irradiation of atoms in deeper layers. This geometry allows atom profiles to be determined at the buried LAO/STO interface.

We show in Fig. 26 random and channeling MEIS spectra using 198.6 keV $He^+$ ions, along with simulations, for a 4 u.c. Tokyo film. The spectra were measured using a low incident ion dose (~1 × $10^{16}$/cm$^2$ per beam spot), and the incident ion beam was rastered in order to minimize beam damage. The two simulations in the random spectrum (Fig. 26a) are for; (i) a fully stoichmetric 4 u.c. LAO film that makes an atomically abrupt interface with STO, and, (ii) an intermixed interface in which optimal agreement with experiment was sought by varying the atom profiles. The abrupt interface simulation clearly does not match experiment. The simulated La peak intensity far exceeds experiment and, as in the case of RBS for the 25 u.c. films, the counts in the valley between the La and Sr peaks are not accounted for. In contrast, the optimized intermixed interface simulation results in much better agreement with experiment. The areal density for La in a stoichiometric 4 u.c. LAO film with an abrupt interface to the substrate is 2.75 × $10^{15}$ at/cm$^2$. However, the La areal density associated with the intermixed simulation in Fig. 26a within the film proper is 1.77 × $10^{15}$ at/cm$^2$. Moreover, the La areal density within the first 1 nm (~2.5 u.c.) of the STO is 2.6 × $10^{14}$at/cm$^2$ , giving rise to a total areal density of 2.03 × $10^{15}$at/cm$^2$ detected in the LAO film and the STO immediately adjacent to the interface. Thus, ~26% of the La is not accounted for and has presumably diffused to deeper levels within the STO, corroborating the RBS results for the 25 u.c. films. The simulation also yields Sr and Ti areal densities within the LAO film proper of 3.9 × $10^{14}$ at/cm$^2$ and 4.0 × $10^{14}$ at/cm$^2$, respectively. The combination of indiffused La and outdiffused Sr account for the counts in the valley between the Sr and La peaks.

These data were taken from a film grown with a relatively low laser energy density on the LAO single crystal target of 0.7 J/cm$^2$. However, similar results were obtained for films grown using intermediate (1.1 J/cm$^2$) and high (1.6 J/cm$^2$) energy densities, as seen in Table 5. There is an increase in the extent of Sr outdiffusion with increasing laser fluence. However, the other atom profiles are nearly the same, suggesting that the intermixing process is not strongly affected by laser fluence.

The channeling spectrum for a typical 4 u.c. Tokyo film is shown in Fig. 26b. With this double alignment of shadowing and blocking of the ion beam, we effectively minimize scattering from the substrate. The simulation in Fig. 26b is for an abrupt interface. In contrast to case of the random spectrum, the abrupt simulation reproduces the La peak intensity more satisfactorily, indicating that the top u.c. of LAO is closer to being stoichiometric in La than the lower three u.c. of LAO. However, the Sr and Ti backscattering peaks in experiment occur at higher ion energies than those from the simulation, indicating closer proximity to the surface. Indeed, the experimental energies for Sr and Ti backscattering reveal that these species are present within the first u.c. of LAO at the surface.

We show in Fig. 27 random and channeling MEIS spectra using 99 keV $H^+$ ions, along with simulations, for a typical 4 u.c. Augsburg film. Again, the abrupt interface simulation does not match experimental at all well, whereas the optimized intermixed simulation matches very well. The simulated La peak is for too intense for the abrupt model, and the counts in the valley between the La and Sr peaks are not accounted for. In contrast, the intermixed simulation reproduces both of these. The La areal densities associated with the intermixed simulation are $1.64 \times 10^{15}$ at/cm$^2$ within the LAO film, and $1.8 \times 10^{14}$ at/cm$^2$ in the top 1.5 nm of STO, and are qualitative similar to the those for the Tokyo film. The Sr and Ti areal densities within the film are $4.7 \times 10^{14}$ at/cm$^2$ and $4 \times 10^{14}$ at/cm$^2$, respectively, which are again comparable to the Tokyo film results. The channeling spectrum (Fig. 27b) also exhibits Sr and Ti peaks at higher ion energies than predicted for an abrupt interface model, and indeed these energies are consistent with some Sr rand Ti being in the top u.c. of LAO.

Additional La-Sr intermixing is promoted at room temperature by increasing the ion beam dose. Fig. 28 shows the effect of prolonged exposure to a collimated beam (not rastered) along the [001] direction in a 4 u.c Augsburg film. Under these irradiation conditions, more than 99% of the proton energy loss in the near-surface region is in the form of electronic excitation and ionization of target atoms. From Fig. 28, we see drop in the La peak, increases in the Ti and Sr peaks, and a quantitative transfer of intensity from the La and Sr peaks to the intervening valley with increasing dose. These results also attest to the relative instability of the interface, and may have implications for the role of energetic ions in the laser plume for driving intermixing.

### 4.3 First principles calculations

#### 4.3.1. Methods, models and notations

Geometrical structures and relative energies of various model interfaces were calculated using the periodic model and several computational methods. We have used a periodic slab model (see Fig. 29a), in which the slab is repeated in the direction perpendicular to the interface (z axis), and each slab is separated from its periodic images by a vacuum gap. For the abrupt configuration, these slabs contain only one atomically flat $TiO_2$/LaO interface. For comparison, the same interface structures have been modeled using a periodic "sandwich" LAO/STO/LAO slab [60], containing two mirror-equivalent $TiO_2$/LaO interfaces (Fig. 29b). In all cases the vacuum gap between the slabs was at least 20 Å wide, depending on the thickness of the LAO film.

The electronic and geometrical structures, as well as the relative energies of the abrupt and several intermixed configurations, were calculated using the Perdew-Burke-Ernzerhof (PBE) density functional [61], a plane-wave basis set, and the projected augmented waves method [62-63] implemented in the Vienna *ab initio* Simulation Package (VASP) [64]. In all cases the in-plane lattice parameter was fixed at the value computed for bulk STO using the PBE functional (3.9447 Å). The total energy was minimized with respect to the internal coordinates of all atoms. For analysis of the electronic structure, the charge density was decomposed over atom-centered spherical harmonics. For comparison, the relative energies of a selected set of interface configurations were calculated using a hybrid B3LYP functional, which combines a three-parameter Becke exchange functional [65] with the correlation functional by Lee, Young, and Parr [66], as implemented in the CRYSTAL code [67] with a Gaussian-type basis set [68].

Finally, the relative energies of diffuse interfaces with different extent of cation intermixing were calculated using the classical shell model [69] and interatomic potentials developed for complex and mixed oxides [70]. In this set of potentials, parametrization of the short-range interaction between various metal ions and an oxide ion is made consistent with the same parameterization of the short-range interaction between oxide ions.

Several slab super cells with lateral sizes ranging from $\sqrt{2}a\times\sqrt{2}a$ to $2\sqrt{2}a\times2\sqrt{2}a$, where $a$ is the STO bulk lattice constant, and k-grids containing up to 34 irreducible k-points were used in the calculations.

We consider specific interface configurations in which metal atoms in the STO and the LAO film intermix. To classify the various configurations, we assign a number to each atomic plane of the slab according to its position relative to the interface, as illustrated in Fig. 29a&b. For example, $Sr_1$ and $La_1$ refer to Sr and La atoms nearest to the interface plane, and $La_1 \Leftrightarrow Sr_1$ denotes the lattice site exchange for these ions, as shown in Fig. 29c.

It is convenient to represent the STO/LAO structures using the values of the formal ionic charges per formula unit in each atomic plane of the unit cell. The $TiO_2$ and SrO layers are thus represented as "0", and the $AlO_2$ and LaO planes are represented as "– " and "+", respectively, as shown in Fig. 29d. We also show in Fig. 29d representations of intermixed configurations with $La_1 \Leftrightarrow Sr_1$ and $Ti_1 \Leftrightarrow Al_2$ site exchange.

### 4.3.2 The ideal $TiO_2$/LaO interface

Our idealized reference system is based on a structurally perfect and atomically abrupt $TiO_2$/LaO interface consisted of a slab of six STO unit cells and between one and four LAO unit cells (Fig. 30a). The $\sqrt{2}\times\sqrt{2}$ lateral cell dimension was used and the size of the supercell along z-axis was set to 64 Å, resulting in a vacuum gap between the periodically repeated slabs of at least 20 Å. The total energy was minimized with respect to internal coordinates of all atoms. The electronic structure of this interface, calculated using the PBE density functional, is similar to what has been reported previously by other groups [60, 71-77]. We show the total one-electron density of states (DOS) calculated for 1 and 3 u.c. LAO in Fig. 30b in such a way the core levels fall at the same energies for the two LAO thicknesses. The band edges are indicated with the vertical dashed lines. As the LAO film thickness increases, the band gap becomes narrower, mainly due to a shift in the top of the valence band to higher energy. The system is insulating for LAO films up to 3 u.c. thick. Fig. 30c shows the DOS projected onto atomic planes parallel to the interface. The bottom twelve plots correspond to the sequence of SrO (S) and $TiO_2$ (T) planes of the STO substrate, and the top eight plots correspond to the LaO (L) and $AlO_2$ (A) planes of the LAO(4 u.c.)/STO(6 u.c.) slab. The positions of the top of the valence band and the bottom of the conduction band do not vary in the STO but shift to higher energy in the LAO with increasing LAO thickness. At 4 u.c. LAO, the top of the LAO VB becomes isoenergetic with the bottom of the CB in STO and the system becomes metallic. This model of the insulator–metal transition is based on an atomically abrupt $TiO_2$/LaO interface and the absence of any structural modification at the interface during LAO film growth.

### 4.3.3 Cation intermixed configurations in the vicinity of the $TiO_2$/LaO interface

To investigate the energetics of cation intermixing, we have calculated the energies required to exchange metal ions between the STO substrate and the LAO film. Taking into account the relative sizes of the ions and their corresponding positions in the perovskite structure, we considered configurations formed by A-site (La⇔Sr) and B-site (Al⇔Ti) exchanges and combinations thereof. We have calculated the cation exchange energies in order to investigate their dependence on the distance of the cations from the plane of the interface and from each other.

The Al⇔Ti, La⇔Sr, and LaAl⇔SrTi exchange energies, calculated with respect to the ideal interface configuration, are summarized in Tables 6 & 7. Here we used a $\sqrt{2}\times\sqrt{2}$ lateral cell, 6 u.c. of STO, and ten irreducible k-points for Brillouin zone integration. Negative values correspond to the total energy gain resulting from the ion exchange. The bold numbers (**1**, **2**, and **3**) refer to the atomic planes relative to the interface from which the exchanged atoms were taken, as indicated in Figs. 29. For example, for 1 u.c. of LAO, the supercell becomes 0.57 eV more stable if an Al atom exchanges with a Ti atom located in the first $TiO_2$ layer. Similarly, the system gains 0.39 eV and 0.34 eV per supercell if the same Al atom exchanges with a Ti atom in the $2^{nd}$ and $3^{rd}$ $TiO_2$ plane away from the interface, respectively.

The data in Table 6 predict that the largest energy gain associated with Al⇔Ti site exchange is obtained when Al goes to a Ti site near the interface and Ti occupies a site in the outermost Al layer in the film. For 1, 2, and 3 u.c. LAO films, such configurations result in overall energy gains of 0.57, 0.76, and 1.07 eV per supercell, respectively. Other configurations involving Al⇔Ti exchanges also lead to significant energy gains of up to 0.8 eV. The dependence of the calculated energies on the k-grid has been investigated for the $Al_1$⇔$Ti_1$ pair. The energies calculated using 20 and 34 irreducible k-points are within 0.01 eV of those reported in Table 6. We emphasize that according to these calculations, Ti atoms in the LAO film favor occupying the most distant sites from the interface for *any* LAO film thickness and for *any* given position of an Al atom inside the STO substrate.

One can argue that the calculated exchange energies could be incorrect because the dipole moment associated with the LAO film results in a saw-like potential across the STO/LAO system in the periodic slab model. Therefore, we also considered a "sandwich" periodic slab of the form LAO/STO/LAO (Fig. 29b) [60] and calculated several Al⇔Ti exchange energies for 1–3 u.c. of LAO on each side of 5.5 u.c. slab of STO. The total dipole moment of this sandwich is zero by construction and the potential outside the slab quickly converges to a constant value, as discussed elsewhere [60]. For each LAO thickness, we considered two mirror-symmetry exchanges in which an Al atom occupies a Ti site near the interface ($Ti_1$) and a Ti atom occupies the outermost Al site in the film ($Al_N$, where $N$ is the LAO film thickness). An example of such an ion exchange configuration for LAO(2 u.c.) film, $Al_2$⇔$Ti_1$, is shown in Fig. 29d.

Similar to the case of the LAO/STO slabs discussed above, the LAO/STO/LAO sandwich becomes more stable when the Al⇔Ti exchange occurs. Moreover, the $Al_N$⇔$Ti_1$ exchange energies (0.57, 0.92, and 1.35 eV for 1, 2, and 3 u.c. thick LAO film, respectively) are close to those obtained using a simpler STO/LAO slab, as summarized in Table 6. These slight differences are attributed to the presence of two $LaO/TiO_2$ interfaces in the LAO/STO/LAO slab compared to one in the STO/LAO slab.

Replacement of an $Al^{3+}$ ion by a $Ti^{4+}$ ion inside the LAO film is expected to result in the appearance of a compensating electron. Similarly, replacement of a $Ti^{4+}$ by an $Al^{3+}$ inside the STO substrate is expected to result in a compensating hole. However, analysis of the electronic states in the vicinity of the band edges suggests that no such electron-hole pairs are formed due to Al⇔Ti site exchange. This result can be understood by examining the layer-projected DOS calculated for the ideal $TiO_2/LaO$ interface (Fig. 30c). Since the CB edge in the STO is lower than that in the LAO, an electron compensating the $Ti^{4+}$ impurity in the LAO will be strongly polarized towards the $TiO_2/LaO$ interface. Similarly, since the VB edge in STO is lower in energy than that in LAO (Fig. 30c), the hole compensating the $Al^{3+}$ impurity in STO is strongly polarized towards the interface. These effects are further strengthened by the electrostatic interaction between the electron and the hole. Thus, the electron and the hole remain in the vicinity of the interface, and the system becomes more energetically stable when they recombine. In other words,

Al⇔Ti site exchange can be considered as an exchange of $Al^{3+}$ and $Ti^{4+}$ ions, which takes place without the formation of carriers.

The effect of the $Al^{3+}$⇔$Ti^{4+}$ exchange on the total energy can be rationalized using schematics shown in Table. 7, where the SrO and $TiO_2$ atomic planes are designated with zeros and LaO and $AlO_2$ planes are designated with plus and minus signs, respectively, corresponding to the formal charges of the formula units in each layer. The coordinate system can be set so that in the case of the ideal $TiO_2$/LaO interface, the dipole moment associated with each LAO unit cell is oriented perpendicular to the interface plane and has a magnitude of $-\boldsymbol{d}_0$. Hence, a √2×√2 cell consisting of $N$ LAO layers would have an overall dipole of $-2N\boldsymbol{d}_0$, with the attendant monotonic increase in one-electron energies with successive LAO layers, as seen in Fig. 31a. However, if $Ti^{4+}$ and $Al^{3+}$ ions undergo site exchange, as illustrated in Table 7, the overall dipole moment vanishes, and the electronic energy becomes the same in all layers, as seen in Fig. 31b. Thus, greater energetic stability at the STO/LAO interface, induced by $Al^{3+}$⇔$Ti^{4+}$ site exchange is linked to the reduction or elimination of the electric field across the interface.

We note that the charge transfer associated with this $Al^{3+}$⇔$Ti^{4+}$ exchange eliminates the polar catastrophe in the same way as the electronic reconstruction at the ideal interface which occurs once the LAO film equals 4 u.c. However, a distinct and important feature of this Al⇔Ti exchange is that it is predicted to be energetically favorable for *any* LAO layer thickness.

To further probe this effect, we investigated the dependence of the energy gain due to the $Al_1$⇔$Ti_1$ exchange on the concentration of the exchanged pairs. We calculated exchange energies for LAO(1 u.c.)/STO(6 u.c.) and lateral cell dimensions of √2×√2, 2×2, and 2√2×2√2. The resulting energies are –0.57, –0.67, and –0.60 eV, indicating that the effects of Al⇔Ti exchanges are nearly additive. This result is consistent with aforementioned conclusion that each $Al_N$⇔$Ti_1$ pair eliminates the $2N\boldsymbol{d}_0$ dipole moment per lateral cell. Importantly, the observed additive effect of Al⇔Ti site exchange reinforces our conclusion that $Al^{3+}$ and $Ti^{4+}$ ions at B sites in both STO and LAO can be modeled using classical shell model and pairwise interatomic potentials.

The energies required to exchange Sr and La atoms in the vicinity of the $TiO_2$/LaO interface have been calculated for LAO(1–3 u.c.) film (see Table 8). In all cases, the La⇔Sr exchanges destabilize the system by approximately 0.4–0.5 eV. The layer-projected DOS for the LAO(3 u.c.)/STO(6 u.c.) slab after an $La_1$⇔$Sr_1$ exchange is shown in Fig. 31c. The overall profile of the DOS is similar to that found for the ideal interface structure. Yet, modifications of the DOS structure near the interface can be seen. These include: (i) a shift to higher binding energy for VB states on the STO side, and (ii) a shift to lower binding energy for VB states on the LAO side. We have placed ellipses in the Fig. 31 a&c to highlight where these subtle changes occur. These shifts are consistent with an ionic character of the exchange: $La^{3+}$⇔$Sr^{2+}$.

La⇔Sr exchange generates an electron which compensates the $La^{3+}$ impurity in the STO, and a hole which compensates the $Sr^{2+}$ impurity in the LAO. However, as long as these impurities are close to each other and the bottom of the CB in the entire system is higher than the top of the VB, these electrons and holes recombine similar to the case of Al⇔Ti. The charge redistribution, induced by the $La^{3+}$⇔$Sr^{2+}$ exchange is shown schematically in Table 7: the dipole moment per supercell increases with respect to that in the ideal interface configuration, which is consistent with the destabilizing effect found for La⇔Sr exchanges. Thus, this result suggests that the La⇔Sr exchanges near the interface can also be modeled using the classical shell model.

Finally, we considered the energy cost of combined or "correlated" LaAl⇔SrTi exchanges, as summarized in Table 8. These exchanges either cost no energy or result in an energy gain of as much as 0.4 eV. We emphasize that for all LAO film thicknesses investigated, correlated LaAl⇔SrTi site exchanges result in the system become more energetically stable than the ideal STO/LAO ($TiO_2$/LaO) interface.

The most energetically favorable configuration considered thus far is obtained by exchanging one Ti atom in the first u.c. of STO with one Al atom in the outermost u.c. of LAO in each $\sqrt{2}\times\sqrt{2}$ lateral cell (see Table 6). The probability of forming this particular configuration decreases rapidly with increasing LAO film thickness. At the same time, the energies calculated for other Al⇔Ti and LaAl⇔SrTi lattice site exchange configurations suggest that complex intermixed quaternary oxide structures can be formed in the process of interface formation. The analysis of such complex configurations and their effect on the electronic structure of the $TiO_2$/LaO interface are considered in this section. When all possible La⇔Sr and Al⇔Ti site exchanges and their various combinations are considered, the number of possible atomic configurations is enormous, and the use of a full quantum mechanical formalism is not feasible. Therefore, we have employed classical potentials to explore the range of possibilities, as described in the next two sections.

### 4.3.4 Classical interatomic potentials for the STO/LAO system

First, we investigate whether the effects of the cation exchange on the energetic stability of STO/LAO can be modeled accurately using the classical shell model and interatomic potentials. This approach provides a simplified description of the system; it takes into account polarizability of ions, but is not designed to model the insulator-metal transition. Therefore, we consider LAO films of up to three unit cells thick, which correspond to the insulating idealized STO/LAO system.

For simplicity, we used the idealized geometry of the interface, in which atoms in both STO and LAO are fixed at their ideal lattice sites within the perovskite structure, and the lattice constants throughout the slab were those of STO. Formal ionic charges and the rigid ion model were used for $Sr^{2+}$, $Ti^{4+}$, $Al^{3+}$, and $La^{3+}$ ions, while the shell model [69] is used for the polarizable $O^{2-}$ ions. The shells of the $O^{2-}$ ions are allowed to adopt the most energetically favorable configuration. Parameters for the short-range interatomic potential were taken from Ref. [70].

The Al⇔Ti, La⇔Sr and LaAl⇔SrTi site exchange energies, calculated with respect to the energy of the idealized abrupt interface, are summarized in Tables 9 & 10. Comparison of these numbers with the results of the *ab initio* calculations (Tables 6 & 8) indicates that the two methods are mutually consistent. La⇔Sr exchanges in isolation destabilize the interface, Al⇔Ti exchanges stabilize the interface, and LaAl⇔SrTi correlated exchanges also stabilize the interface, although not as much as Al⇔Ti exchanges. Due to the details of the parameterization [70], the values of the energies calculated using the classical interatomic potentials are much larger than those calculated quantum mechanically. Nevertheless, it is possible to quantify the extent of consistency between the two methods. To do so, the classical exchange energies $\Delta E^{cl}_i$ for each of the $N$ configurations considered ( $\Delta E_i^{cl} \equiv E_i^{cl} - E_0^{cl}$ , where $E_0^{cl}$ is the energy of the abrupt interface and $E^{cl}_i$ is the energy of the $i_{th}$ configuration) were scaled by a factor α. The root mean square (RMS) of the difference between the scaled classical exchange energies ( $\alpha \Delta E_i^{cl}$ ) and the analogously defined *ab initio* exchange energies ( $\Delta E_i^{DFT}$ ) was then minimized with respect to α, yielding the optimal RMS value ($W$) of 0.27 eV. Here $W$ is defined as

$$W^2 = \frac{1}{N} \sum_i (\alpha \Delta E_i^{cl} - \Delta E_i^{DFT})^2 \qquad (9)$$

This exercise validates the approach based on classical potentials.

#### 4.3.5. A quaternary La-Al-Sr-Ti oxide interfacial phase

To model more complex intermixed interface structures, we defined a “window” of the lattice sites available for site exchanges. This window includes the three STO u.c. nearest to the interface and the entire LAO film, as shown in Fig. 32a. All possible single and double Al-Ti and La-Sr exchanges and their various combinations within this window were considered. In what follows, we discuss the specific example of the 3 u.c. LAO film and the √2×√2 lateral cell. The resulting total number of calculated distinct atomic configurations is ~65,000. The formation energy of each of these configurations was calculated with respect to the ideal interface model using the shell model and energy scaling, as discussed above.

The energy distribution for these structures is shown in Fig. 32b. In order to construct this figure, each configuration $i$ with energy $\alpha \Delta E^{cl}_i$ was represented by a delta function positioned at $\alpha \Delta E^{cl}_i$. The delta functions were then broadened by convolution with a Gaussian-type function using a dispersion $\sigma = kT$, where $T$ was set to 300 K. The full set of configurations can be decomposed into two groups – those for which the interface dipole is zero and those for which the interface dipole is nonzero, as marked in the figure and illustrated in Table 7.  It is apparent that a significant fraction of the configurations are more stable than the ideal $TiO_2$/LaO interface structure. A much larger fraction of the zero-dipole structures are more stable than the abrupt configuration, revealing that elimination of the interface dipole, as achieved by intermixing, is a significant stabilizing factor in interface formation.

In order to select statistically significant configurations, each contribution to Fig. 32b, was weighted with the Boltzmann factor $\exp[-\alpha(E-E_0)/kT]$, where $E$ and $E_0$ are the energies of the configuration in question and the abrupt configuration, respectively, and $\alpha$ is the scaling factor discussed above. Here the temperature was set to 1000 K, which corresponds to typical growth and processing conditions. The Boltzmann weighted distributions are shown in Figure 32c. These clearly demonstrate that the energies of statistically relevant interface configurations are significantly lower than that of the abrupt interface. Thus, these calculations predict that the abrupt interface is highly thermodynamically unstable, and that significant thermodynamic stabilization is achieved through intermixing.

We can also us these results to estimate the atom profiles across the plane of the interface. The number density of all metal ions was calculated using their geometrical positions in each configuration along with the corresponding Boltzmann factors. The resulting composition profiles are shown in Fig. 33. In general, all four cation species tend to intermix and form an interface that is best described as a complex quaternary oxide with concentration gradients normal to the interface. In the case considered here, the intermixing window was restricted to the STO(3 u.c.)/LAO(3 u.c.) region near the interface and our results predict that thermodynamics drives the entire region to be heavily intermixed.

We note that the chemical compositions profiles for A and B lattice sites are different. For the A sites, the calculated concentrations of Sr and La decrease gradually with distance from the interface. In the case of the B sites, however, the composition has a peak at the greatest distance from the interface atomic plane within the intermixing window. These profiles arise from thermodynamic considerations alone and do not account for kinetic effects induced by growth and subsequent annealing.

Finally, we make a few comments regarding the statistically most significant configurations obtained using the shell model calculations. Just over 49 % of all the configurations we considered are more thermodynamically stable than the idealized interface structure. The interface dipole is zero in approximately 7.7% of all configurations. However, among the most stable 10 % of all configurations, 29% have no dipole, which further supports the thesis that thermodynamically driven intermixing decreases the electric field across the interface. Furthermore, in the full set of configurations, as many as 73.5% are obtained by simultaneously exchanging two La⇔Sr pairs and two Al⇔Ti pairs within the intermixing “window” of 12 atomic planes (Fig. 32a). Moreover, this fraction increases to 81.3% among the most stable 10% of all structures. Additionally, configurations in which one La⇔Sr and two Al⇔Ti sites are exchanged simultaneously constitute 17.7 % of the most stable 10% of all structures. These numbers clearly show that the most stable structures are heavily intermixed.

In order to corroborate the results of these shell model calculations, we have selected several inequivalent, low-energy, intermixed configurations and have calculated their energies relative to that of the abrupt interface using DFT. The structure of the intermixing window for each of these configurations is shown in Table 11 for the $\sqrt{2}\times\sqrt{2}$ lateral cell. (These lateral cell dimensions result in two independent metal ions within each atomic plane.) The total energy of each system was minimized with respect to all internal coordinates using a plane-wave basis set and the PBE density functional. Then, a Gaussian-type basis set and the B3LYP functional were used to calculate the energies of these configurations for their optimal geometrical structures. It is clear from the relative energies shown at the bottom of the table that there is significant stabilization of the interface brought about by extensive intermixing, in agreement with the results of the classical shell model calculations. The layer-projected DOSs for the three intermixed LAO(3 u.c.)/STO(6 u.c.) configurations are further discussed in Section 5 in conjunction with the band offset at the interface.

## 5. Implications for electronic structure

We now consider the relationship(s) between atom profiles and electronic structure at the LAO/STO interface. As mentioned in Section 1, the ostensible reason this interface of two band insulators exhibits conductivity at all is because charge is transferred from LAO to STO to eradicate the polar discontinuity. In order to eliminate the associated electrostatic potential, which increases with film thickness, it has been proposed that electronic charge transfers from LAO to $TiO_2$-terminated STO at a critical thickness of 4 u.c. and a confining potential localizes the charge near the interface. This confining potential is defined by the heterojunction band offset on the LAO side, and a steep potential gradient in the near-interface portion of the STO side. The large bandgap difference between LAO films ($E_g$ = 5.84 – 6.33 eV, depending on thickness) [78] and bulk STO ($E_g$ = 3.25 eV) [79] results in a wide range of possible band offset values. The numerical values of the band offsets and the extent of band bending at the interface determine the properties of the confining potential that is required to keep itinerant electrons within the interfacial region. First-principles calculations predict valence band offsets (VBO) ranging from -0.15 [80] to -0.90 [73] with the minus sign signifying that the LAO valence band maximum (VBM) is deeper in binding energy than that of the STO. Theory also predicts sharp band bending on the STO side of the interface [76], as well as a sizeable electric field (0.24 V/Å) within the LAO film below 4 u.c. [60]. An idealized and perfectly abrupt interface was assumed in these calculations.

On the experimental front, second harmonic generation (SHG) studies are consistent with polarization on the STO side of the interface, but do not yield a numerical value for the electric field [81]. Additionally, XPS investigation for MBE-grown LAO on STO(001) did not detect any measurable electric field in the LAO [14]. In contrast, a soft x-ray XPS investigation produced evidence for a small amount of band bending (~0.25 eV) on the STO side of the interface of LAO/STO grown by pulsed laser deposition (PLD) [82], whereas a hard x-ray XPS study concluded that band bending in the STO did not occur [83].

We have carried out a set of XPS measurements and density functional theory (DFT) calculations to determine the band offset and band bending within both the LAO and the STO at this interface. Our experiments were carried out using two high-energy resolution spectrometers. The Gamma Data/Scienta SES 200 analyzer described in Section 4.1, was used to measure films grown on 10 mm x 10 mm STO substrates and to carry out angle-dependent measurements. A Physical Electronics Quantum 2000 analyzer was used to measure films grown on 5 mm x 5 mm substrates. Both spectrometers utilize monochromatic AlK$\alpha$ x-ray sources, but the focusing is different in the two. X-rays resulting from 500W of electrical power were focused into an area of dimensions ~9 mm x ~1 mm on the sample surface in the SES 200. X-rays resulting from 100W of electron beam power were focused into a spot of diameter ~100 μm in the Quantum 2000, and the beam was rastered over an area of dimensions given by ~1.3 x ~0.2 $mm^2$. The binding energy scales in both spectrometers were calibrated using noble metal standards such that the energy scale is accurate to within ±0.02 eV over a range of 0 to 500 eV. As check of consistency in energy dispersion for the two analyzers, the binding energy difference between the Ti $2p_{3/2}$ and Sr $3d_{5/2}$ peaks for $TiO_2$-terminated STO(001) was measured to be 324.41(3) eV on both instruments. The energy resolution was ~0.45 eV for both spectrometers.

The 4 u.c. LAO films grown on 10 mm x 10 mm STO substrates were fastened to sample holders using spring-loaded metal clips which made electrical contact to the conductive interfacial plane and thus did not retain the positive photoemission-induced charge that normally accumulates on the surfaces of insulators. No low-energy electron flood gun was needed for these samples, and the resulting binding energies are accurate on an absolute scale, allowing determination of the position of the Fermi level relative to the band edges. 4 u.c. samples grown on 5 mm x 5 mm substrates and all 25 u.c. films did charge during measurements and a low-energy electron flood gun had to be used to neutralize the accumulated positive charge. The flood gun energy and current were set so the peaks exhibited constant, albeit incorrect in an absolute sense, binding energies. However, the binding energies differences used to determine band offsets are accurate to within a few hundredths of an eV. The binding energy scale was calibrated using noble metal standards such that the energy scale is accurate to within ±0.02 eV over a range of 0 to 500 eV.

We measured the VBO, which we define as $\Delta E_V = E_V^{STO} - E_V^{LAO}$, by referencing the VBM of LAO and STO, while separated and in intimate contact, to core orbitals for the various elements [84-85]. The first step was to determine the binding energy differences between core peaks and the VBM for bulk or bulk-like LAO and STO. To this end, we used bulk Nb-doped STO(001) and 25 u.c. films of LAO/STO(001). Representative spectra for a Tokyo film are shown in Fig. 34; the spectra for Augsburg films are essentially identical. The VBM values were determined by extrapolating the leading edge of the VB to the energy axis in order to account for instrumental broadening [86-87]. The Al 2p peak is an unresolved spin-orbit (SO) doublet with some asymmetry. The Sr 3d and Ti 2p spectra are well-resolved SO pairs and the more intense higher angular momentum features, marked by vertical arrows, were used. The La 4d line shape is highly complex. In addition to SO splitting, this spectrum exhibits final-state “shake” features in which valence charge rearrangement accompanies 4d core photoionization. A complete theoretical description of the La 4d spectrum in LAO is in progress, but is not required to accurately measure the LAO/STO VBO. What is needed is a consistent method of fitting the overall line shape so that one principal feature can be measured to within a few hundredths of an eV for all films. We have chosen the more intense of the two peaks in the La $4d_{5/2}$-derived feature at ~ 102.5 eV for this purpose. In eV, the resulting energy differences between the core orbitals and their associated VBM values are: La $4d_{5/2}$ - 99.60(5) (Augsburg) and 99.52(5) (Tokyo), Al 2p - 71.16(5) (Augsburg) and 71.13(4) (Tokyo), Sr $3d_{5/2}$ -130.54(3), and Ti $2p_{3/2}$ - 455.95(3). The full widths at half maximum (FWHM) of these core states are sensitive to band bending in the near-surface region over which photoelectrons are collected because core-level binding energies shift in energy by the same amount as the valence band when band bending occurs. Band bending results in an increase in FWHM since the core level now consists of a manifold of peaks whose binding energies change with depth. The spectra in Fig. 34 exhibit the same narrow FWHM values as those we have measured for the same elements in non-semiconducting oxides (such as bulk LAO, $Al_2O_3$ , $TiO_2$ and undoped STO) where band bending cannot occur because of the absence of free carriers. Accordingly, we conclude that the Nb:STO and 25 u.c. LAO surfaces are in a near flat-band state.

We then measured core-level binding energy differences between film and substrate elements in 4 u.c. LAO/STO films to determine $\Delta E_V$. In order to check for internal consistency, we took four pairs of core orbitals, each pair consisting of one orbital from the substrate and one from the film, to determine $\Delta E_V$. The four resulting expressions are,

$$\Delta E_V = (E_{La4d5/2} - E_V)_{LAO} - (E_{Sr3d5/2} - E_V)_{STO} + (E_{Sr3d5/2} - E_{La4d5/2})_{HJ} \quad (10)$$

$$\Delta E_V = (E_{Al2p} - E_V)_{LAO} - (E_{Sr3d5/2} - E_V)_{STO} + (E_{Sr3d5/2} - E_{Al2p})_{HJ} \quad (11)$$

$$\Delta E_V = (E_{La4d5/2} - E_V)_{LAO} - (E_{Ti2p3/2} - E_V)_{STO} + (E_{Ti2p3/2} - E_{La4d5/2})_{HJ} \quad (12)$$

$$\Delta E_V = (E_{Al2p} - E_V)_{LAO} - (E_{Ti2p3/2} - E_V)_{STO} + (E_{Ti2p3/2} - E_{Al2p})_{HJ} \quad (13)$$

In each equation, the first two terms were measured for 25 u.c. LAO/STO and Nb:STO, respectively, and the third term was measured for 4 u.c. LAO/STO heterojunctions (HJ). In this convention, a positive (negative) VBO signifies that the STO VBM is more (less) deeply bound than that of the LAO (i.e. $\Delta E_V = E_V^{STO} - E_V^{LAO}$). An energy level diagram is shown in Fig. 35.

We show in Fig. 36 representative spectra for a Tokyo film, and Table 12 summarizes the VBO values for representative samples from each lab before and after surface cleaning. The values derived from each pair of core levels are well within experimental error for each film. In all cases, the VBO is near zero or slightly positive. There is a slight increase in VBO upon *in-situ* oxygen plasma surface cleaning which rids the surface of adventitious carbon. The experimental VBOs differ markedly from theoretical predictions based on an abrupt interface [73, 80]. Moreover, in all cases, the FWHM values for the La $4d_{5/2}$, Al 2p, Sr $3d_{5/2}$ and Ti $2p_{3/2}$ peaks in the 4 u.c. films are less than 0.1 eV larger than those for the bulk STO and 25 u.c. LAO reference surfaces, revealing that little or no electric field can be detected by XPS on either side of the interface. Segal et al. [14] made a similar observation for the La $4d_{5/2}$ and Al 2p peak widths in MBE-grown LAO/STO and came to the same conclusion.

Absolute Ti $2p_{3/2}$ and Sr $3d_{5/2}$ binding energies measured as a function of take-off angle ($\theta_t$) were used to determine the position of the STO conduction band minimum (CBM) relative to the Fermi level ($E_F$) in 4 u.c. LAO/STO as a function of depth. We used the formulae $E_F - E_C = E_g + (E_{Ti2p3/2} - E_V)_{STO} - (E_{Ti2p3/2})_{HJ}$ and $E_F - E_C = E_g + (E_{Sr3d5/2} - E_V)_{STO} - (E_{Sr3d5/2})_{HJ}$, where $E_g$ = 3.25 eV, the bandgap of STO. The probe depth ranges from ~1.5 nm (~4 u.c.) at $\theta_t = 19^o$ to ~4.5 nm (~11 u.c.) at $\theta_t = 90^o$. Representative spectra for a Tokyo film are shown in Fig. 37a, and $E_F - E_C$ vs. probe depth is plotted for both Tokyo and Augsburg films in Fig. 37b. The Fermi level is constant at ~0.05 (0.25) eV below the CBM for all depths probed for the Augsburg (Tokyo) film. To within experimental error (±0.06 eV), there is no detectable band bending within the STO in either film, contrary to what was predicted by theory for an idealized interface [76]. Corroborating this result, the Ti $2p_{3/2}$ FWHM, 1.06 eV (Fig. 36), is very similar to that of nearly flat-band Nb-doped STO(001), 1.00 eV (Fig. 34). Sharp band bending within the probe depth would result in a measurable broadening of this peak due to the different electronic environments therein. For instance, if the STO transitions from being intrinsic ($E_F - E_C =$ 0.70 eV, the charge neutrality level in pure STO, see ref. [88]) ~1.2 nm away from the interface to degenerate *n*-type ($E_F - E_C =$ -0.20 eV) at the interface, the composite Ti $2p_{3/2}$ peak FWHM would be 1.35 eV, which clearly exceeds the experimental value of 1.06 eV. If the potential drop is 0.3 eV rather than 0.9 eV, the FWHM would be 1.05 eV, which matches what is observed. Therefore, 0.3 eV of band bending cannot be ruled out on the basis of the peak width. However, a 0.3 eV potential change should be detectable using absolute core binding energies, because the latter averaged over the band bending region would shift ~0.15 eV relative to that in the interface layer, and a shift of this magnitude would be readily detectable in Fig. 37a. However, no such shift is observed.

It is known that LAO/STO heterojunctions exhibit a high degree of light sensitivity. Photocarriers generated by room light persist for several hours and heterojuctions must be stored in the dark for 24 hours prior to transport experiments in order to measure intrinsic effects. Moreover, x-ray induced persistent photoconductivity has been reported with the use of a much more intense x-ray source [83]. However, we find no dependence of the Ti $2p_{3/2}$ binding energies and FWHM on x-ray anode power level from our normal SES 200 operating value of 500W down to 70W, ruling out any x-ray induced anomalies, at least over this range of x-ray intensity. These results are shown in Fig. 38. Additionally, positive photoemission induced charge was conducted away from the surface by direct electrical contact, precluding perturbations to band bending created by any photocurrents that may be present [89].

Another noteworthy result is that there is no evidence for $Ti^{3+}$ in our Ti 2p spectra for 4 u.c. LAO/STO, despite the fact that ~3 nm of STO is probed at normal emission, and the interfaces are conductive. It is commonly thought that if the 2DEG exists, each electron therein will effectively reduce one structural $Ti^{4+}$near the interface to $Ti^{3+}$ because the bottom of the STO conduction band is largely Ti 3d derived [83]. Indeed, Ti $2p_{3/2}$ spectra for heavily (6 at. %) Nb-doped STO(001) bulk crystals exhibit a feature shifted ~2 eV to lower binding energy relative to the normal $Ti^{4+}$ lattice peak with a peak area of ~6% of the total [90]. We interpret the absence of such a $Ti^{3+}$ feature in our spectra to more itinerant electron delocalization normal to the interface than would occur if indeed a 2DEG exists at this interface. In this scenario, the itinerant electron concentration (electrons per unit volume) is low to result in a detectable $Ti^{3+}$ peak. This result and its interpretation are consistent with the absence of band bending in the near-interface STO described above.

The present results reveal that: (i) the underlying STO at the LAO/STO interface is *n*-type despite the fact that the STO substrate was not intentionally doped, and, (ii) no band bending can be detected. The numerical value of the $Ti^{4+}$ $2p_{3/2}$ binding energy reported in ref. [83] (~459.4 eV) also indicates *n*-type behavior in the underlying STO, and its invariance with collection angle also reveals negligible band bending. In contrast, Yoshimatsu et al. [82] concluded that ~0.25 eV of downward band bending occurs as the LAO thickness increased from 1 to 4 u.c., leading to 2DEG formation directly at the interface for the higher thicknesses. However, their conclusions were critically dependent upon the way in which they assigned the position of the VBM relative to the Fermi level. Their method was subsequently challenged because it ignored the effect of instrumental broadening in deducing the position of the VBM [82, 86-87, 91]. When instrumental broadening is accounted for, the spectra published in ref. [82] reveal that the underlying STO is *n*-type for all LAO thicknesses from 1 to 6 u.c.

To understand these results from first principles, we have calculated the valence band offset and band bending for both abrupt and intermixed atom configurations. As described in Section 4.3, we considered several intermixed interface structures with specific Al⇔Ti and La⇔Sr site exchanges that reduced the overall dipole. Exchanges of this kind lead to a considerable energy gain relative to the abrupt interface configuration (up to 0.9 eV for $\sqrt{2}\times\sqrt{2}$ in-plane supercells of 2 u.c.LAO/6 u.c. STO and up to 1.4 eV for 3 u.c. LAO). Moreover, band bending within the LAO, predicted to be sizeable on the basis of *ab initio* calculations of ideal interfaces [60], is reduced or eliminated when intermixing is modeled. This effect is illustrated in Fig. 39 for three distinct, low-energy, intermixed configurations of LAO(3 u.c.)/STO(6 u.c.) discussed at the end of Sec. 4.3 and described in the last three columns of Table 11. Here we show the densities of states (DOS) projected onto individual atomic planes. Inspection of the layer-resolved DOS in Fig. 39a&b reveals that there is negligible electric field in either material, primarily because intermixing has eliminated the dipole in the LAO and has not imparted a dipole to the STO. The VBO for these configurations is ~0.2 eV, in reasonable agreement with experiment. The DOS shown in Fig. 39c is obtained for a higher energy configuration in which a small dipole moment across the interface exists and is due to Al and La impurities in the STO substrate (see Table 11). The VBO is ~0.4 eV for this configuration. In contrast, the perfectly abrupt heterojunction (Fig. 31a) yields a VBO of >1.0 eV, which is quite far from experiment. Thus, intermixing is clearly required to account for the experimental band offset.

Finally, we consider the cause of *n*-type doping in the underlying STO. As described in Sections 3 & 4, RBS, STEM/EELS, ToF-STIM, ARXPS and MEIS results for LAO/STO cannot be accounted for without La diffusion into the STO during growth. La (a shallow donor) in the STO can account for at least some the itinerant electrons if La substitutes for Sr and is electrically activate. La doping in the STO lattice may also impact electron mobility, at least at low temperature. Tufte and Chapman [92] showed many years ago that the low-temperature mobility in bulk Nb-doped STO crystals exhibits values similar to those observed for LAO/STO [5, 93-94]. The dependence of low-temperature mobility on carrier concentration revealed that ionized impurity scattering is most likely the primary determinant of mobility in *n*-STO, even down to 2K. This study also showed that polar optical phonon scattering is most likely responsible for the low mobilities observed at higher temperatures. Likewise, Ohta et al. [2] have shown that bulk Nb- and La-STO doped exhibit room-temperature mobilities similar to those measured for LAO/STO interfaces. These similarities suggest that the conductivity in LAO/STO is due at least in part to La doping of the underlying STO. Additionally, Dubroka et al. [95] have very recently combined transport measurements with infrared ellipsometry to show that the frequency dependence of the electron mobility at LAO/STO interfaces grown at Augsburg is similar to that exhibited by Nb-doped STO, and that the conductive plane is considerably wider than originally thought. Indeed, these authors found that modeling their ellipsometric data yielded a depth-dependent carrier concentration that falls off with distance from the interface into the STO to a depth of at least 11 nm. This result is consistent with La doping being the source of the carriers over depths corresponding to the depth of the La diffusion tail in STO.

## 6. Summary and Outlook

Using several independent analytical methods, we have shown that there is a strong tendency for the LAO/STO interface, as prepared by on-axis PLD in laboratories external to ours, to intermix rather than form an atomically abrupt configuration. Although not described herein, the same kinds of measurements carried out on analogous films prepared by off-axis PLD at PNNL yield very similar results [96-97]. Moreover, first principles calculations carried out with classical and quantum mechanical potentials show that interfacial intermixing is energetically preferable to abruptness, and that forming an intermixed interface represents a thermodynamically favorable process at 1000K, which is essentially the growth temperature. Although the intermixing is approximately correlated, which means that A-site exchanges (La⇔Sr) occur to the same extent as B-site exchanges (Al⇔Ti), there is preferential diffusion of La into the STO, which leads to n-type doping of the STO and the likely formation of itinerant electrons within the STO. These results call into question the popular interpretation of conductivity based on the formation of a 2DEG.

These results also go against the grain of conventional wisdom on the LAO/STO system, which is that cation disorder at the interface, if it occurs at all, is a negligible effect. The conventional view has been supported by HAADF-STEM images of the interface which often give the appearance of an abrupt interface because of the high extent of Z contrast the technique affords. However, as we have shown here, very careful EELS measurements, as well as volume-averaging experiments with more sensitivity to low elemental concentrations than EELS, must accompany the STEM investigation.

While there is a clear energetic tendency for intermixing to occur, it remains to be seen if metastable abrupt interfaces can be formed using growth techniques such as MBE, in which the energies of the incident atoms are much lower (~0.1 eV) than that of the incident ions in PLD (~10 eV or higher). Until such time as abrupt interfaces can be unambiguously demonstrated, it seems prudent that the community of materials scientists working on this and other complex oxide interfaces would stop defaulting to the more straightforward atomically abrupt description of the interface and embrace more realistic physical models, as difficult as they are to deal with. Then and only then can our understanding of the functional properties of this (and other) fascinating complex oxide materials systems be based on realistic physical models of the composition and structure at the interface.

## Acknowledgements

The authors are indebted to the groups of Jochen Mannhart and Harold Hwang for providing the samples used in this study. The PNNL and UCL work was supported by the US Department of Energy, Office of Science, Division of Materials Sciences and Engineering, and was performed in the Environmental Molecular Sciences Laboratory, a national scientific user facility sponsored by the Department of Energy's Office of Biological and Environmental Research and located at PNNL. The electron microscopy work is supported by Argonne National Laboratory under the Office of Basic Energy Sciences, U.S. Department of Energy grants No. DE-AC02-06CH11357 Digital Synthesis FWP (AS and JMZ) and No. DE-AC02-05CH11231 (QR). Electron microscopy was carried out at the Frederick Seitz Materials Research Laboratory Central Facilities, University of Illinois, which are partially supported by the U.S. Department of Energy under grants DE-FG02-07ER46453 and DE-FG02-07ER46471 and the National Center for Electron Microscopy, Lawrence Berkeley Lab, which is supported by the U.S. Department of Energy under grant DE-AC02-05CH11231. P.V.S. acknowledges the additional financial support from Royal Society of London and thanks Alex Demkov, Ricardo Grau-Crespo and Michael Finnis for many useful discussions.

## References

[1] E. Dagatto and Y. Tokura, MRS Bull. 33 (2008) 1037.

[2] S. Ohta, T. Nomura, H. Ohta, K. Koumoto, J. Appl. Phys. 97 (2005) 034106.

[3] Y. Tokura, Y. Taguchi, Y. Okada, Y. Fujishima, T. Arima, K. Kumagai, Y. Iye, Phys. Rev. Lett. 70 (1993) 2126.

[4] T. S. Santos, S. J. May, J. L. Robertson, A. Bhattacharya, Phys. Rev. B 80 (2009) 155114.

[5] A. Ohtomo and H. Y. Hwang, Nature 427 (2004) 423.

[6] S. Thiel, G. Hammerl, A. Schmehl, C. W. Schneider, J. Mannhart, Science 313 (2006) 1942.

[7] D. A. Muller, Nat. Mat. 8 (2009) 263.

[8] N. Nakagawa, H. Y. Hwang, D. A. Muller, Nat. Mat. 5 (2006) 204.

[9] C. L. Jia, S. B. Mi, M. Faley, U. Poppe, J. Schubert, K. Urban, Phys. Rev. B 79 (2009) 081405.

[10] P. R. Willmott, S. A. Pauli, R. Herger, C. M. Schleputz, D. Martoccia, B. D. Patterson, B. Delley, R. Clarke, D. Kumah, C. Cionca, Y. Yacoby, Phys. Rev. Lett. 99 (2007) 155502.

[11] A. S. Kalabukhov, Y. A. Boikov, I. T. Serenkov, V. I. Sakharov, V. N. Popok, R. Gunnarsson, J. Borjesson, N. Ljustina, E. Olsson, D. Winkler, T. Claeson, Phys. Rev. Lett. 103 (2009) 146101.

[12] R. Pentcheva and W. E. Pickett, J. Phys: Cond. Mat. 22 (2010) 043001.

[13] J. Mannhart and D. G. Schlom, Science 327 (2010) 1607.

[14] Y. Segal, J. H. Ngai, J. W. Reiner, F. J. Walker, C. H. Ahn, Phys. Rev. B 80 (2009) 241107(R).

[15] S. A. Chambers, Adv. Mater. 22 (2010) 219.

[16] T. Higuchi, Masters, Masters Thesis, University of Tokyo, 2009.

[17] S. Thiel, Ph.D. Thesis, University of Augsburg, 2009.

[18] W. K. Chu, J. W. Mayer, M. A. Nicolet, *Rutherford Backscattering Spectrometry* (Academic Press, New York, 1978).

[19] L. C. Feldman and J. W. Mayer, *Fundamentals of Surface and Thin Film Analysis* (North Holland, New York, 1986).

[20] S. A. Chambers, C. M. Wang, S. Thevuthasan, T. Droubay, D. E. McCready, A. S. Lea, V. Shutthanandan, C. F. Windisch, Thin Solid Films 418 (2002) 197.

[21] V. Shutthanandan, S. Thevuthasan, Y. Liang, E. M. Adams, Appl. Phys. Lett. 80 (2002) 1803.

[22] V. Shutthanandan, A. A. Saleh, R. J. Smith, Surf. Sci. Rep. 450 (2000) 204.

[23] S. Thevuthasan, C. H. F. Peden, M. H. Engelhard, D. R. Baer, G. S. Herman, W. Jiang, Y. Liang, W. J. Weber, Nucl. Instrum. Methods Phys. Res A 420 (1999) 81.

[24] M. Mayer, (Max-Planck-Institut für Plasmaphysik, Garching, Germany, 2008), p. SIMNRA User's Guide 6.04.

[25] M. Mayer, Proceedings of the 15th International Conference on the Application of Accelerators in Research and Industry 475 (1999) 541.

[26] E. Niehuis and T. Grehl, in *IM Publications and SurfaceSpectra Limited*, edited by J. C. Vickerman and D. Briggs, (Manchester, U.K, 2001), p. p753.

[27] M. Watanabe, D. B. Williams, M. G. Burke, Proc. Inter. Conf. on Solid-Solid Phase Trans. Inorg. Matls. 2 (2005) 431.
[28] C. J. Howard, B. J. Kennedy, B. C. Chakoumakos, J. Phys.: Condens. Matter 12 (2000) 349.
[29] R. F. Egerton, Ultramicroscopy 107 (2007) 575.
[30] A. Ohtomo, D. A. Muller, J. L. Grazul, H. Y. Hwang, Nature 419 (2002) 378.
[31] J. Verbeeck, S. Bals, A. N. Kravtsova, D. Lamoen, M. Luysberg, M. Huijben, G. Rijnders, A. Brinkman, H. Hilgenkamp, D. H. A. Blank, G. Van Tendeloo, Phys. Rev. B 81 (2010) 085113.
[32] D. A. Muller, N. Nakagawa, A. Ohtomo, Nature (2004) 657.
[33] F. M. F. Degroot, J. Faber, J. J. M. Michiels, M. T. Czyzyk, M. Abbate, J. C. Fuggle, Phys. Rev. B 48 (1993) 2074.
[34] R. Byrdson, H. Sauer, W. Engel, J. Phys.: Condens. Matter 4 (1992) 3429.
[35] Z. L. Zhang, W. Sigle, W. Kurtz, M. Ruhle, Phys. Rev. B 66 (2002) 214112.
[36] A. Fujimori, I. Hase, M. Nakamura, H. Namatame, Y. Fujishima, Y. Tokura, M. Abbate, F. M. F. Degroot, M. T. Czyzyk, J. C. Fuggle, O. Strebel, F. Lopez, M. Domke, G. Kaindl, Phys. Rev. B 46 (1992) 9841.
[37] M. Abbate, F. M. F. Degroot, J. C. Fuggle, A. Fujimori, O. Strebel, F. Lopez, M. Domke, G. Kaindl, G. A. Sawatzky, M. Takano, Y. Takeda, H. Eisaki, S. Uchida, Phys. Rev. B 46 (1992) 4511.
[38] C. S. Fadley, Prog. Surf. Sci 16 (1984) 275.
[39] C. S. Fadley, J. Electron Spectrosc. Relat. Phenom. 178-179 (2010) 2.
[40] D. P. Woodruff, J. Electron Spectrosc. Relat. Phenom. 178-179 (2010) 186.
[41] J. J. Rehr, W. Schattke, F. J. G. de Abajo, R. D. Muino, M. A. Van Hove, J. Electron Spectrosc. Relat. Phenom. 126 (2002) 67.
[42] S. A. Chambers, Advances in Physics 40 (1991) 357.
[43] S. A. Chambers and T. T. Tran, Phys. Rev. B 47 (1993) 13023.
[44] T. T. Tran and S. A. Chambers, J. Vac. Sci. Technol. B 11 (1993) 1459.
[45] S. A. Chambers and V. S. Sundaram, Appl. Phys. Lett. 57 (1990) 2342.
[46] S. A. Chambers and V. S. Sundaram, J. Vac. Sci. Technol. B 9 (1991) 2256.
[47] S. A. Chambers, T. R. Greenlee, C. P. Smith, J. H. Weaver, Phys. Rev. B 32 (1985) 4245.
[48] S. A. Chambers, J. Vac. Sci. Technol. A 7 (1989) 2459.
[49] S. Tanuma, C. J. Powell, D. R. Penn, Surf. Interface Anal. 11 (1988) 577.
[50] M. P. Seah and W. A. Dench, Surf. Interface Anal. 1 (1979) 2.
[51] S. A. Chambers and T. T. Tran, Surf. Sci. 314 (1994) L867.
[52] S. A. Chambers, T. T. Tran, T. A. Hileman, T. A. Jurgens, Surf. Sci. 320 (1994) L81.
[53] S. A. Chambers, Y. Gao, Y. Liang, Surf. Sci. 339 (1995) 297.
[54] S. Y. Tong, H. C. Poon, D. R. Snider, Phys. Rev. B 32 (1985) 2096.
[55] M.-L. Xu, J. J. Barton, M. A. Van Hove, Phys. Rev. B 39 (1989) 8275.
[56] J. F. van der Veen, Surf. Sci. Rep. 5 (1985) 199.
[57] R. M. Tromp, M. Copel, M. C. Reuter, M. H. Vonhoegen, J. Speidell, R. Koudijs, Rev. Sci. Instrum. 62 (1991) 2679.
[58] J. Ziegler, Stopping and Range of Ions in Matter www.srim.org.

[59] L. C. Feldman, J. W. Mayer, P. S., *Materials Analysis by Ion Channeling; Submicron Crystallography* (Academic Press, 1982).
[60] J. Lee and A. A. Demkov, Phys. Rev. B 78 (2008) 193104.
[61] J. P. Perdew, K. Burke, M. Ernzerhof, Phys. Rev. Lett. 77 (1996) 3865.
[62] P. E. Blochl, Phys. Rev. B 50 (1994) 17953.
[63] G. Kresse and D. Joubert, Phys. Rev. B 59 (1999) 1758.
[64] G. Kresse and J. Furthmuller, Phys. Rev. B 54 (1996) 11169.
[65] A. D. Becke, J. Chem. Phys. 98 (1993) 5648.
[66] C. T. Lee, W. T. Yang, R. G. Parr, Phys. Rev. B 37 (1988) 785.
[67] V. R. Saunders, R. Dovesi, R. Roetti, R. Orlando, C. M. Zicovich-Wilson, N. M. Harrison, K. Doll, B. Civalleri, I. Bush, P. D'Arco, M. Llune, CRYSTAL2003 User's Manual (2003).
[68] CRYSTAL09 basis sets http://www.crystal.unito.it/Basis_Sets/Ptable.html.
[69] B. G. Dick and A. W. Overhauser, Phys. Rev. 112 (1958) 90.
[70] T. S. Bush, J. D. Gale, C. R. A. Catlow, P. D. Battle, J. Mater. Chem. 4 (1994) 831.
[71] J. M. Albina, M. Mrovec, B. Meyer, C. Elsasser, Phys. Rev. B 76 (2007) 165103.
[72] Z. C. Zhong and P. J. Kelly, Europhys. Lett. 84 (2008) 27001.
[73] Z. S. Popovic, S. Satpathy, R. M. Martin, Phys. Rev. Lett. 101 (2008) 256801.
[74] H. H. Chen, A. M. Kolpak, S. Ismail-Beigi, Phys. Rev. B 79 (2009) 161402.
[75] U. Schwingenschlogl and C. Schuster, Chem. Phys. Lett. 467 (2009) 354.
[76] K. Janicka, J. P. Velev, E. Y. Tsymbal, Phys. Rev. Lett. 102 (2009) 106803.
[77] R. Pentcheva and A. E. Pickett, J. Phys: Cond. Mat. 22 (2010) 043001.
[78] E. Cicerrella, J. L. Freeouf, L. F. Edge, D. G. Schlom, T. Heeg, J. Schubert, S. A. Chambers, J. Vac. Sci. Technol. A 23 (2005) 1676.
[79] K. van Bentham, C. Elsasser, R. H. French, J. Appl. Phys. 90 (2001) 6156.
[80] R. Pentcheva and W. E. Pickett, Phys. Rev. B 78 (2008) 205106.
[81] N. Ogawa, K. Miyano, M. Hosoda, T. Higuchi, C. Bell, Y. Hikita, H. Y. Hwang, Phys. Rev. B 80 (2009) 081106.
[82] K. Yoshimatsu, R. Yasuhara, H. Kumigashira, M. Oshima, Phys. Rev. Lett. 101 (2008) 026802.
[83] M. Sing, G. Berner, K. Goss, A. Muller, A. Ruff, A. Wetscherek, S. Thiel, J. Mannhart, S. A. Pauli, C. W. Schneider, P. R. Willmott, M. Gorgoi, F. Schafers, R. Claessen, Phys. Rev. Lett. 102 (2009) 176805.
[84] E. A. Kraut, R. W. Grant, J. W. Waldrop, S. P. Kowalczyk, Phys. Rev. Lett. 44 (1980) 1620.
[85] E. A. Kraut, R. W. Grant, J. W. Waldrop, S. P. Kowalczyk, Phys. Rev. B 28 (1983) 1965.
[86] S. A. Chambers, T. Droubay, T. C. Kaspar, M. Gutowski, M. van Schilfgaarde, Surf. Sci. 554 (2004) 81.
[87] S. A. Chambers, T. Droubay, T. C. Kaspar, M. Gutowski, J. Vac. Sci. Technol. B 22 (2004) 2205.
[88] J. Robertson, J. Vac. Sci. Tech. B 18 (2000 ) 1785.
[89] M. H. Hecht, J. Vac. Sci. Technol. B 8 (1990) 1018.
[90] M. Castell and R. G. Egdell, (private communication).
[91] S. A. Chambers, Phys. Rev. Lett. 102 (2009) 199703.

[92] O. N. Tufte and P. W. Chapman, Phys. Rev. 155 (1967) 796.
[93] C. Bell, S. Harashima, Y. Hikita, H. Y. Hwang, Appl. Phys. Lett. 94 (2009) 222111.
[94] T. Fix, F. Schoofs, J. L. MacManus-Driscoll, M. G. Blamire, Phys. Rev. Lett. 103 (2009) 166802.
[95] A. Dubroka, M. Rossle, K. W. Kim, V. K. Malik, L. Schultz, S. Thiel, C. W. Schneider, J. Mannhart, G. Herranz, O. Copie, M. Bibes, A. Barthelemy, C. Bernhard, Phys. Rev. Lett. 104 (2010) 156807.
[96] T. C. Droubay, L. Qiao, T. C. Kaspar, M. H. Engelhard, V. Shutthanandan, S. A. Chambers, Appl. Phys. Lett. (submitted).
[97] L. Qiao, T. C. Droubay, V. Shutthanandan, Z. Zhu, P. V. Sushko, S. A. Chambers, J. Phys.: Condens. Matter (submitted).

Table 1. Summary of PLD growth parameters

| specimen | working distance (mm) | laser energy (mJ/pulse) | spot size ($mm^2$) | repetition rate (Hz) | $O_2$ pressure (Torr) | substrate temperature (Celcius) |
|---|---|---|---|---|---|---|
| Tokyo | 55 | 39.2 | 2.4, 3.5, 5.6 | 4 | $1 \times 10^{-5}$ | 700 |
| Augsburg | 50 | 450 | 48 | 1 | $8 \times 10^{-6}$ | 770 |

Table 2. Thickness and composition of 25 u.c. LAO/STO(001) films from RBS at 2 MeV.

| sample | areal density ($1 \times 10^{15}$ atoms/$cm^2$) | film thickness (Å) | La (at. %) | Al at. %) | O (at. %) |
|---|---|---|---|---|---|
| Augsburg | 70 | ~80 | 22.0(7) | 18(1) | 60 |
| Tokyo | 74 | ~84 | 18.0(6) | 20(1) | 62 |

Table 3. La atom profiles for 25 u.c. LAO/STO(001) films from RBS at 2 MeV.

| Layer number | Augsburg areal density ($1 \times 10^{15}$ atoms/$cm^2$) | Augsburg layer thickness (Å) | Augsburg La (at. %) | Tokyo areal density ($1 \times 10^{15}$ atoms/$cm^2$) | Tokyo layer thickness (Å) | Tokyo La (at. %) |
|---|---|---|---|---|---|---|
| 1 | 67 | ~76 | 18.5 | 70 | ~80 | 20.5 |
| 2 | 10 | ~11 | 2.5 | 10 | ~11 | 2.5 |
| 3 | 100 | ~110 | 0.3 | 50 | ~55 | 0.3 |
| 4 | 700 | ~800 | 0.05 | 200 | ~225 | 0.17 |
| 5 | -- | -- | -- | 700 | ~800 | 0.1 |

Table 4. La atom profiles for Tokyo 25 u.c. LAO/STO(001) film from RBS at 3 MeV.

| Layer number | areal density ($1 \times 10^{15}$ atoms/$cm^2$) | layer thickness (Å) | La (at. %) |
|---|---|---|---|
| 1 | 70 | ~76 | 20.5 |
| 2 | 10 | ~11 | 2.5 |
| 3 | 100 | ~110 | 0.4 |
| 4 | 200 | ~225 | 0.1 |
| 5 | 170 | ~190 | 0.05 |

Table 5. Areal densities ($10^{15}$ at/cm$^2$) for La, Sr and Ti in 4 u.c. Tokyo films grown with different laser fluences from MEIS

| | low (0.7 J/cm$^2$) | medium (1.1 J/cm$^2$) | high (1.6 J/cm$^2$) |
|---|---|---|---|
| La in the LAO film | 1.77 | 1.57 | 1.69 |
| La in the first *n* Å of the STO | 0.26 (*n* = 10 Å) | 0.38 (*n* = 15 Å) | 0.39 (*n* = 16Å) |
| La total | 2.03 | 1.95 | 2.08 |
| Sr in the LAO film | 0.39 | 0.55 | 0.63 |
| Ti in the LAO film | 0.40 | 0.41 | 0.37 |

Table 6. Al⇔Ti exchange energies (in eV) calculated with respect to the ideal interface reference structure using a periodic slab model and a √2×√2 lateral cell with *N* u.c. of LAO. The bold numbers **1**, **2**, and **3** refer to the corresponding atomic planes relative to the plane of the interface (see Fig. 29a); n/a stands for not applicable. Positions of Ti atoms in the LAO film are shown.

| *N* | Al in **1** of STO | | | Al in **2** of STO | | | Al in **3** of STO | | |
|---|---|---|---|---|---|---|---|---|---|
| | Ti in **1** | Ti in **2** | Ti in **3** | Ti in **1** | Ti in **2** | Ti in **3** | Ti in **1** | Ti in **2** | Ti in **3** |
| 1 | –0.57 | n/a | n/a | –0.39 | n/a | n/a | –0.34 | n/a | n/a |
| 2 | –0.25 | –0.76 | n/a | 0.01 | –0.50 | n/a | 0.07 | –0.45 | n/a |
| 3 | –0.34 | –0.54 | –1.07 | –0.08 | –0.28 | –1.08 | –0.01 | –0.23 | –0.77 |

Table 7. Schematics of selected Al⇔Ti, La⇔Sr, and LaAl⇔SrTi site exchange configurations and associated dipole moments for a √2×√2 supercell. See also Fig. 29. The first and second columns indicate the atomic plane and its distance from the interface, respectively.

| | | ideal | | $Al_3 \Leftrightarrow Ti_1$ | | $Al_1 \Leftrightarrow Ti_1$ | | $La_1 \Leftrightarrow Sr_1$ | | $La_1Al_1 \Leftrightarrow Sr_1Ti_1$ | |
|---|---|---|---|---|---|---|---|---|---|---|---|
| $AlO_2$ | 3 | – | – | 0 | – | – | – | – | – | – | – |
| LaO | 3 | + | + | + | + | + | + | + | + | + | + |
| $AlO_2$ | 2 | – | – | – | – | – | – | – | – | – | – |
| LaO | 2 | + | + | + | + | + | + | + | + | + | + |
| $AlO_2$ | 1 | – | – | – | – | 0 | – | – | – | 0 | – |
| LaO | 1 | + | + | + | + | + | + | 0 | + | 0 | + |
| $TiO_2$ | 1 | 0 | 0 | – | 0 | – | 0 | 0 | 0 | – | 0 |
| SrO | 1 | 0 | 0 | 0 | 0 | 0 | 0 | + | 0 | + | 0 |
| $TiO_2$ | 2 | 0 | 0 | 0 | 0 | 0 | 0 | 0 | 0 | 0 | 0 |
| SrO | 2 | 0 | 0 | 0 | 0 | 0 | 0 | 0 | 0 | 0 | 0 |
| Dipole | | $-6\boldsymbol{d}_0$ | | 0 | | $-4\boldsymbol{d}_0$ | | $-8\boldsymbol{d}_0$ | | $-6\boldsymbol{d}_0$ | |

Table 8. La⇔Sr and LaAl⇔SrTi exchange energies (in eV) calculated with respect to the ideal interface structure using the periodic slab model and the √2×√2 lateral cell with *N* u.c. of LAO. Bold numbers **1**, **2**, and **3** refer to the corresponding atomic planes with respect to the plane of the interface (see Fig. 29a); n/a stands for not applicable. Positions of Sr and Ti atoms in the LAO film are shown.

| *N* | La in **1** of STO | LaAl in **1** of STO | | |
|---|---|---|---|---|
| | Sr in **1** of LAO | SrTi in **1** | SrTi in **2** | SrTi in **3** |
| 1 | 0.55 | –0.28 | n/a | n/a |
| 2 | 0.44 | –0.01 | –0.28 | n/a |
| 3 | 0.42 | –0.26 | –0.14 | –0.41 |

Table 9. Al⇔Ti exchange energies (in eV) calculated with respect to the ideal interface structure with *N* u.c. of LAO using the classical shell model and interatomic potentials. Other details are the same as described in Table 8.

| *N* | Al in **1** of STO | | | Al in **2** of STO | | | Al in **3** of STO | | |
|---|---|---|---|---|---|---|---|---|---|
| | Ti in **1** | Ti in **2** | Ti in **3** | Ti in **1** | Ti in **2** | Ti in **3** | Ti in **1** | Ti in **2** | Ti in **3** |
| 1 | –3.89 | n/a | n/a | –2.12 | n/a | n/a | 0.30 | n/a | n/a |
| 2 | –6.03 | –6.23 | n/a | –4.13 | –4.65 | n/a | –1.63 | –2.24 | n/a |
| 3 | –5.90 | –8.15 | –8.40 | –3.83 | –6.48 | –6.83 | –1.20 | –3.98 | –4.43 |

Table 10. La⇔Sr and LaAl⇔SrTi exchange energies (in eV) calculated with respect to the ideal interface structure with *N* u.c. of LAO using the classical shell model and interatomic potentials. Other details are the same as described in Table 8.

| *N* | La in **1** of STO | LaAl in **1** of STO | | |
|---|---|---|---|---|
| | Sr in **1** of STO | SrTi in **1** | SrTi in **2** | SrTi in **3** |
| 1 | 3.71 | –2.46 | n/a | n/a |
| 2 | 3.17 | –5.14 | –2.44 | n/a |
| 3 | 2.98 | –5.16 | –4.99 | –2.45 |

Table 11 Schematics of selected lowest-energy configurations for LAO(2-3 u.c.)/STO(6 u.c.) ($\sqrt{2}\times\sqrt{2}$ lateral cell) slabs. ΔE (in eV) are the energies relative to that of the abrupt interface. See also Fig. 29. First and second columns indicate an atomic plane and its distance from the interface, resepctively.

| | | LAO(2 u.c.) | | | | LAO(3 u.c.) | | | | | |
|---|---|---|---|---|---|---|---|---|---|---|---|
| $AlO_2$ | 3 | | | | | – | 0 | – | 0 | – | 0 |
| LaO | 3 | | | | | + | + | + | + | + | + |
| $AlO_2$ | 2 | – | 0 | – | 0 | – | – | – | – | – | – |
| LaO | 2 | + | + | + | + | + | + | + | + | + | + |
| $AlO_2$ | 1 | – | – | 0 | – | 0 | – | 0 | – | 0 | – |
| LaO | 1 | + | + | 0 | 0 | 0 | 0 | 0 | 0 | 0 | 0 |
| $TiO_2$ | 1 | 0 | – | – | 0 | 0 | 0 | 0 | – | 0 | – |
| SrO | 1 | 0 | 0 | + | + | 0 | 0 | + | + | 0 | + |
| $TiO_2$ | 2 | 0 | 0 | 0 | – | 0 | – | – | 0 | 0 | 0 |
| SrO | 2 | 0 | 0 | 0 | 0 | + | + | 0 | 0 | 0 | 0 |
| $TiO_2$ | 3 | 0 | 0 | 0 | 0 | – | 0 | 0 | 0 | – | 0 |
| SrO | 3 | 0 | 0 | 0 | 0 | 0 | 0 | 0 | 0 | + | 0 |
| ΔE(PBE) | | –0.76 | | –0.85 | | –1.34 | | –1.40 | | –1.23 | |
| ΔE(B3LYP) | | –1.06 | | –0.87 | | –1.41 | | -------- | | -------- | |

Table 12 - Valence band offsets for 4 u.c. LAO/STO(001) heterojunctions determined using eqns. 7-10 (in eV – number in parentheses is the uncertainty in the last digit)

| Specimen | Eqn. 7 | Eqn. 8 | Eqn. 9 | Eqn. 10 | average |
|---|---|---|---|---|---|
| Tokyo - AR | -0.19(8) | -0.04(7) | -0.09(8) | -0.06(7) | -0.06(10) |
| Tokyo - clean | +0.12(8) | +0.15(7) | +0.14(8) | +0.17(7) | +0.14(10) |
| Augsburg - AR | +0.26(8) | +0.17(7) | +0.24(8) | +0.15(7) | +0.20(10) |
| Augsburg - clean | +0.30(8) | +0.25(7) | +0.41(8) | +0.39(7) | +0.34(10) |

AR – as received. Clean – after oxygen plasma cleaning.

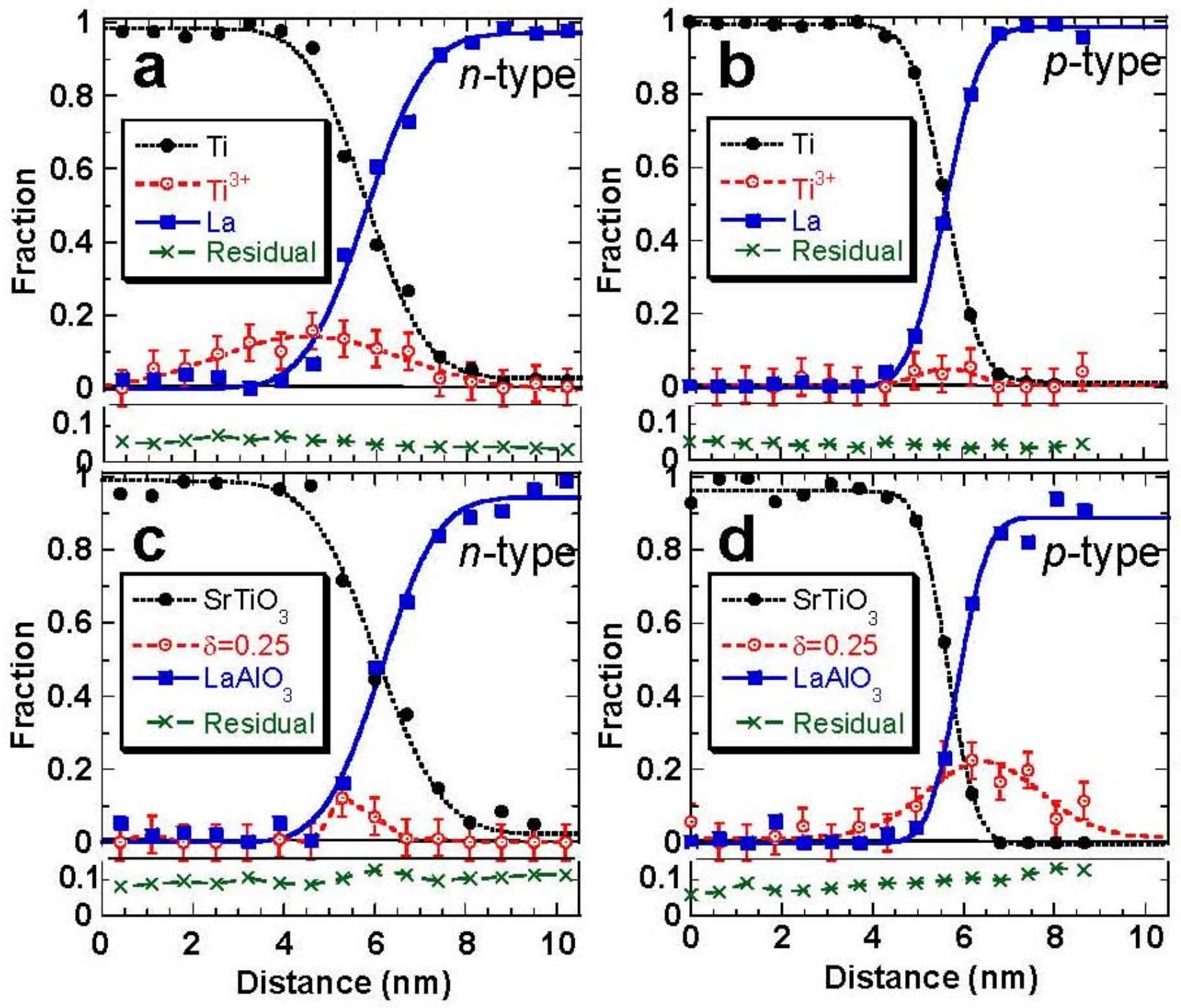

Fig. 1. Elemental distributions across the interface for LAO/$TiO_2$-terminated STO (a & c) and LAO/SrO-terminated STO(001) (b & d) interfaces, based on spatially-resolved Ti L-edge (a & b) and O *K*-edge (c & d) electron energy loss spectral profiles obtained in HRTEM. The quantity δ is a measure of the O vacancy concentration in the STO layer ($SrTiO_{3-\delta}$), and can be determined from the O K-edge loss spectra. Taken from ref. [8].

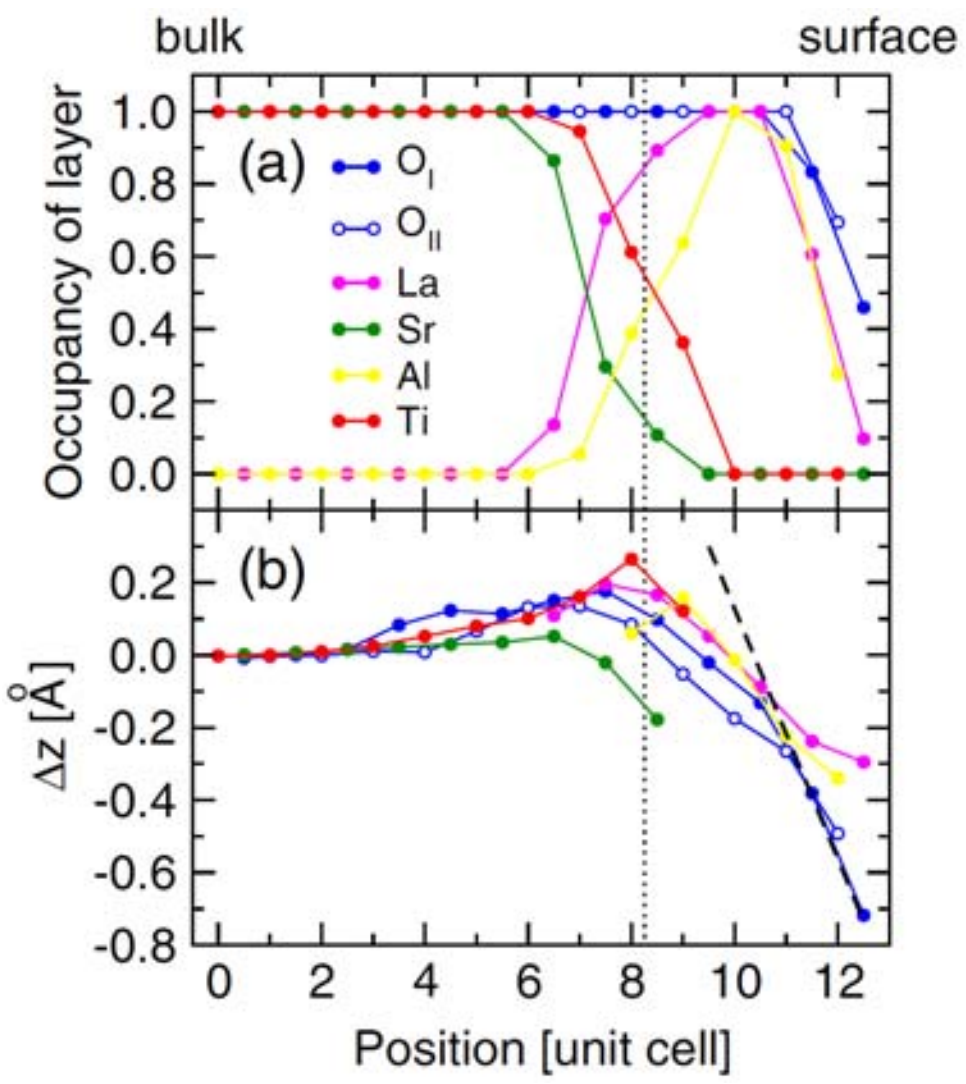

Fig. 2. Elemental distributions (a) and lattice dilations (b) across the LAO/$TiO_2$-terminated STO(001) interface, as determined from surface x-ray diffraction. Taken from ref. [10].

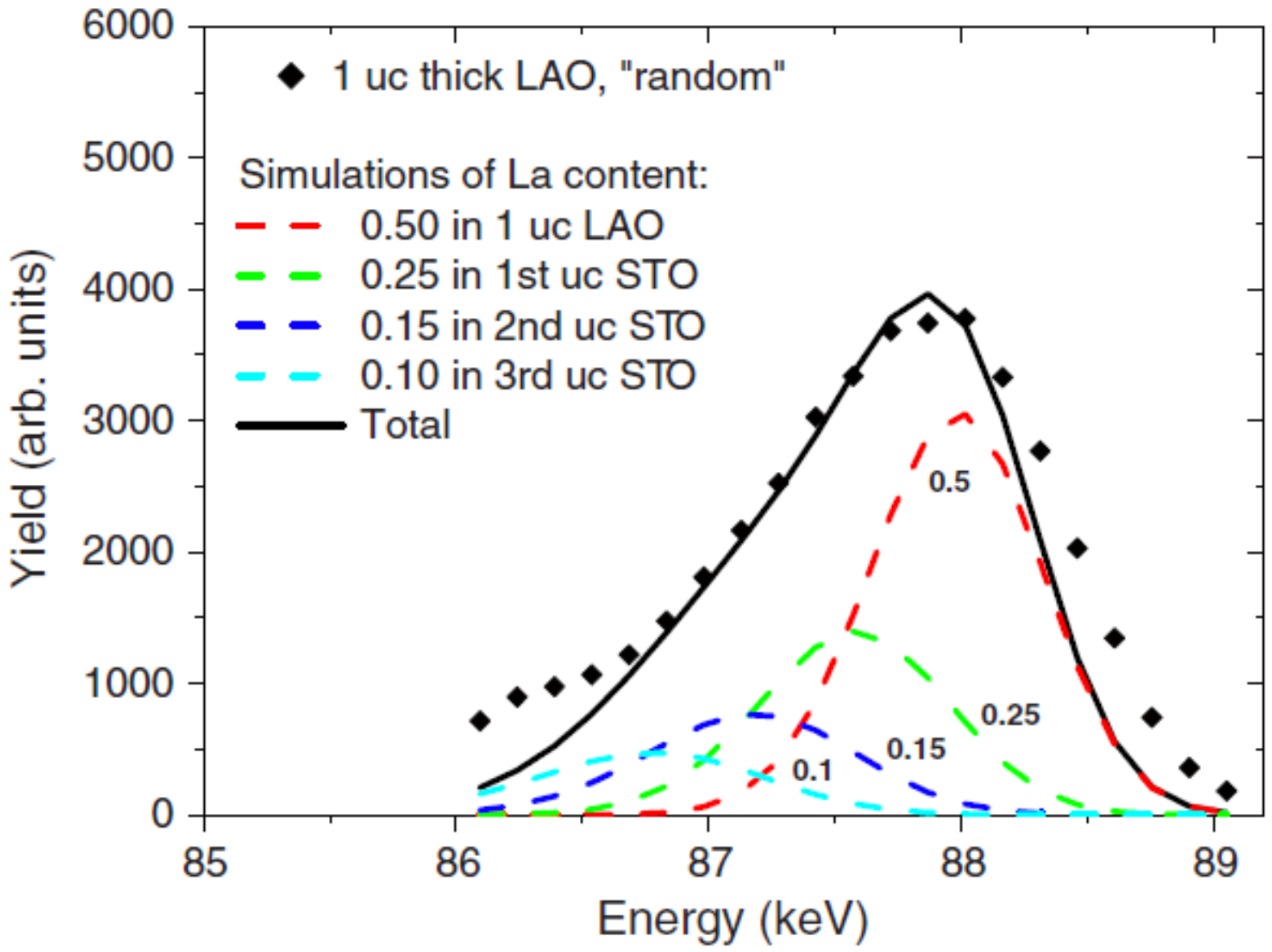

Fig. 3. Medium-energy ion scattering spectrum measured in the random geometry for 1 u.c. LAO/$TiO_2$-terminated STO(001), along with an optimized simulation broken down into contributions from specific LAO and STO unit cells. Taken from ref. [11]

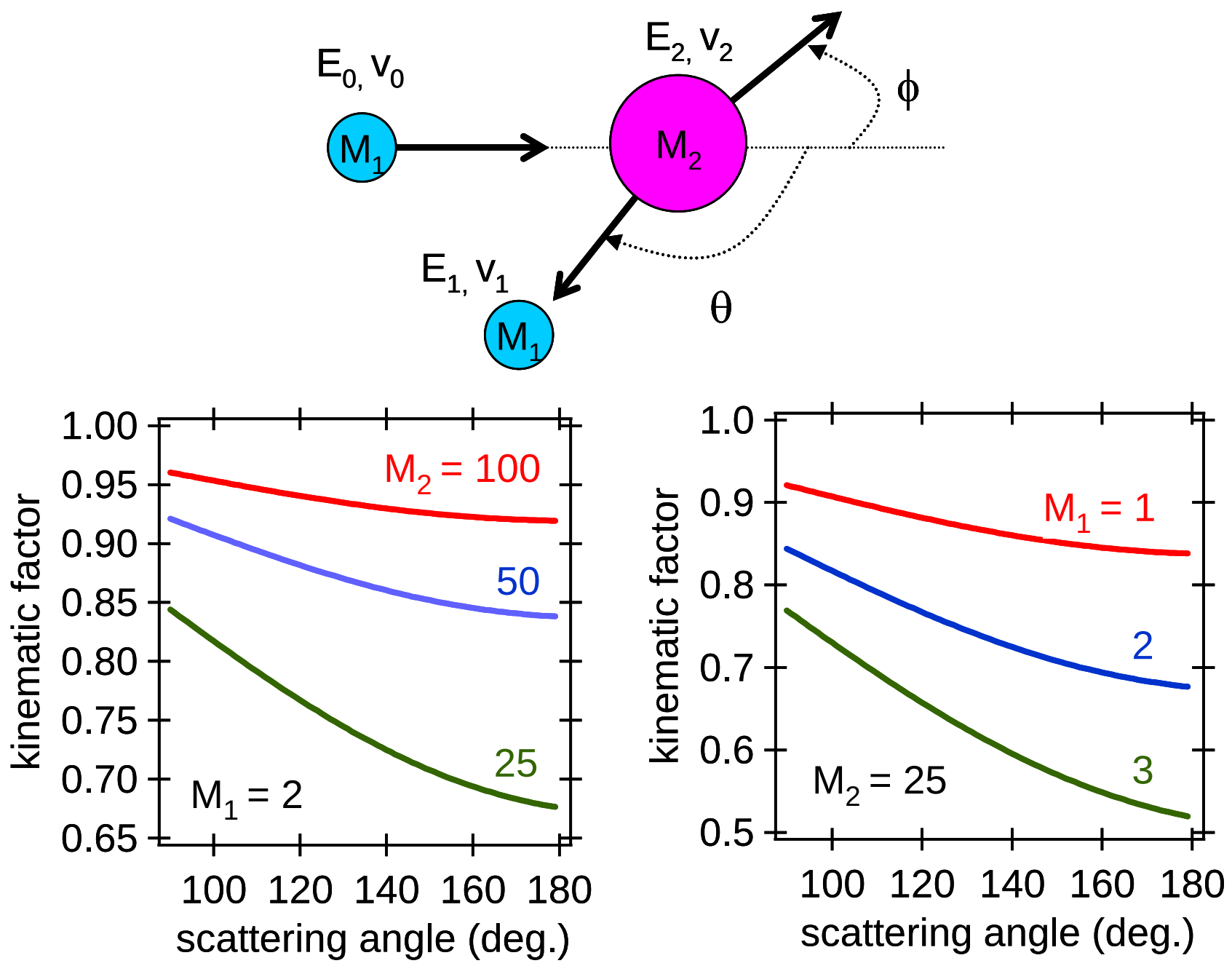

Fig. 4. Schematic diagram of the Rutherford scattering process (top) and plots of the RBS kinematical factor (eq. 1) for a fixed incident particle mass and a range of target nuclei masses (lower left), as well as a range of incident particle masses and a fixed target nucleus mass (lower right).

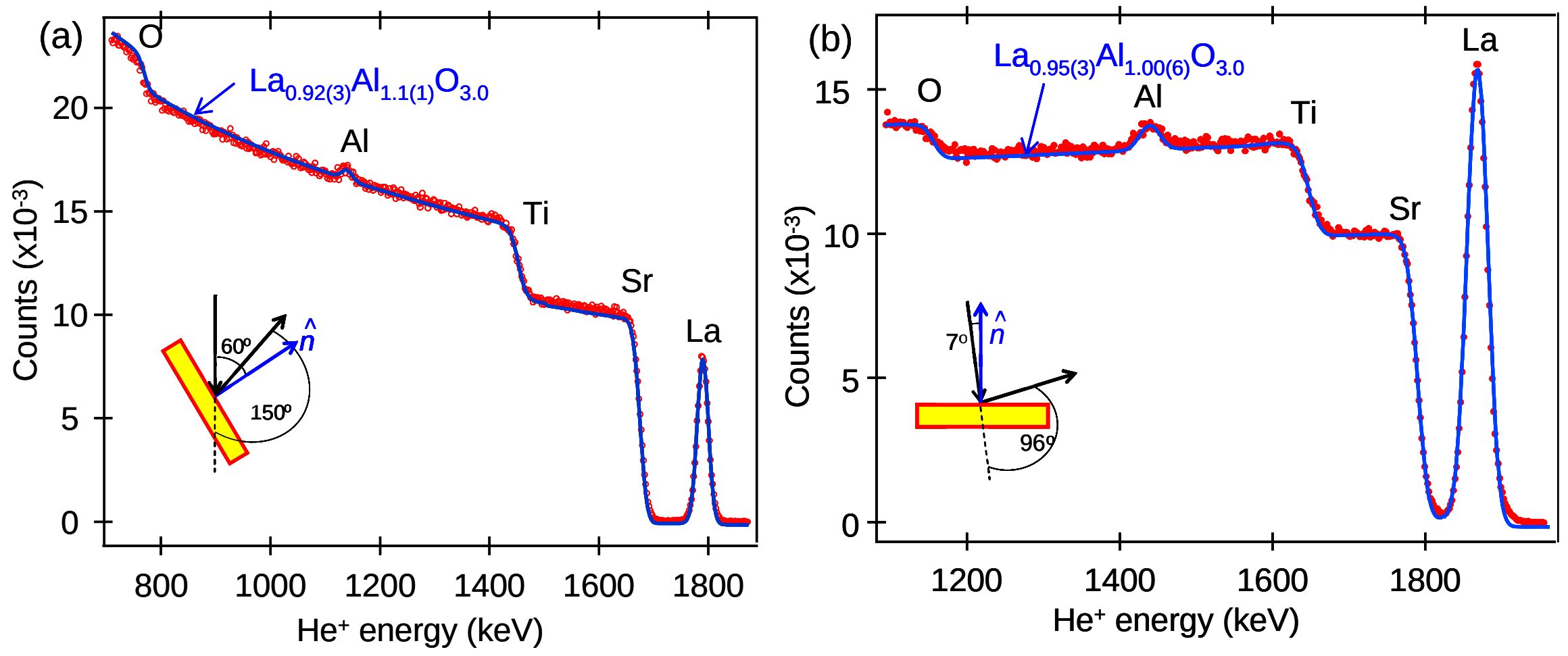

Fig. 5. Experimental RBS spectra using 2 MeV He$^+$ for 25 u.c. LAO/STO(001) grown at Augsburg for scattering angles of 150° (a) and 96° (b), along with optimized SIMNRA simulations to obtain the film composition.

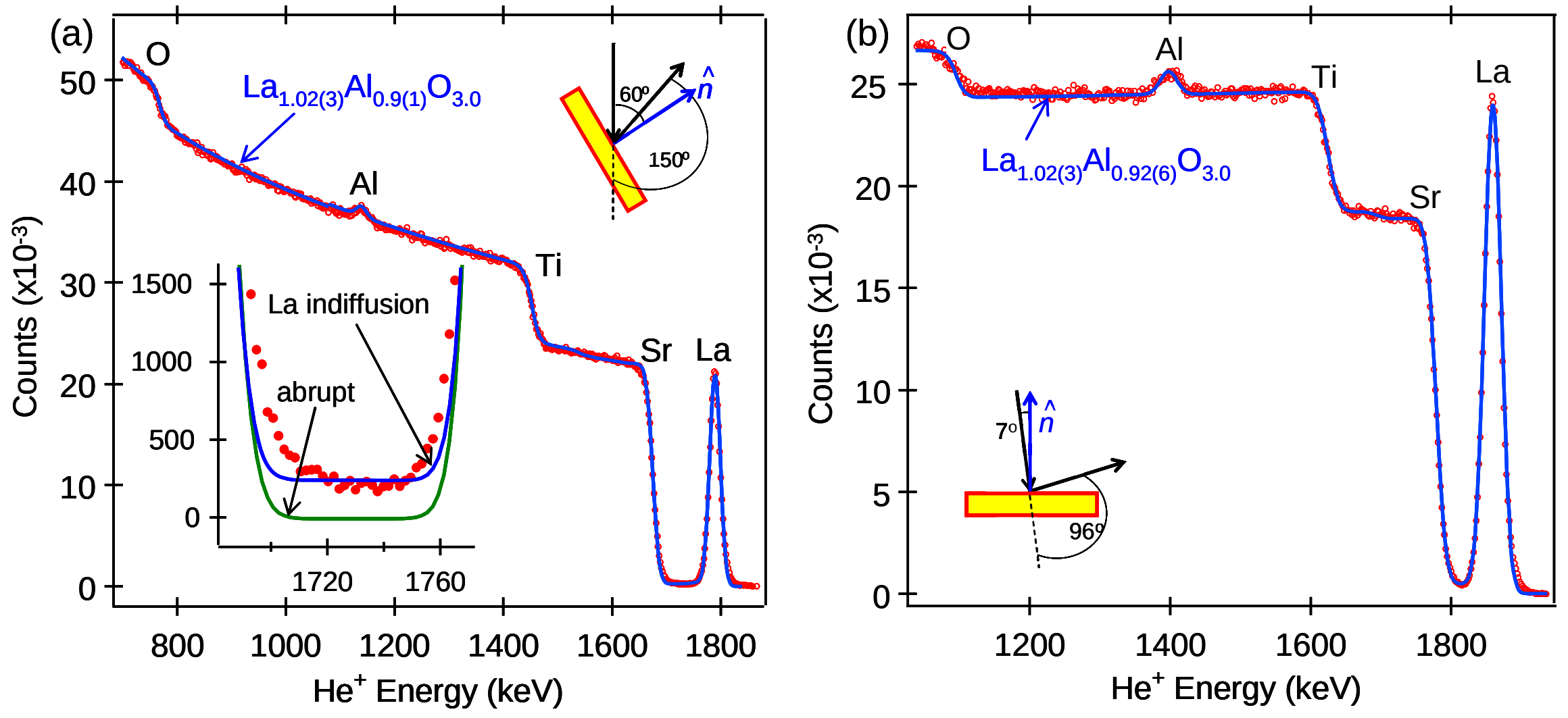

Fig. 6. Same as Fig. 5, but for a film grown at Tokyo using a laser fluence of 0.7 J/cm$^2$. Inset in (a) – the valley between the La and Sr peaks along with abrupt and intermixed interface simulations.

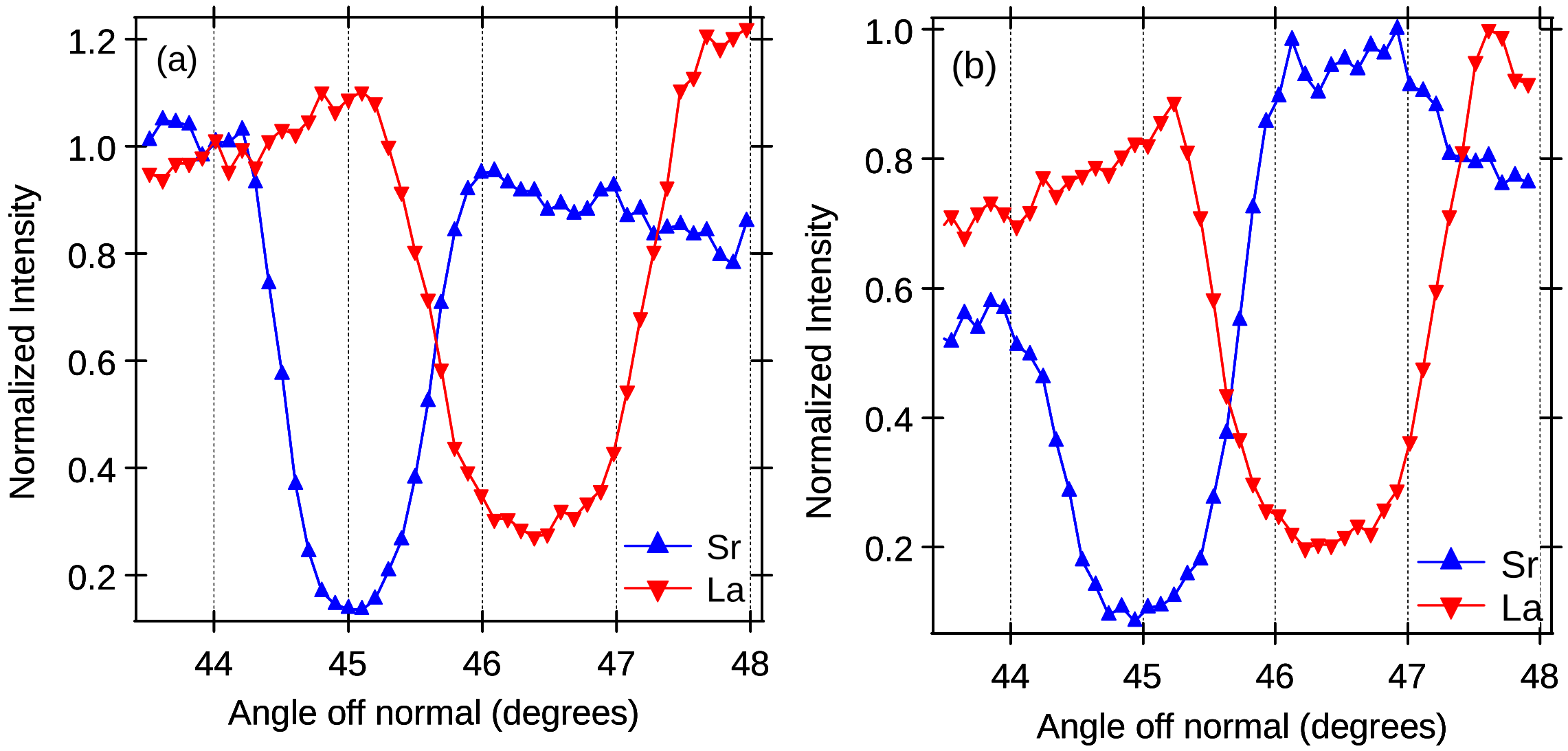

Fig. 7. RBS rocking curves near [101] for the La and Sr peaks using 2 MeV $He^+$ for 25 u.c. LAO/STO(001) grown at Augsburg(a) and Tokyo (b).

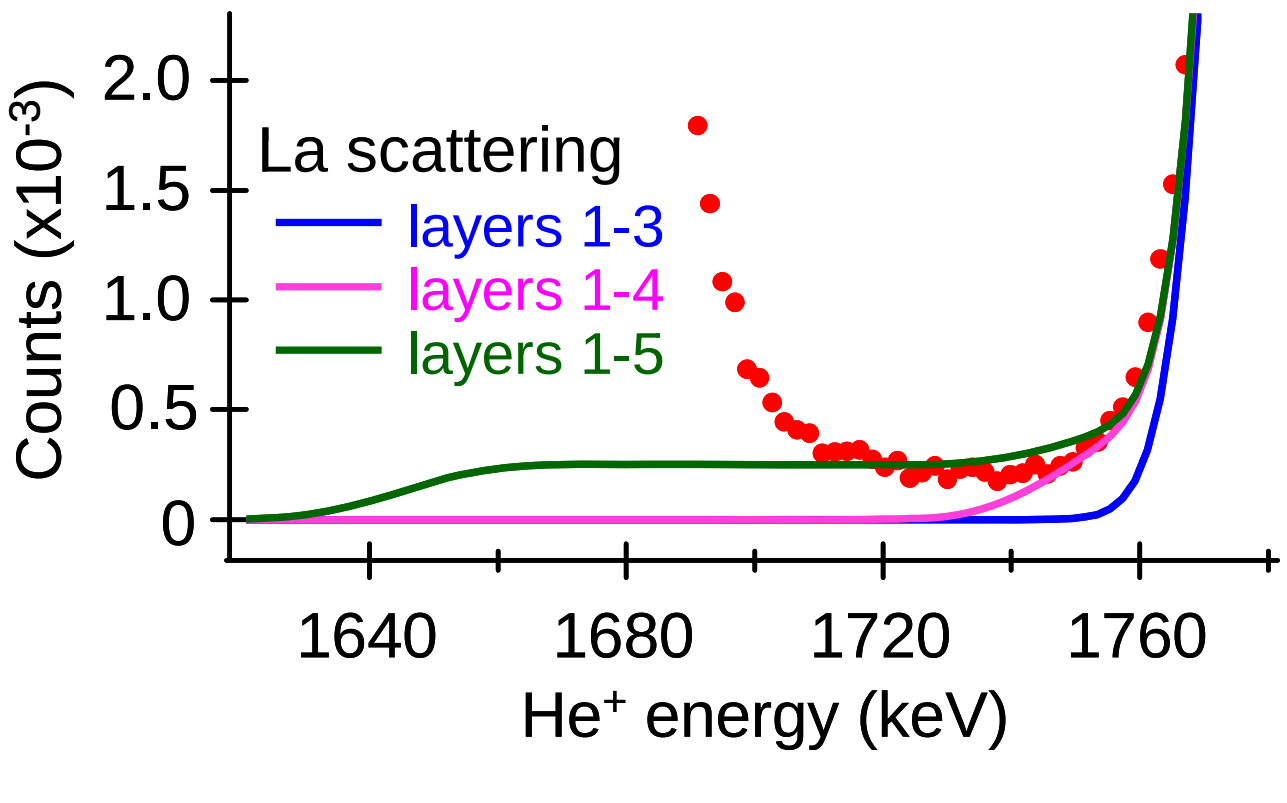

Fig. 8. The valley between the La and Sr RBS peaks taken from Fig. 6a along with simulations based on La indiffusion into the STO showing the contributions of the individual layers defined in Table 3.

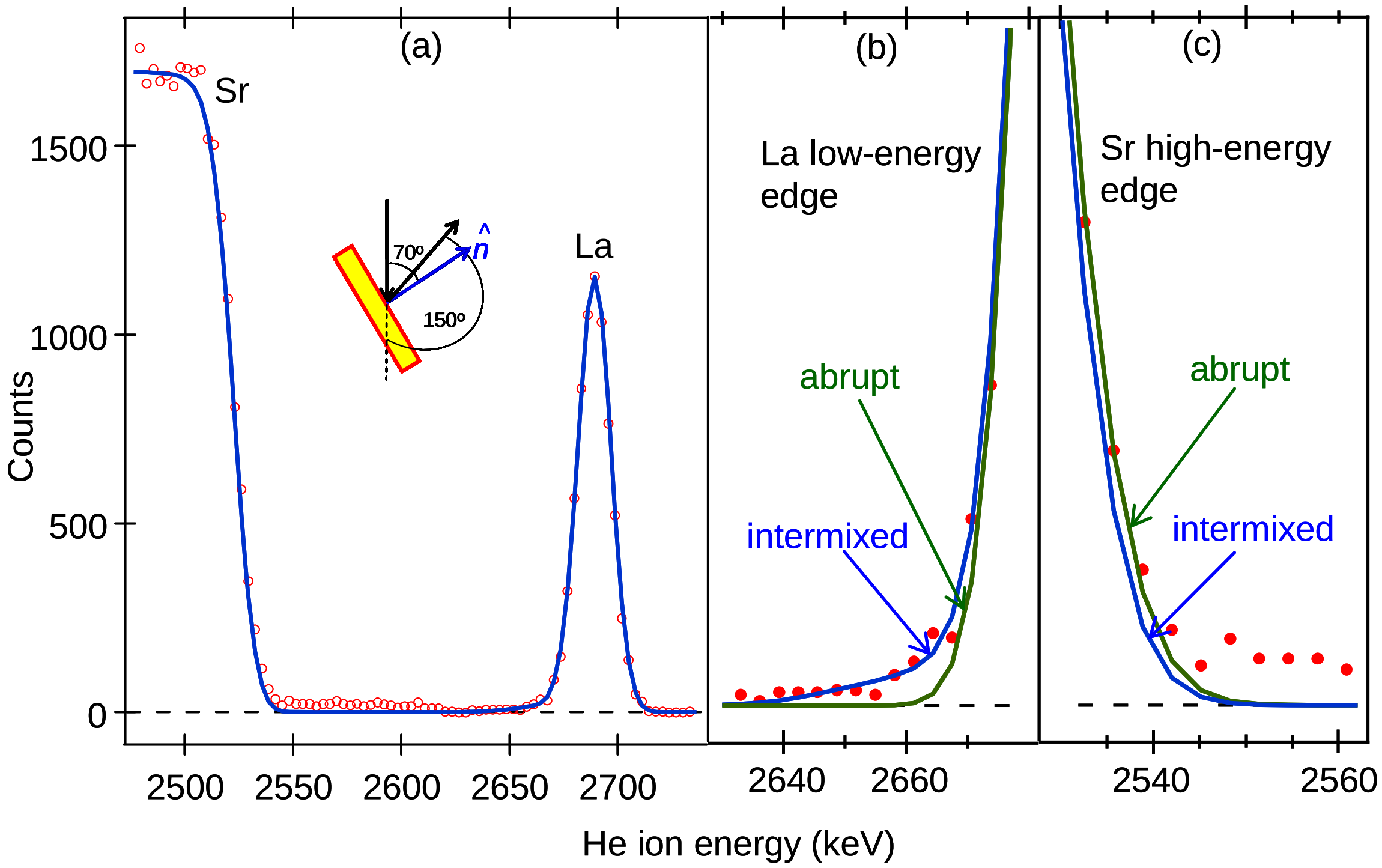

Fig. 9. Experimental RBS spectrum using 3 MeV He$^+$ for 25 u.c. LAO/STO(001) grown at Tokyo using a scattering angle of 150$^o$ (a), along with an expanded view of the low-energy side of the La peak (b), and the high-energy side of the Sr peak (c). Also shown are optimized SIMNRA simulations in which La and Sr interdiffusion have been modeled, along with abrupt-interface simulations.

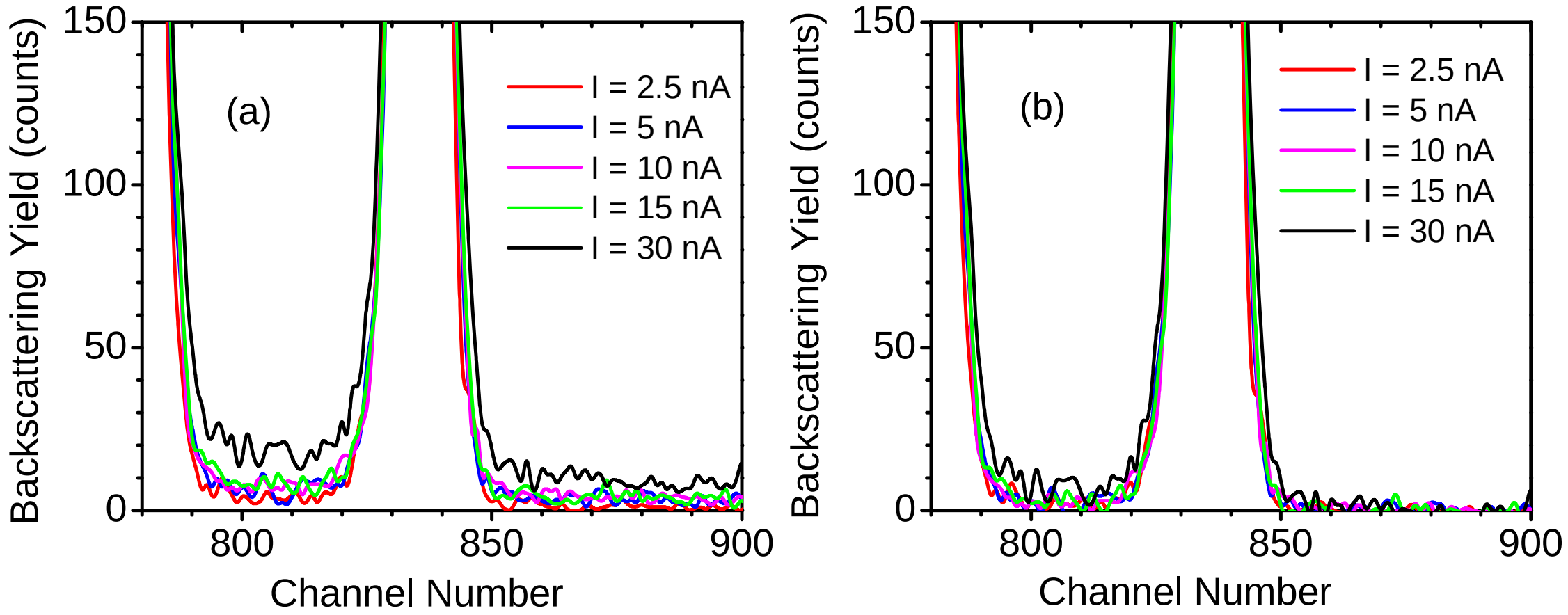

Fig.10. (a) Experimental RBS spectra using 2 MeV He$^+$ for 25 u.c. LAO/STO(001) grown at Tokyo for a scattering angle of 150$^o$ with various incident beam currents to illustrate the effect of pulse pile up. (b) Spectra from (a) after SIMNRA correction for pulse pile up.

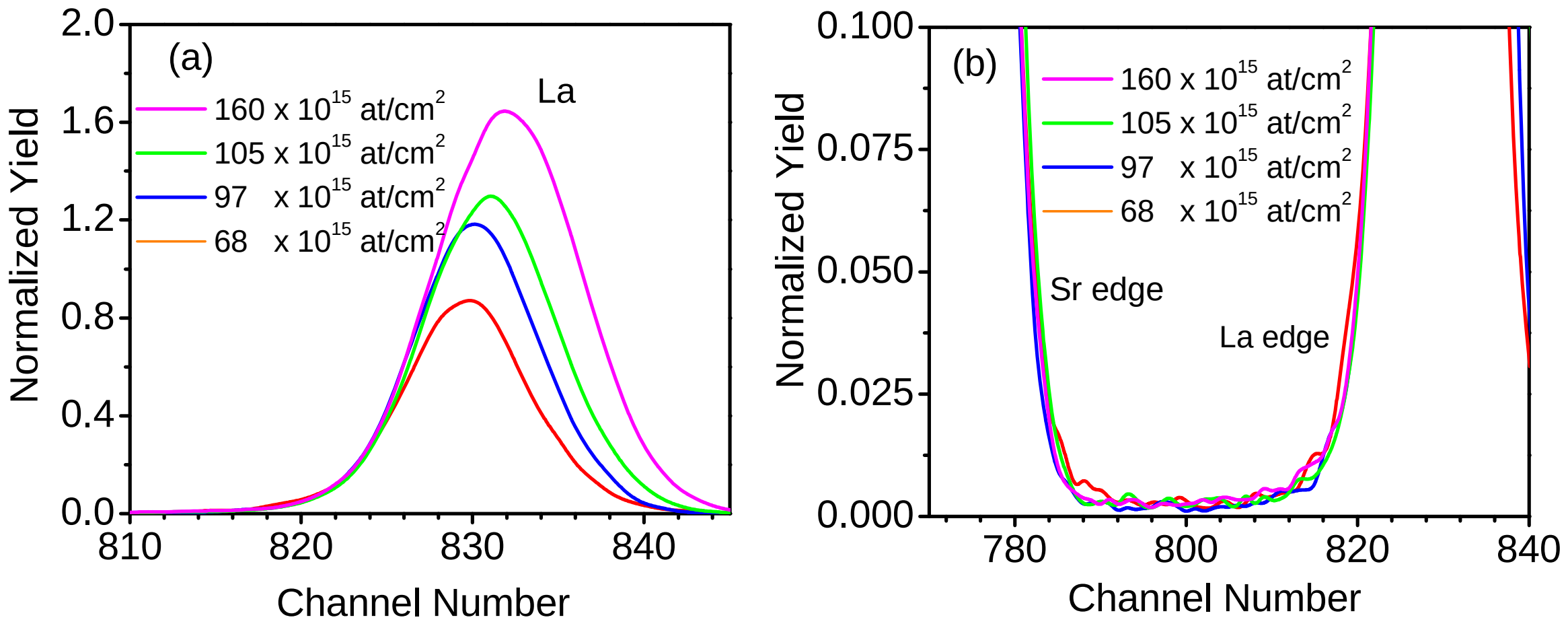

Fig.11. RBS La peak (a) and valley between the La and Sr peaks (b) at 2 MeV $He^+$ for LAO/STO(001) films grown at Tokyo at a scattering angle of 150° for various LAO film thicknesses.

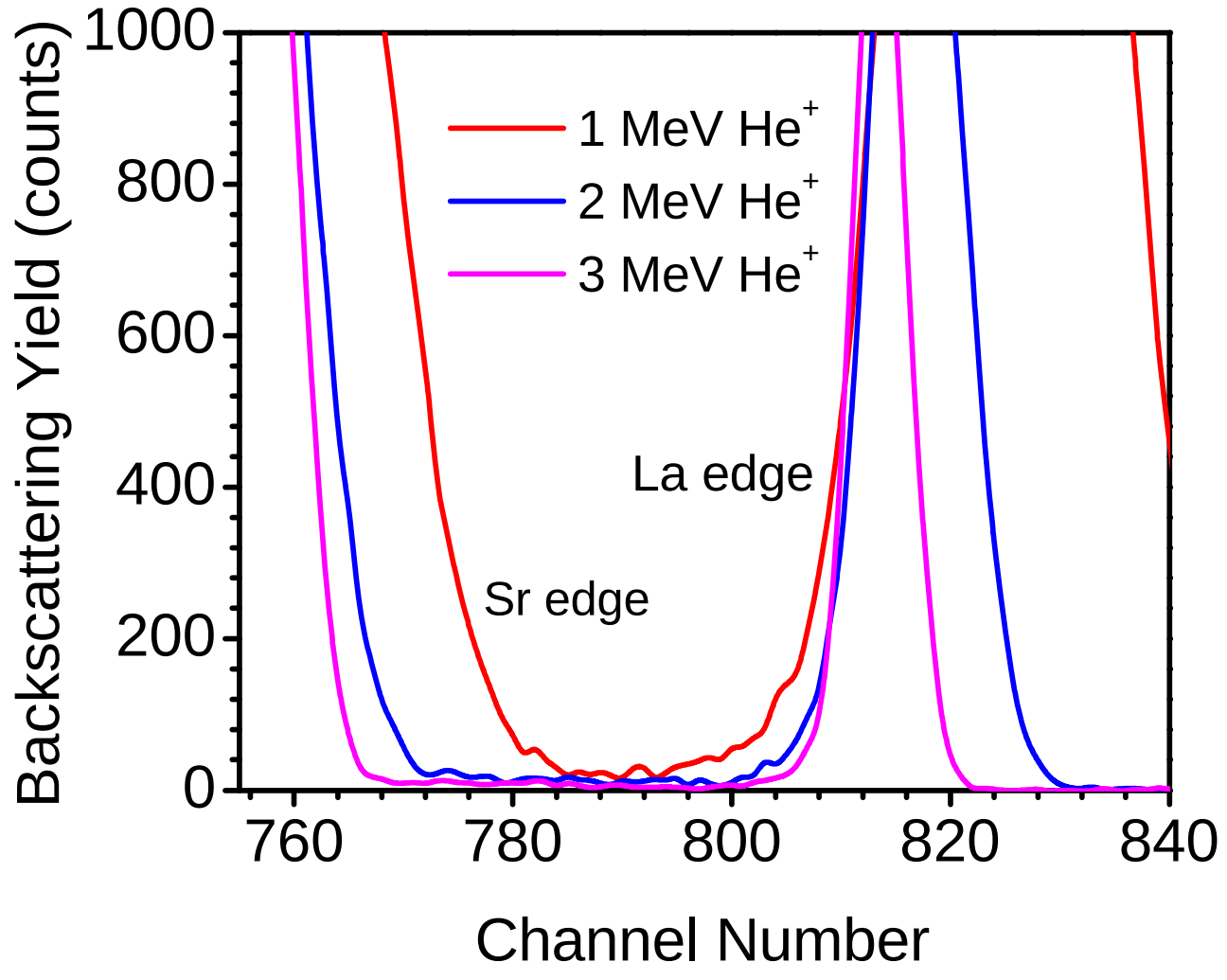

Fig.12. Valley between the La and Sr RBS peaks for 25 u.c. LAO/STO(001) grown at Tokyo at a scattering angle of 150° as a function of incident beam energy.

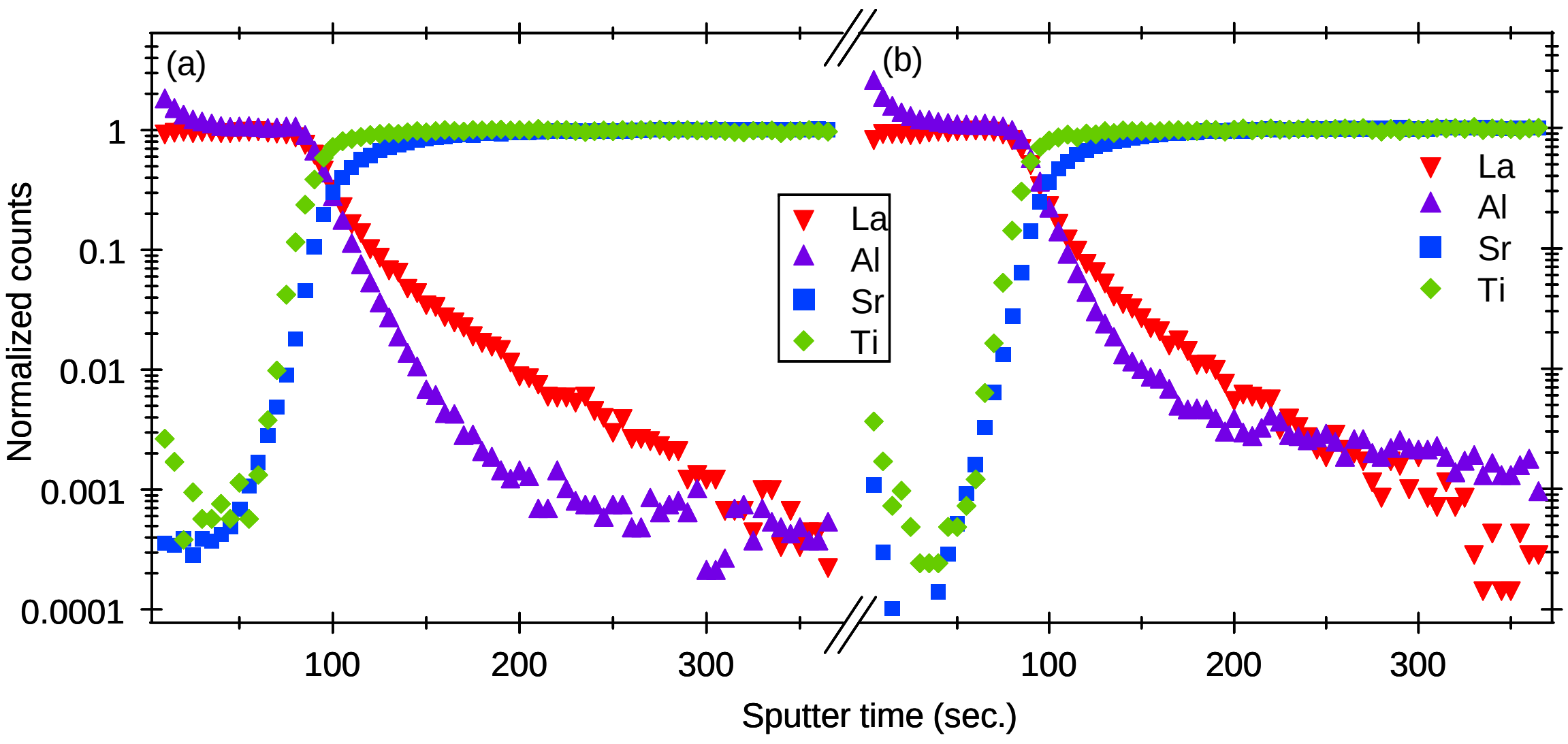

Fig.13. ToF-SIMS depth profiles for 25 u.c. LAO/STO(001) from Augsburg (a) and Tokyo at 0.7 J/cm$^2$ laser fluence (b). The count rates have been normalized to unity near the surface for La and Al, and deep in the bulk for Sr and Ti.

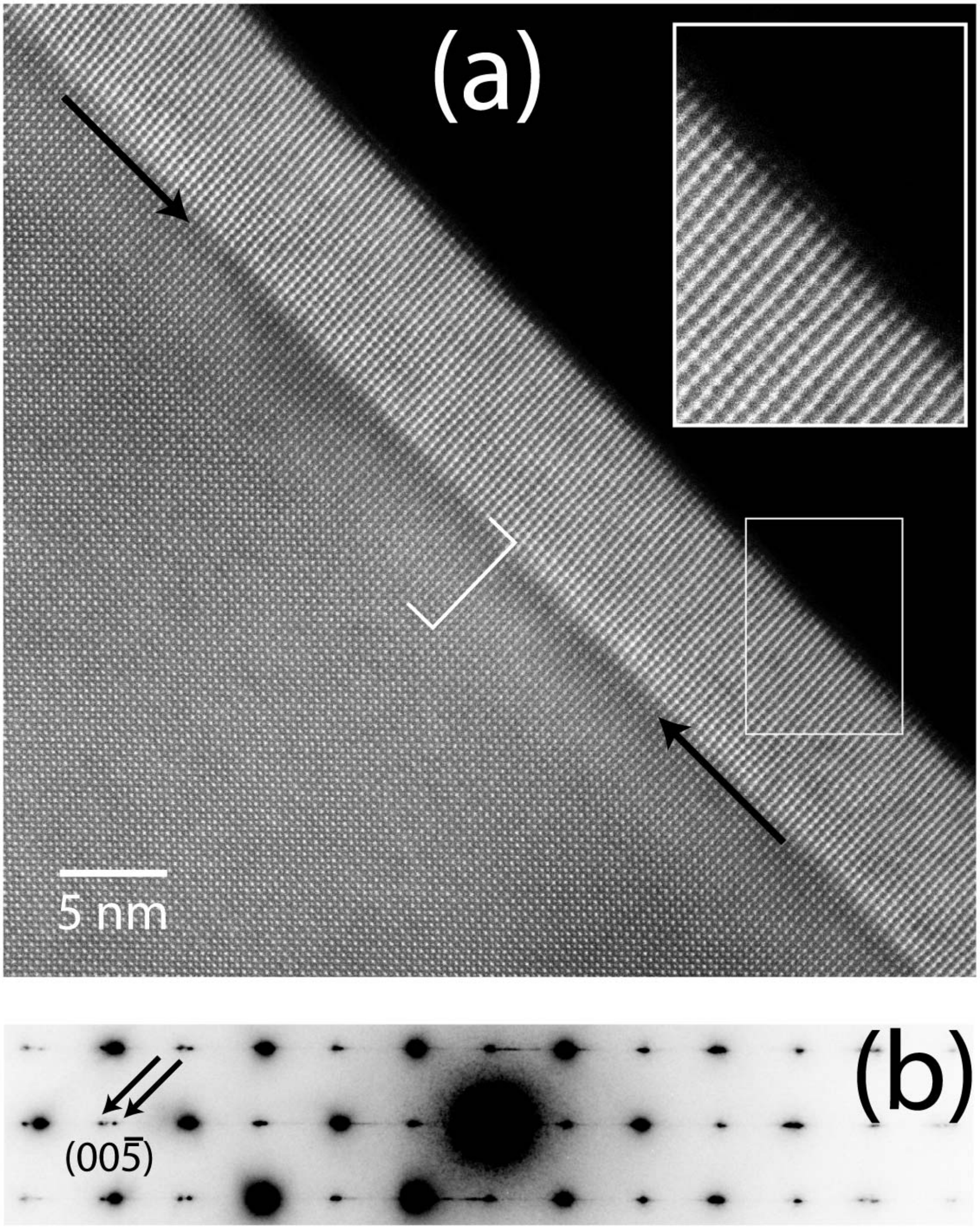

Fig.14. (a) Representative HAADF STEM image from a 25 u.c. Augsburg film. The LAO film is coherent with the underlying STO substrate with a slight misorientation. Arrows mark a region of reduced contrast in the substrate which was found throughout the sample. (b) The nano-area electron diffraction reveals out-of-plane compressive strain and in-plane coherence with the substrate for the film

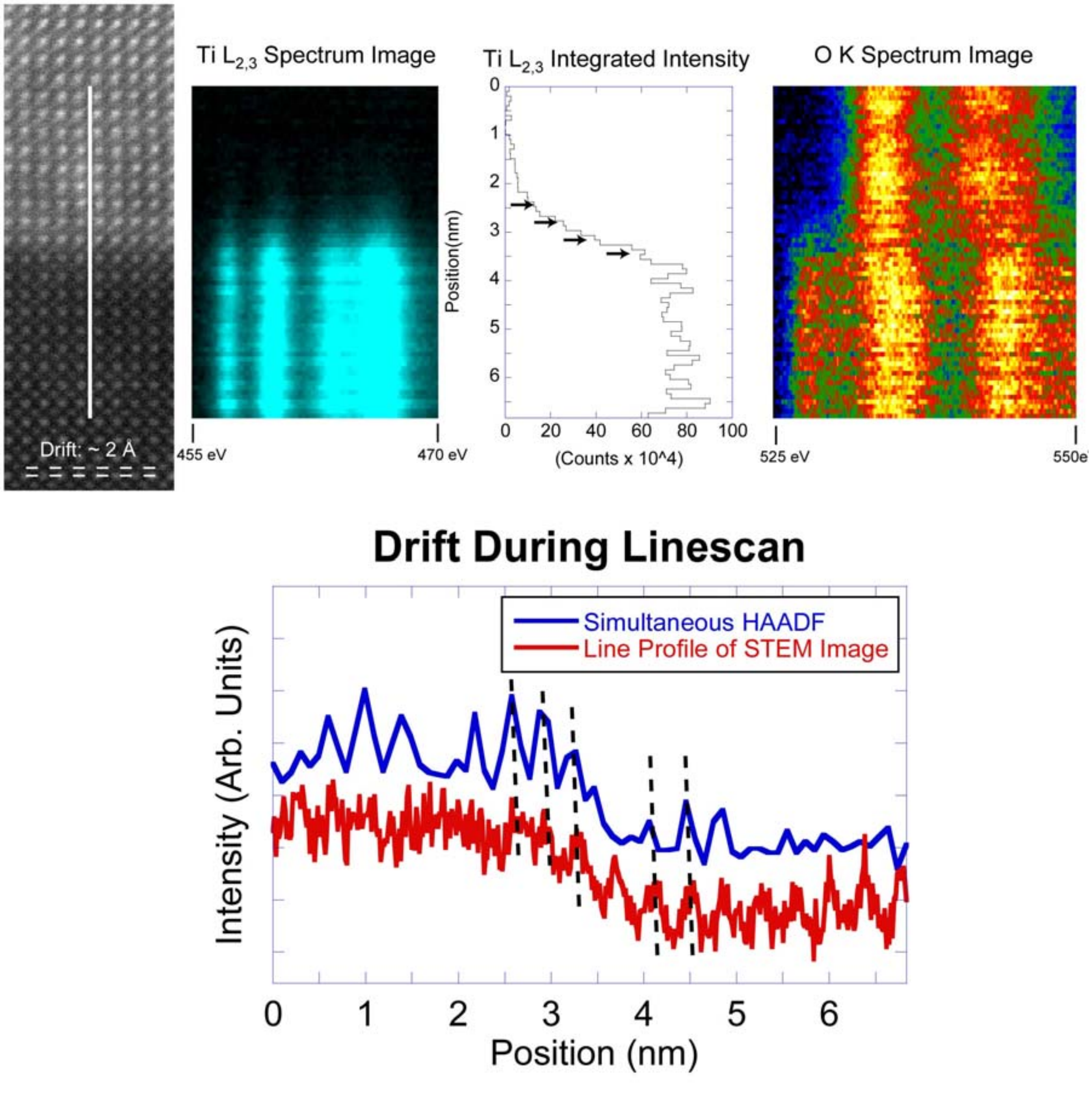

Fig. 15. Ti $L_{2,3}$ EELS line scan from a 25 u.c. Augsburg film revealing Ti outdiffusion into the LAO film up to 3 u.c. Also shown is the O K-edge profile.

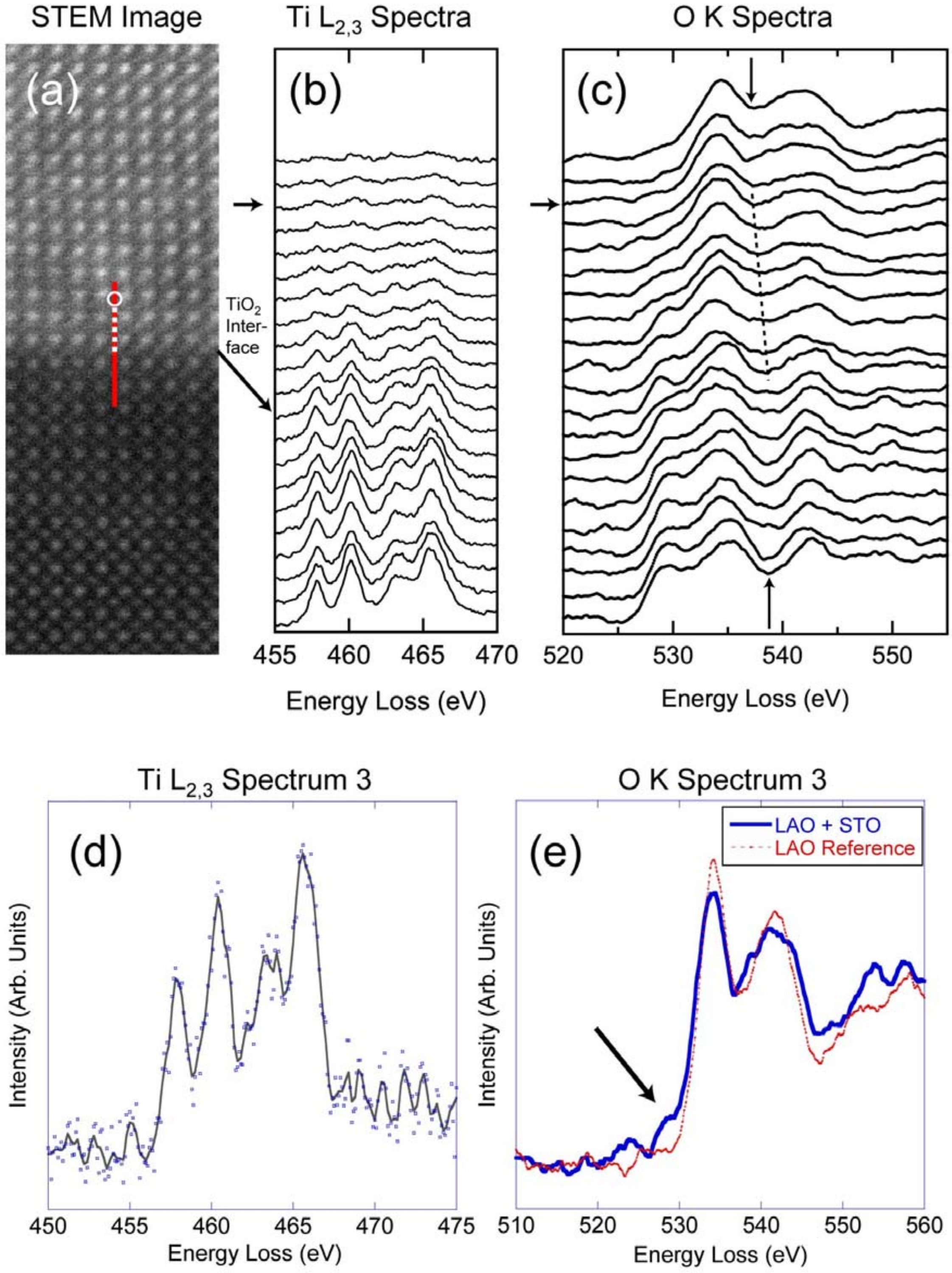

Fig. 16. Individual Ti $L_{2,3}$ (b) and O K (c) loss spectra taken every 1 Ǻ along the line shown in the HAADF image in (a) for a 25 u.c. Augsburg film. Ti $L_{2,3}$ (d) and O K (e) loss spectra measured within the 3$^{rd}$ u.c. of LAO out from the interface showing the clear presence of Ti, along with a weak O K-edge pre-edge feature characteristic of STO.

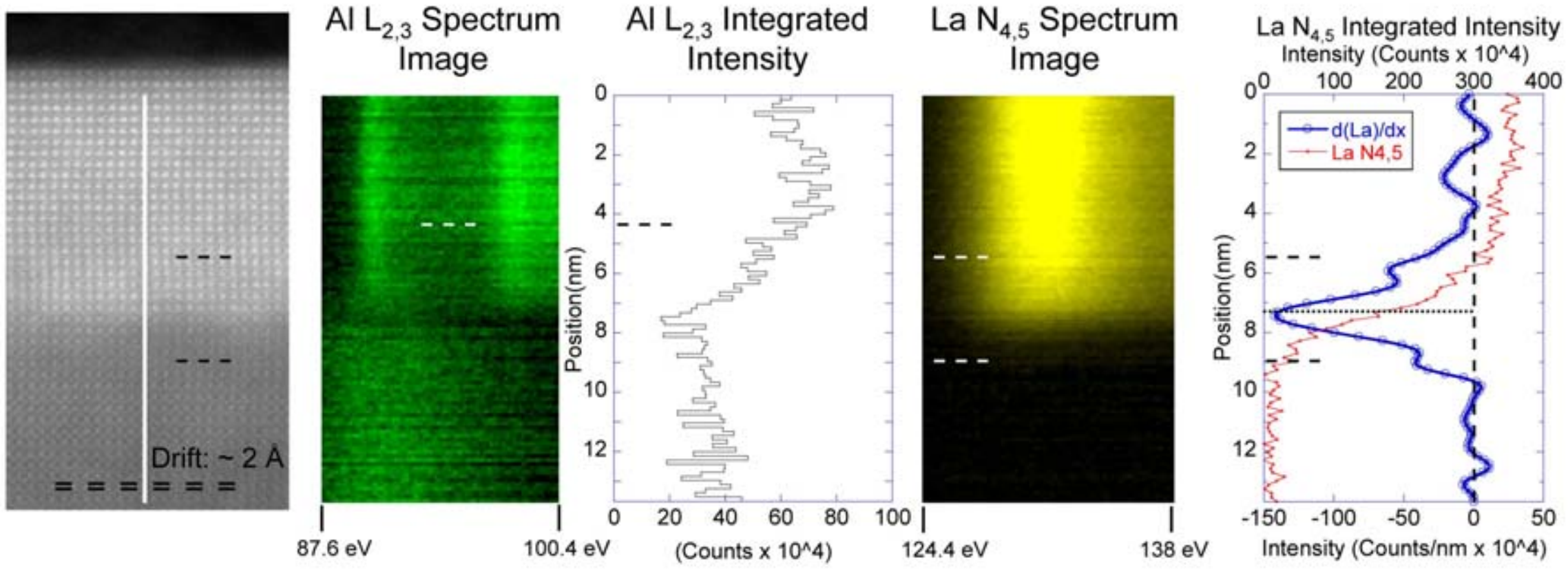

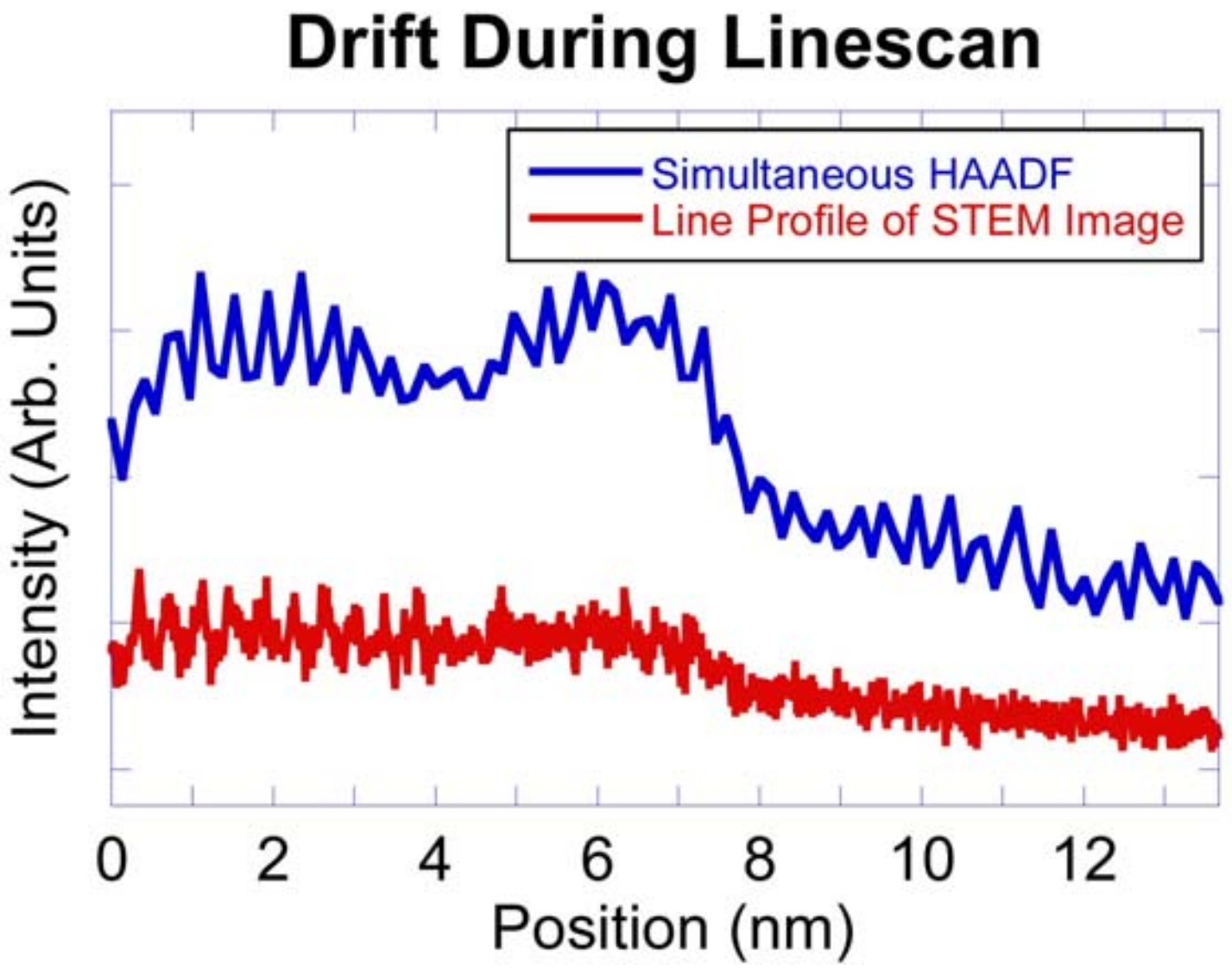

Fig. 17. La $N_{4,5}$ EELS line scan and its derivative from a 25 u.c. Augsburg film showing La diffusion into the STO film up to 4 u.c. Also shown is the Al $L_{2,3}$ spectral image, which interferes with the Sr $M_{4,5}$ and the Ti $M_1$ loss regions on the STO side of the interface.

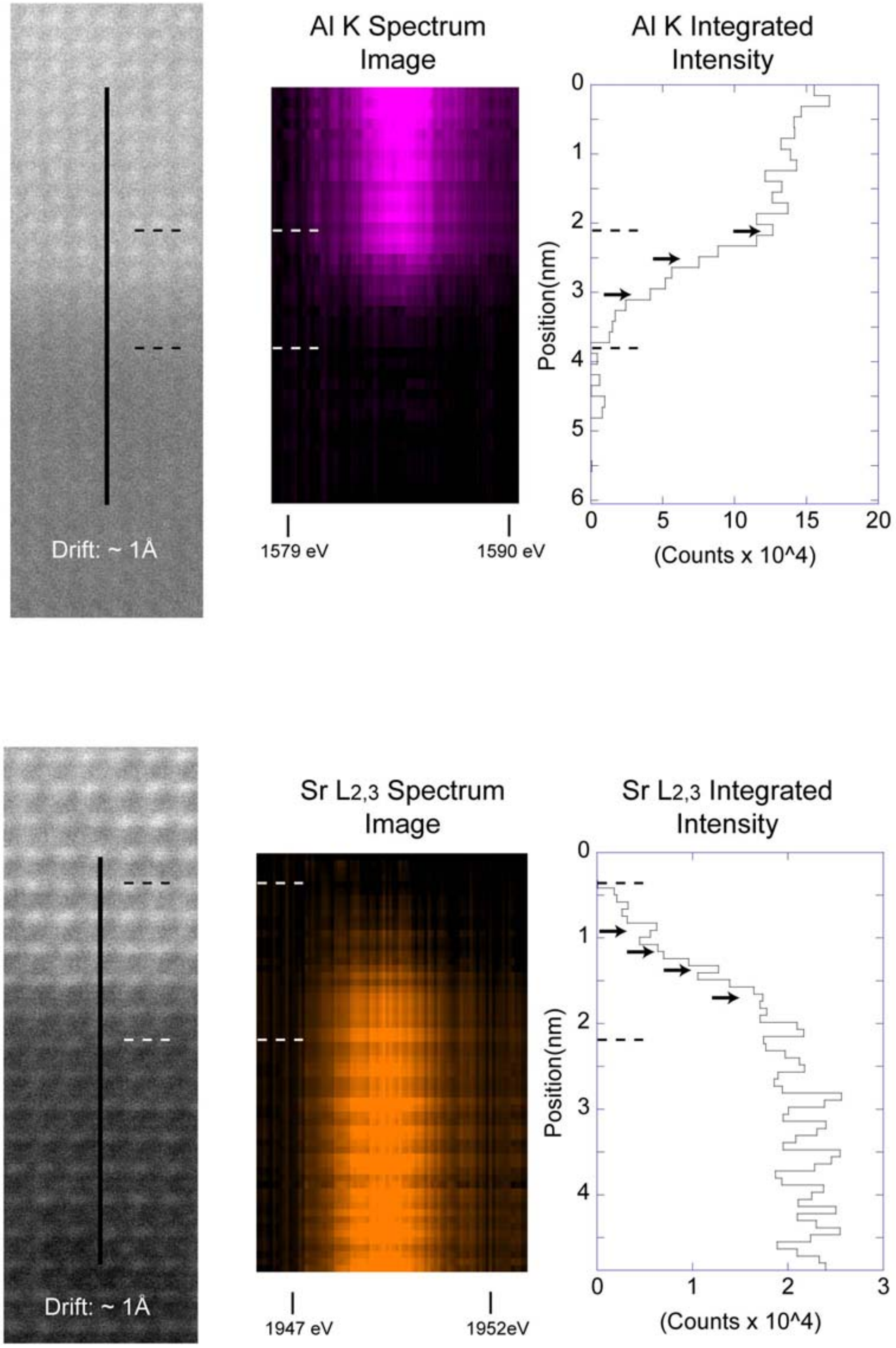

Fig. 18. Al K and Sr $L_3$ EELS line scans and integrated intensities from a 25 u.c. Augsburg film revealing diffusion lengths of ~ 2 u.c. for Al into the STO and ~4 u.c. for Sr into the LAO.

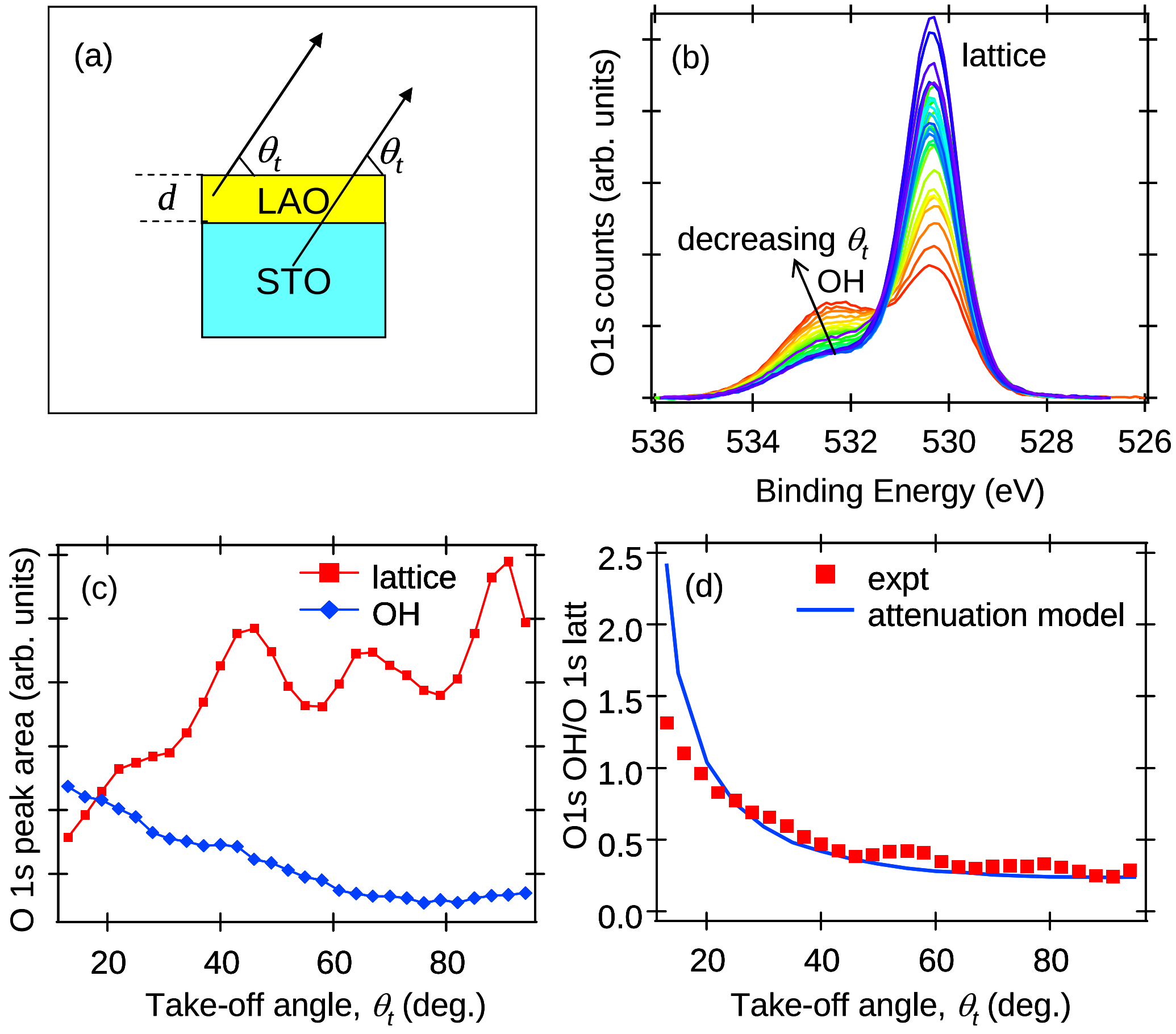

Fig. 19. APXPS schematic (a); O 1s spectra for a range of take-off angles ($\theta_t$) for 4u.c. LAO/STO(001) (b); O1s lattice and OH peak areas vs. $\theta_t$ (c); Measured and calculated O 1s:OH peak area ratio vs. $\theta_t$ (d).

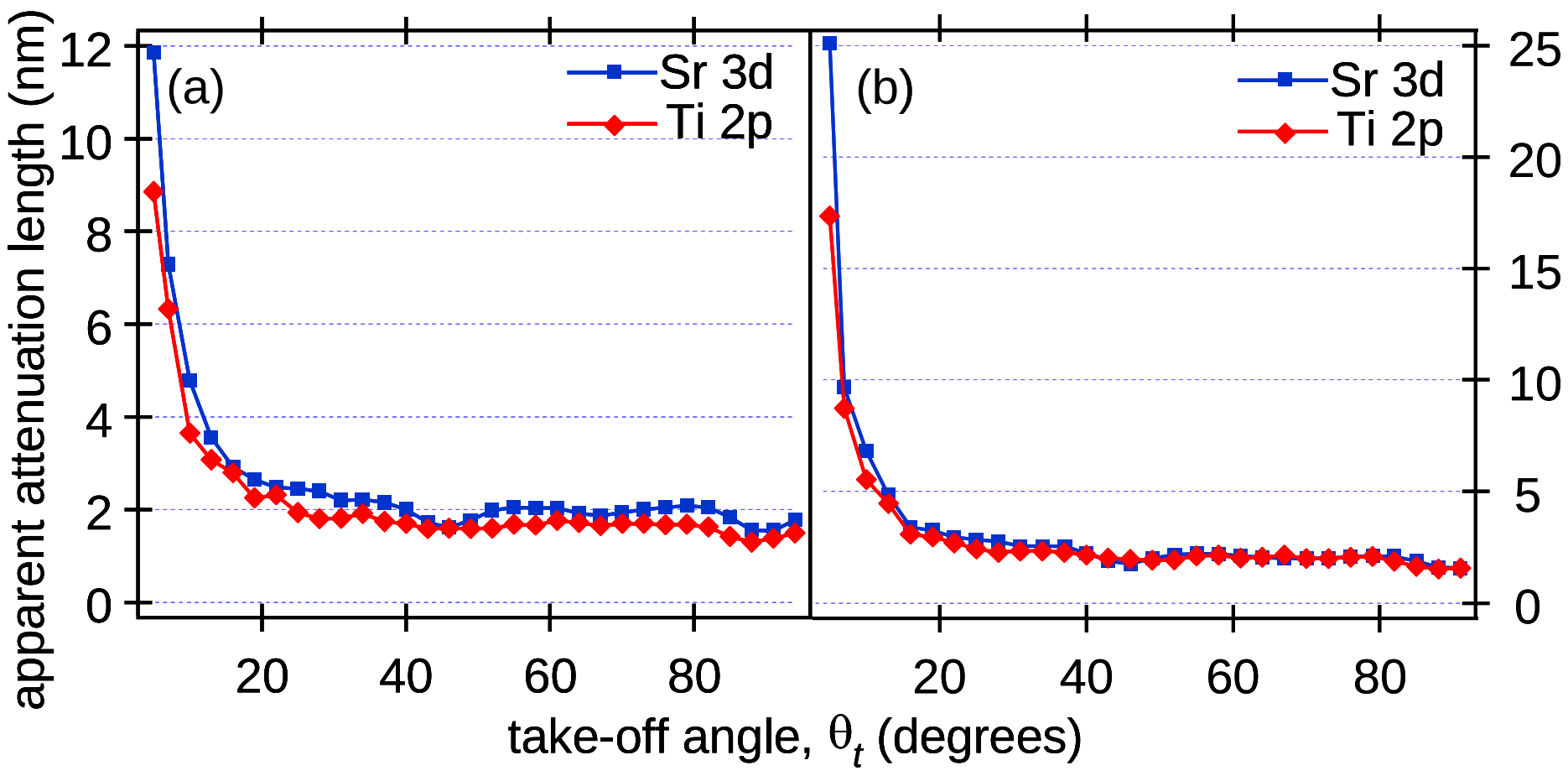

Fig. 20. Plots of $[d\sin\theta_t]/\{\ln[I_{STO}(\theta_t)] - \ln[I_{4uc}(\theta_t)]\}$ vs. $\theta_t$ for Sr 3p and Ti 2p for 4 u.c. LAO/STO(001) from Augsburg (a) and Toyko (b). This quantity should be constant with $\theta_t$ if the interface is abrupt, and should yield reasonable values of the electron attenuation lengths, $\lambda$.

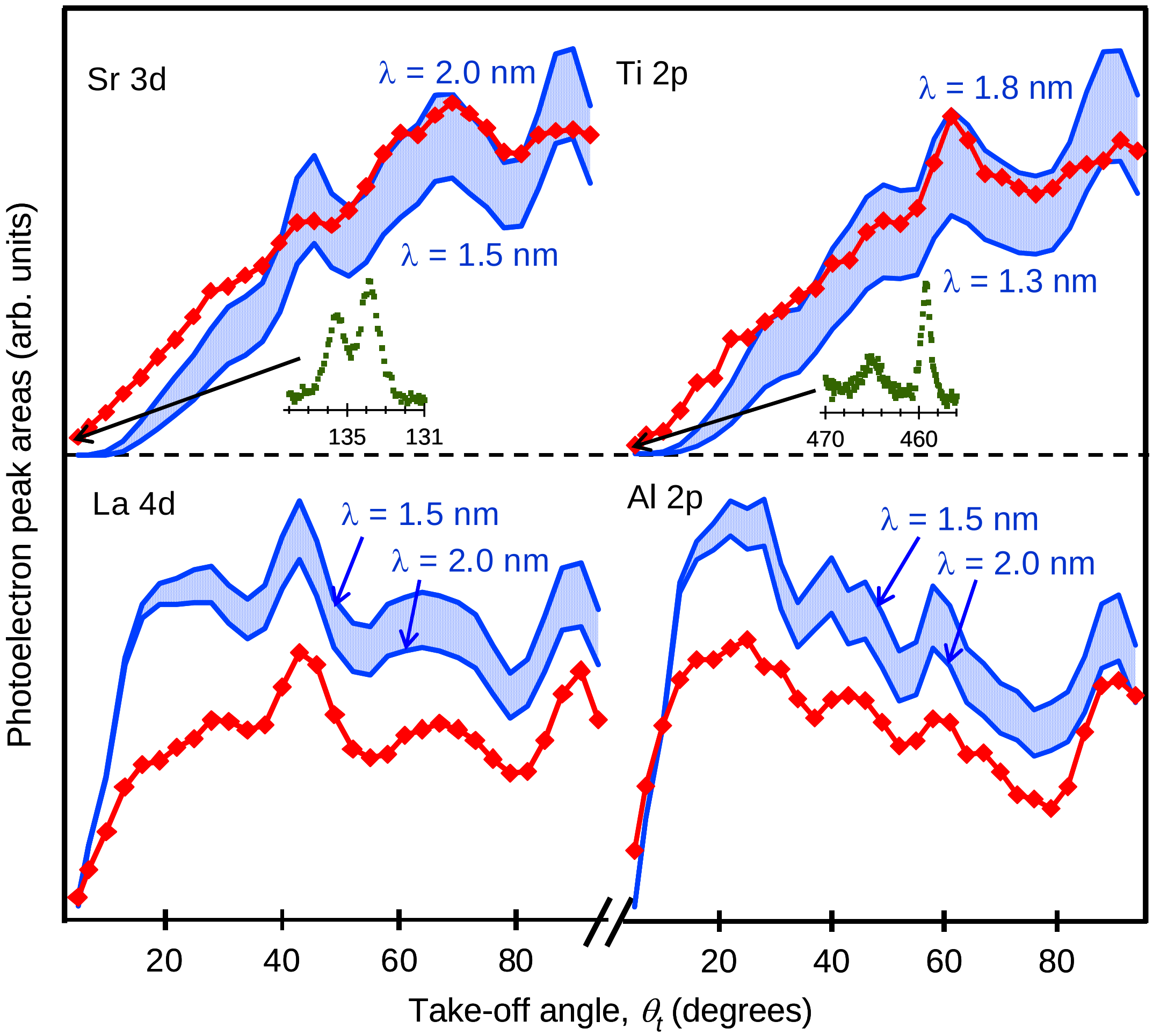

Fig. 21. Polar scans in the (100) azimuthal plane of Sr 3d, Ti 2p, La 4d and Al 2p intensities for 4 u.c. LAO/STO(001) from Augsburg (diamonds), along with analogous scans for bulk STO (top) and a 25 u.c. LAO film from Augsburg (bottom), scaled by factors of $\exp(-d/\lambda \sin\theta_t)$ (top) and $[1 - \exp(-d/\lambda \sin\theta_t)]$ (bottom) (solid curves connected by hatch). A reasonable range of $\lambda$ values has been included, as discussed in the text

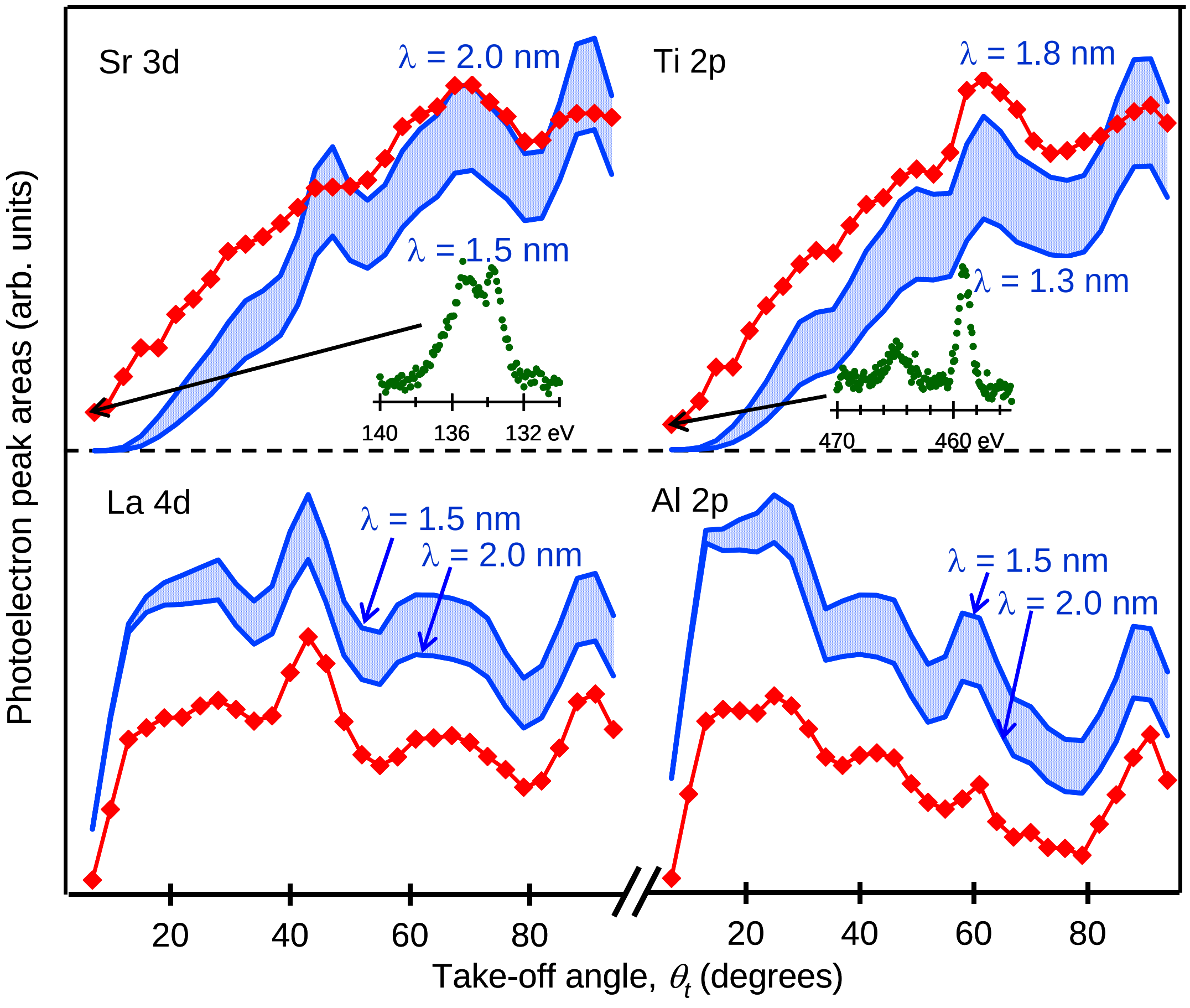

Fig. 22. Same as 21, except for 4 u.c. LAO/STO(001) from Tokyo grown at 0.7 J/cm$^2$ laser fluence.

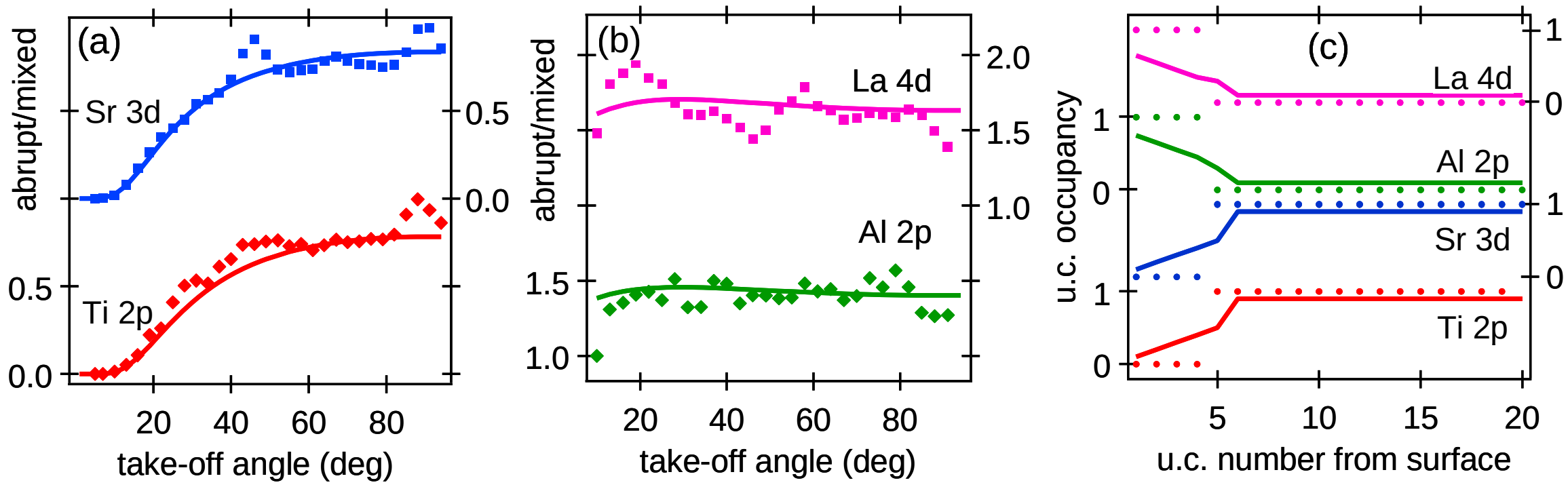

Fig. 23. $I_{4uc}(\theta_t)/I_{STO}(\theta_t)\exp(-d/\lambda\sin\theta_t)$ vs. $\theta_t$ (a), and $I_{4uc}(\theta_t)/I_{LAO}(\theta_t)[1 - \exp(-d/\lambda\sin\theta_t)]$ vs. $\theta_t$ (b) for Sr 3d & Ti 2p (a) and La 4d & Al 2p (b) photoemission from 4 u.c. LAO/STO(001) from Augsburg. Also shown are model predictions of these ratios based on an intermixed interface with atom profiles shown as solid curves in (c). The associated abrupt-interface atom profiles as shown as dots in (c).

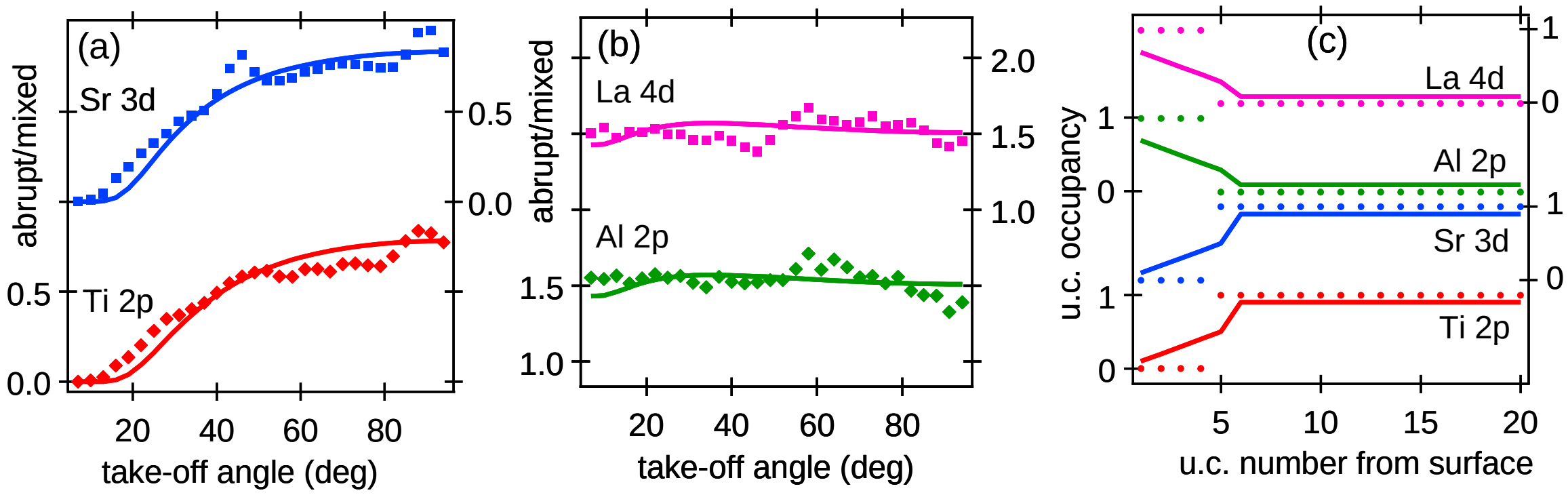

Fig. 24. Same as 23, except for 4 u.c. LAO/STO(001) from Tokyo grown at 0.7 J/cm$^2$ laser fluence.

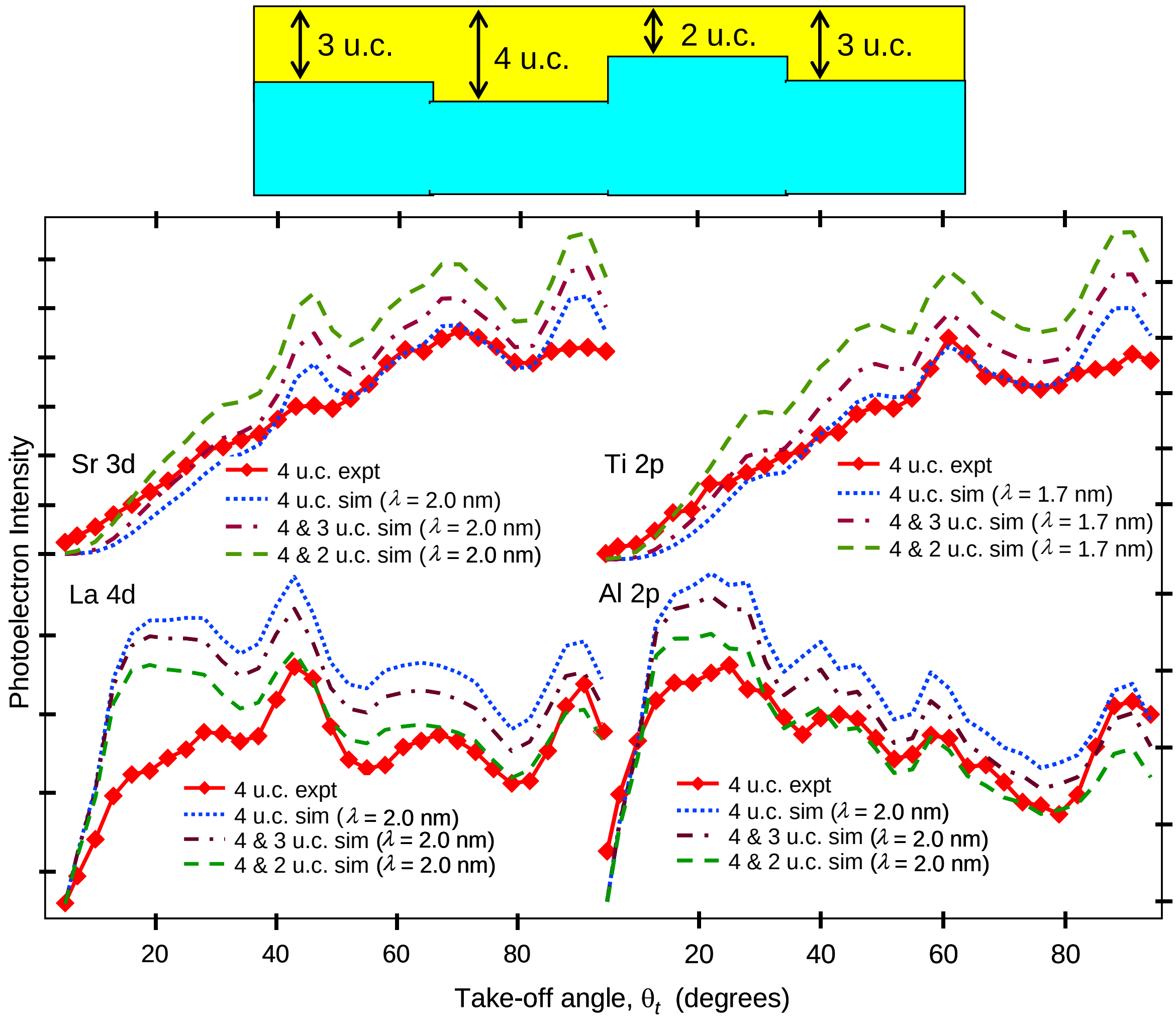

Fig. 25. Effect of interface roughness on ARXPS data for 4 u.c. LAO/STO(001) from Augsburg.

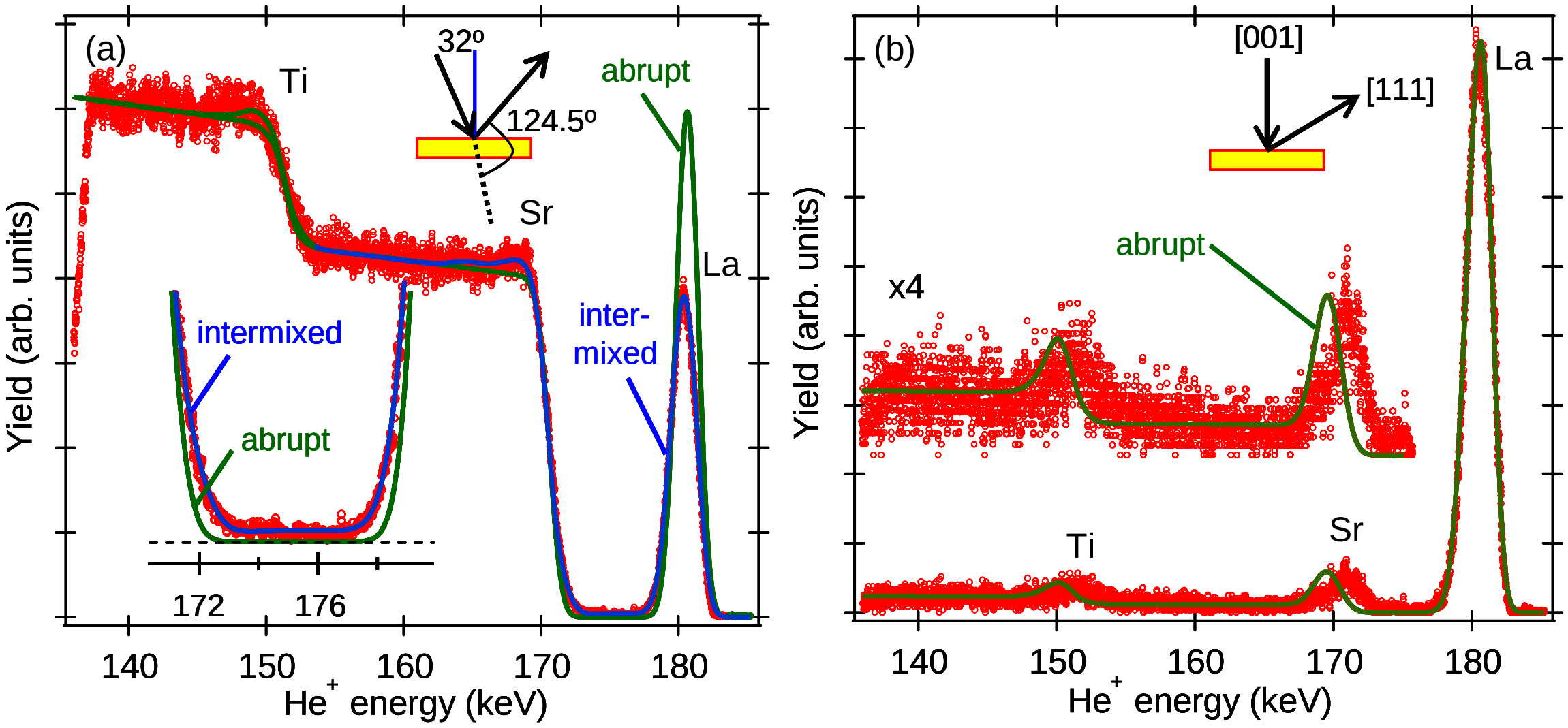

Fig. 26. 198.6 keV $He^{+}$ MEIS spectra in random (a) and aligned (b) geometries, along with simulations for abrupt and intermixed interface models, for a 4 u.c.Tokyo film grown at 0.7 J/cm$^2$ laser fluence. The random spectrum (a) was measured with the incident beam 32° off [001] and a backscattering angle of 124.5°. The incident and backscattered beams were aligned along [001] and [111], respectively, in (b).

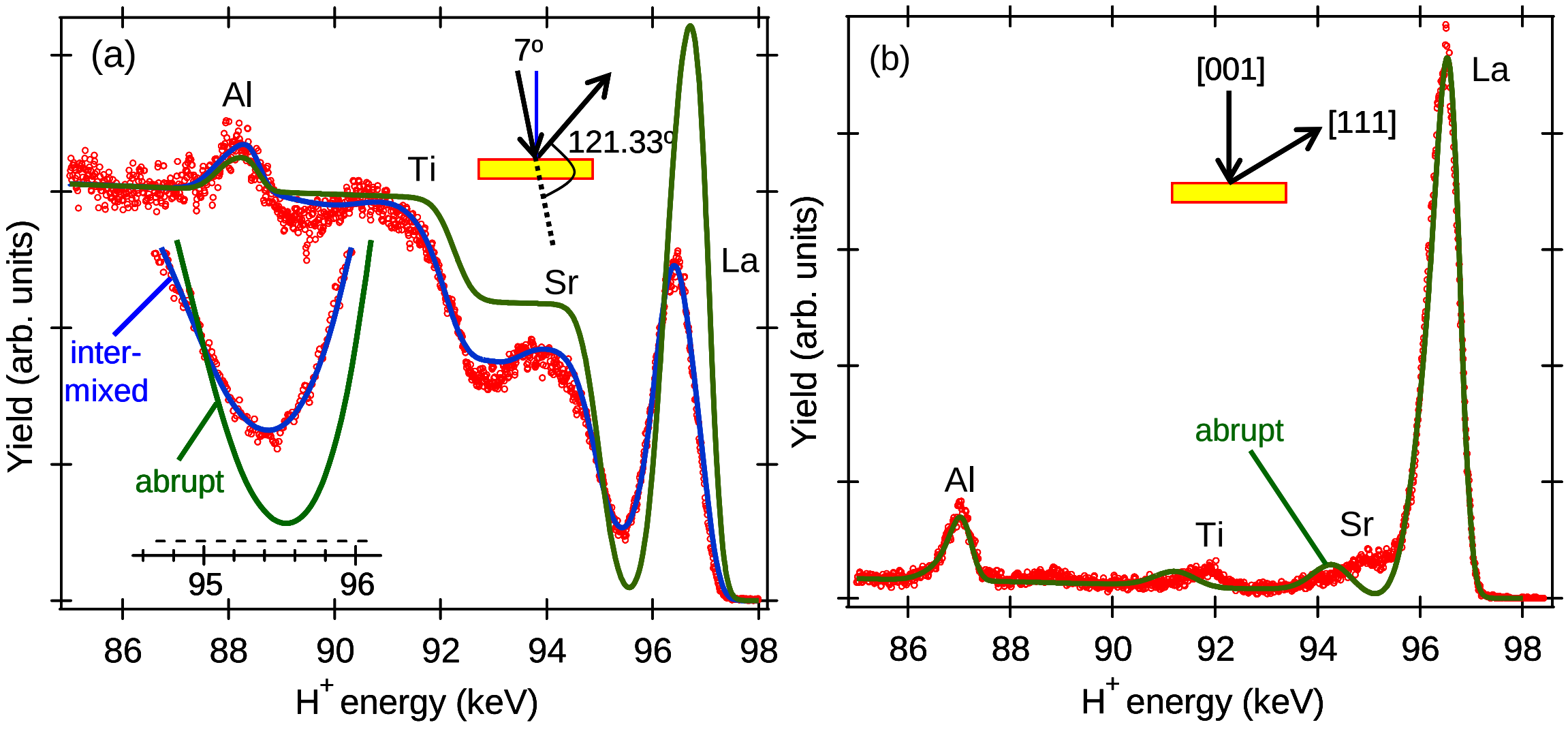

Fig. 27. 99 keV $H^{+}$ MEIS spectra in random (a) and aligned (b) geometries, along with simulations for abrupt and intermixed interface models, for a 4 u.c.Augsburg film. The random spectrum (a) was measured with the incident beam 7° off [001] and a backscattering angle of 121.3°. The incident and backscattered beams were aligned along [001] and [111], respectively, in (b).

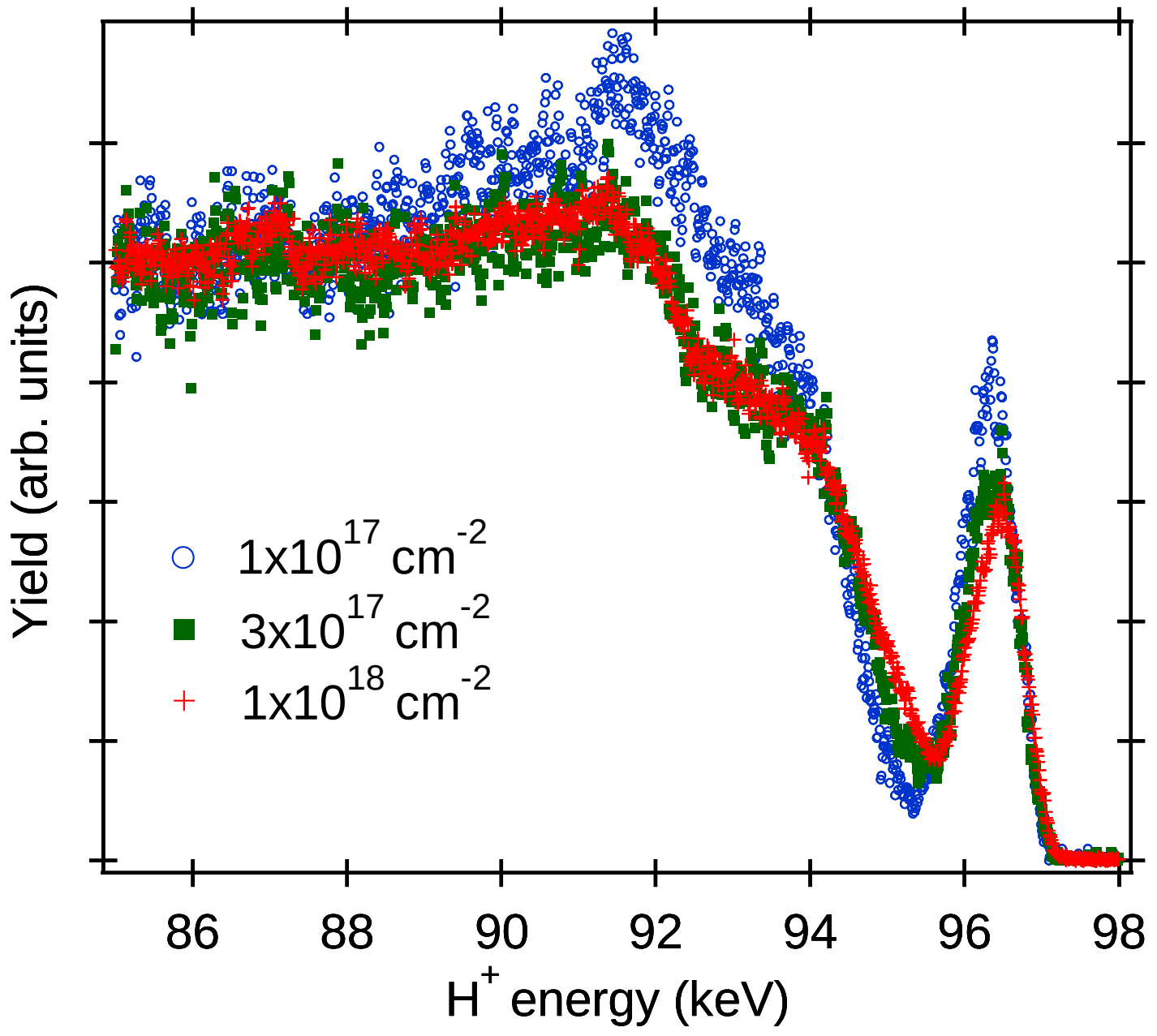

Fig. 28. 99 keV $H^+$ MEIS spectra as a function of total ion dose for a 4 u.c. Tokyo film.

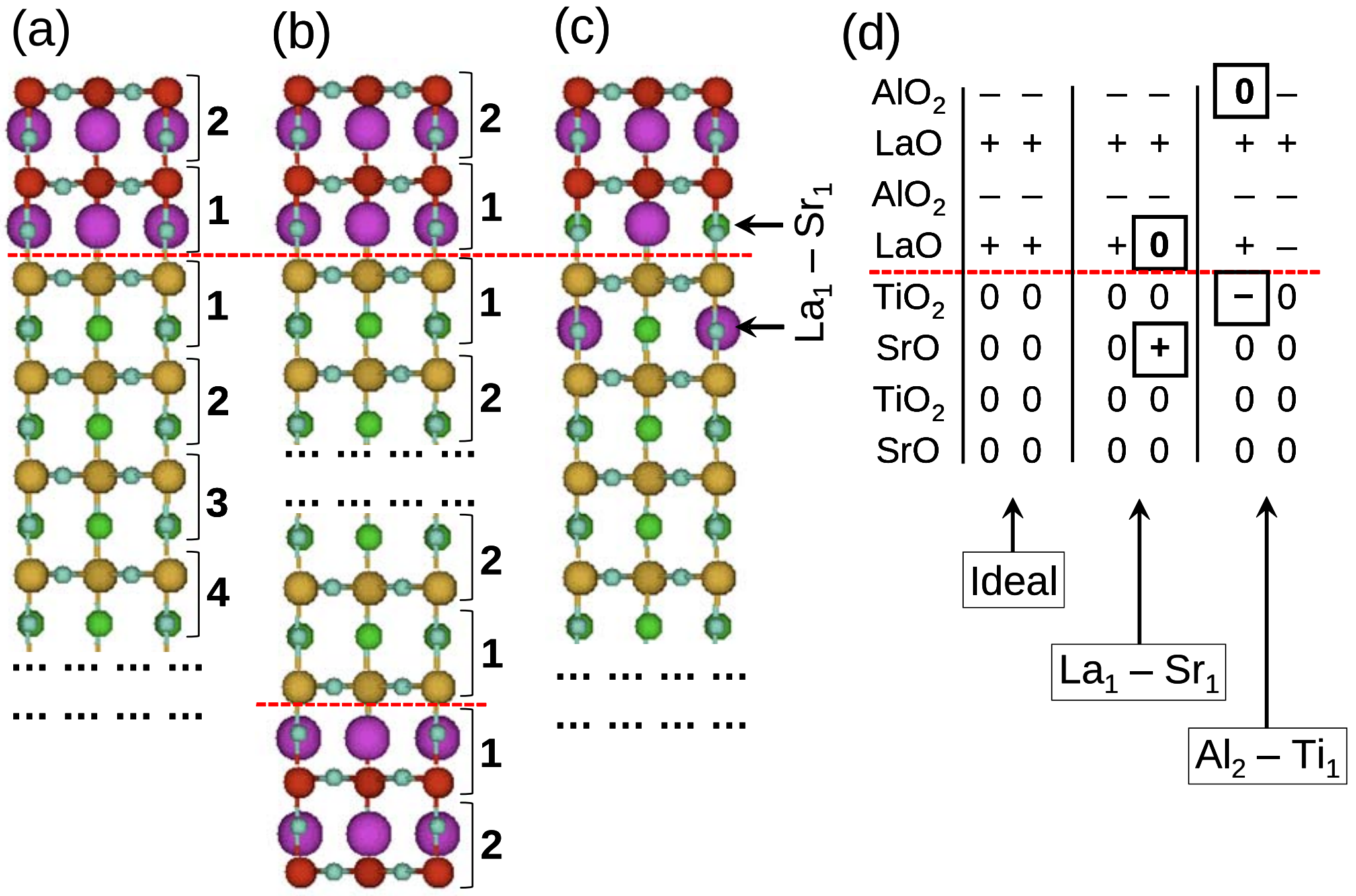

Fig. 29. Idealized interface represented using a periodic slab model (a) and a periodic "sandwich" model (b). Bold numbers show positions of the unit cells from the interface plane. (c) A La–Sr lattice site exchange configuration. (d) Schematics for the LAO/STO interfaces: "+" and "–" indicate LaO and $AlO_2$ units; SrO and $TiO_2$ units are indicated with "0".

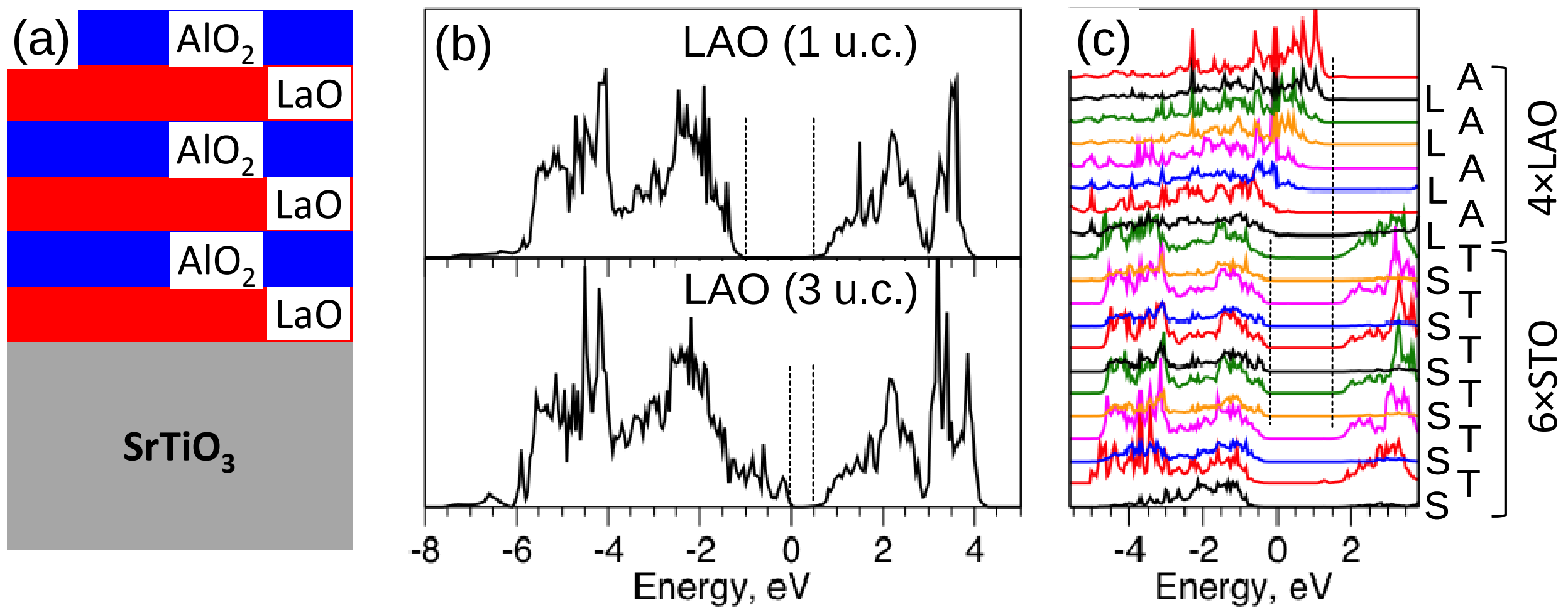

Fig. 30. (a) Schematic of the idealized $SrTiO_3(001)/LaAlO_3$ interface. (b) Total DOS calculated for one- and three unit cell thick LAO films. (c) Layer-projected DOS calculated for LAO(4 u.c.)/STO(6 u.c.). The 12 bottom plots correspond to the SrO and $TiO_2$ layers in the STO and 8 upper plots correspond to the LaO and $AlO_2$ layers in the LAO.

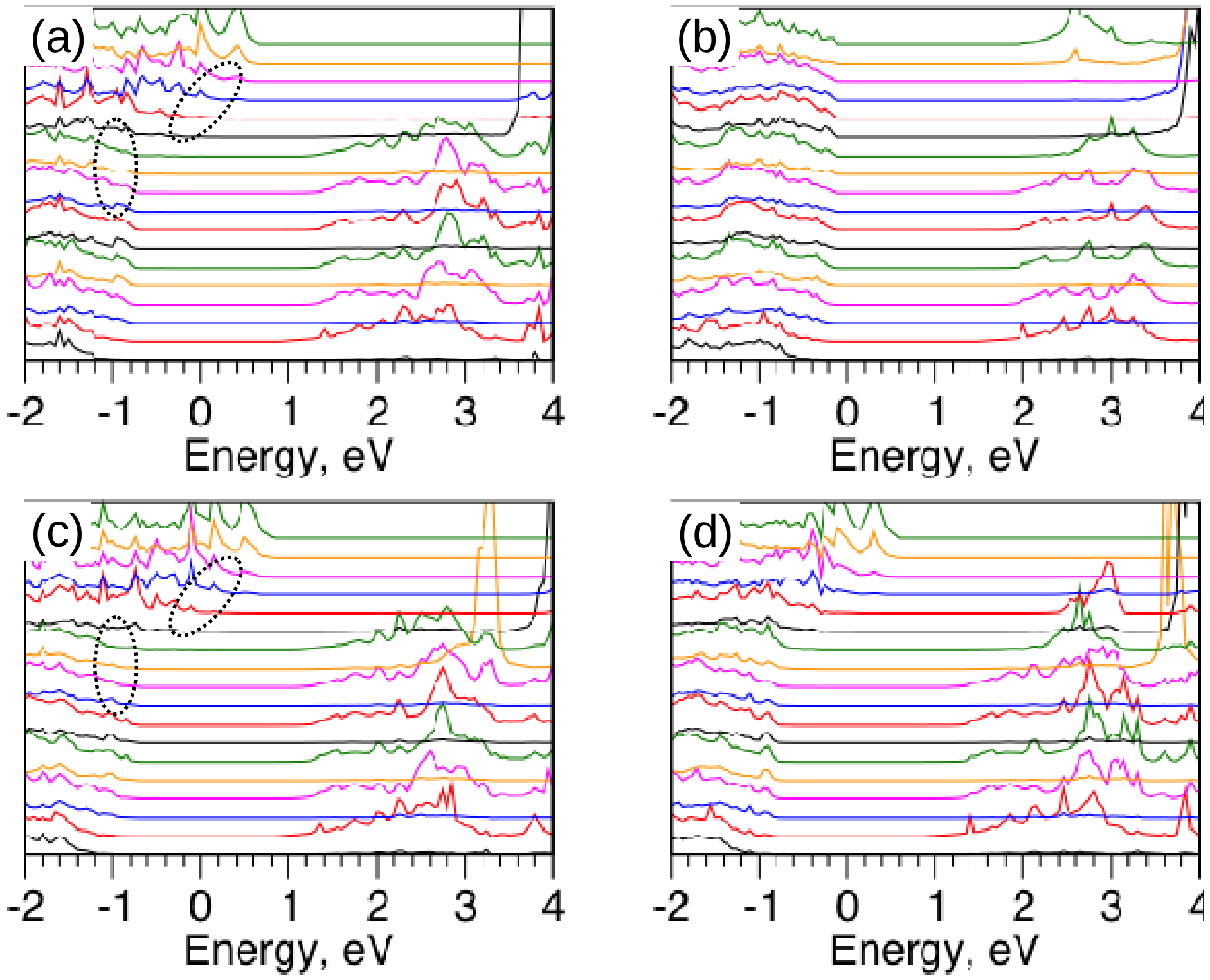

Fig. 31. Layer-projected DOS calculated for the idealized LAO/STO interface (a) and interfaces with Al3 ⇔ Ti1 (b), La1 ⇔ Sr1 (c), and La1Al1 ⇔ Sr1Ti1 (d) exchanges per √2×√2 lateral cell. In each plot the bottom 12 DOS correspond to STO(6 u.c.) and the top 6 DOS – to LAO(3 u.c.). Dashed lines highlight the similarities and differences in the DOS character.

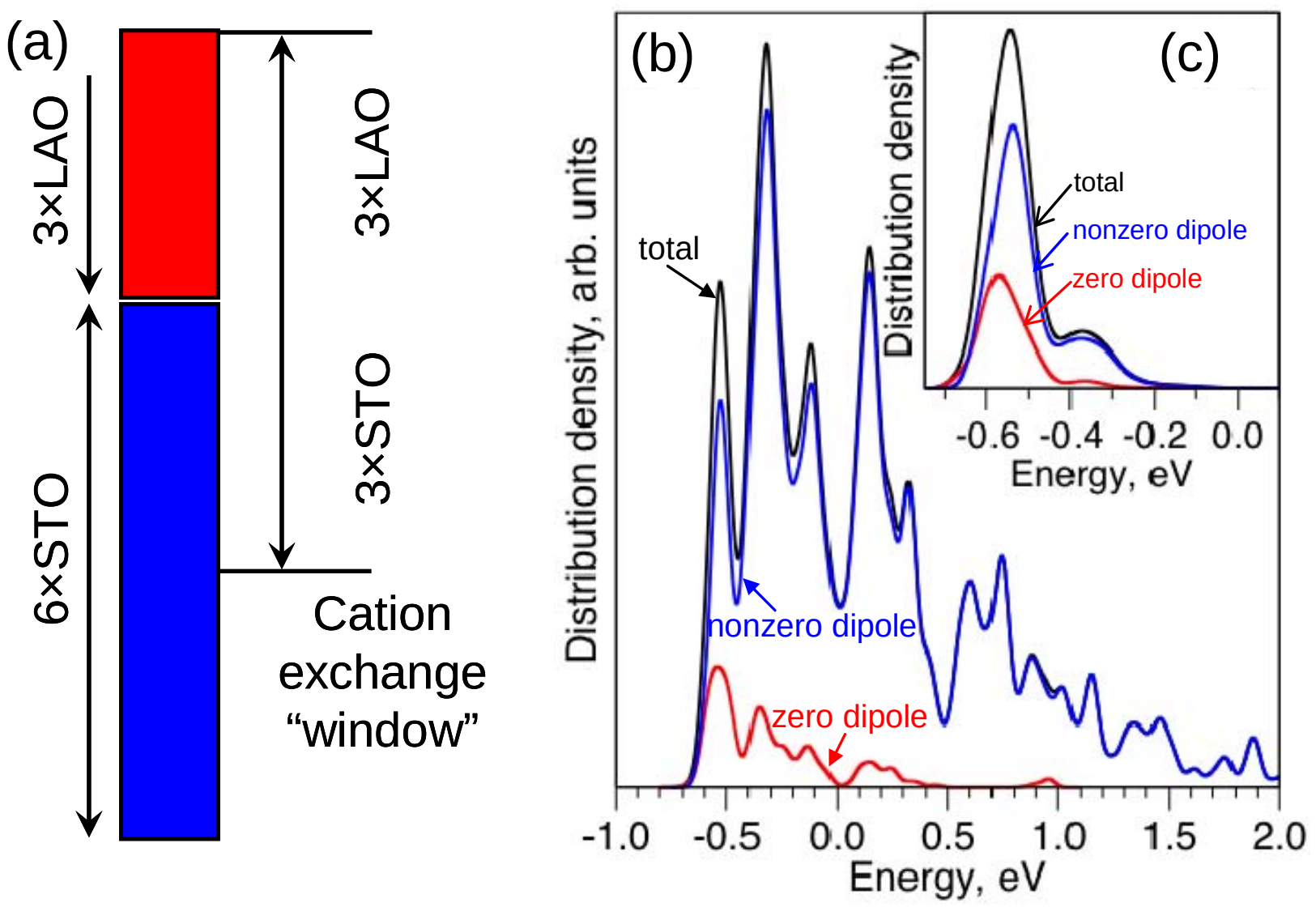

Fig. 32. (a) Structural diagram showing the "intermixing window" used to construct complex intermixed configurations. (b) Distribution of the La-Al-Sr-Ti intermixed configurations with respect to the idealized interface configuration calculated using the classical shell model for the STO(6 u.c.)/LAO(3 u.c.) interface, a √2√×2 lateral cell, and a STO(3 u.c.)/LAO(3 u.c.) intermixing window. (c) The same distribution weighted by the Boltzmann factor with T = 1000 K.

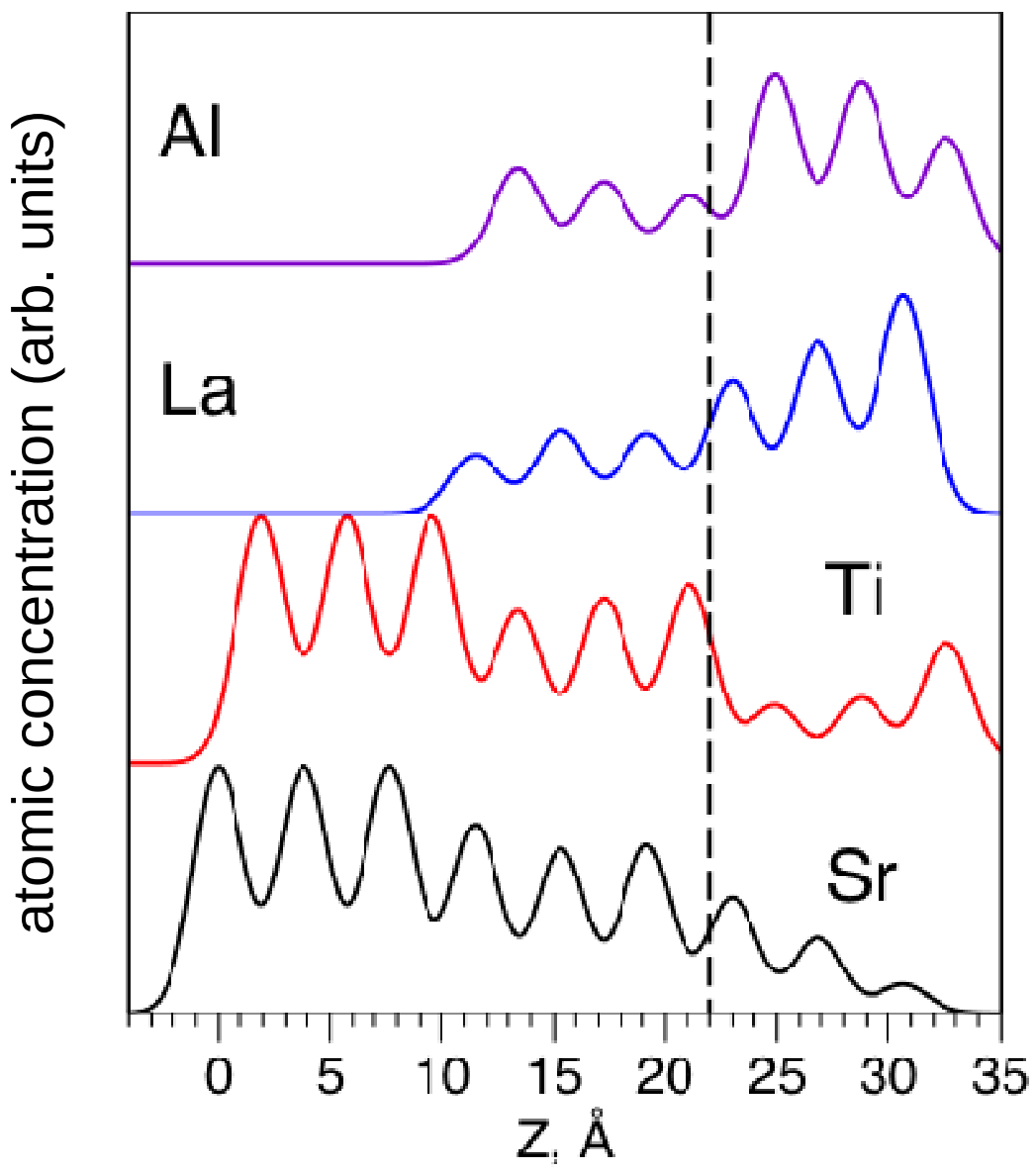

Fig. 33. Atomic concentration profiles for all four metal atoms in the direction perpendicular to the interface plane calculated for STO(6 u.c.)/LAO(3 u.c.) with a √2√×2 lateral cell, and a STO(3 u.c.)/LAO(3 u.c.) intermixing window, using the classical shell model. The vertical dashed line indicates the position of the $TiO_2$/LaO interface for a perfectly abrupt model.

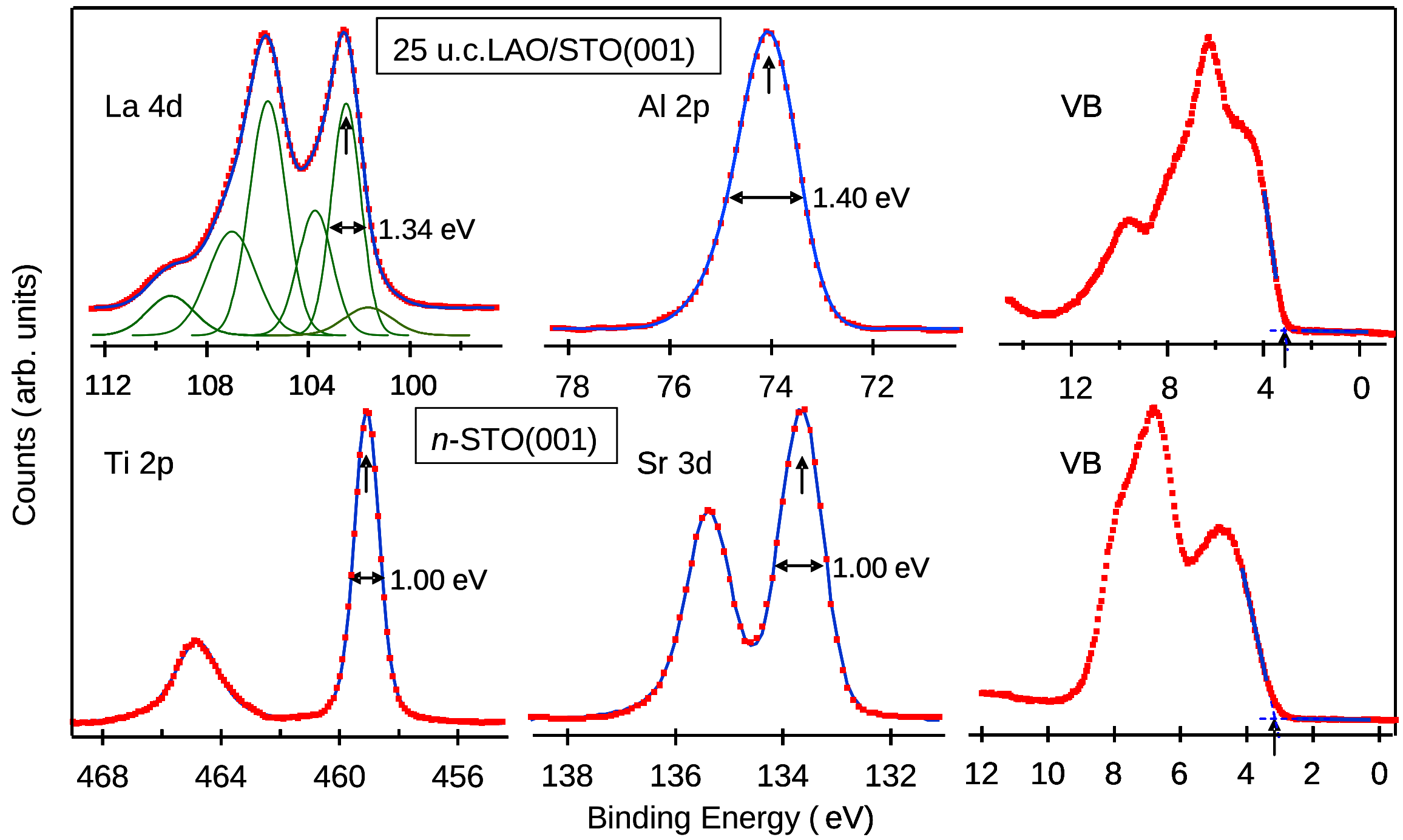

Fig. 34. Core-level and valence band photoelectron spectra for 25 u.c. LAO/ STO(001) (Tokyo) and Nb-doped bulk STO(001). Vertical arrows mark the spectral features used to determine the band offsets and band bending.

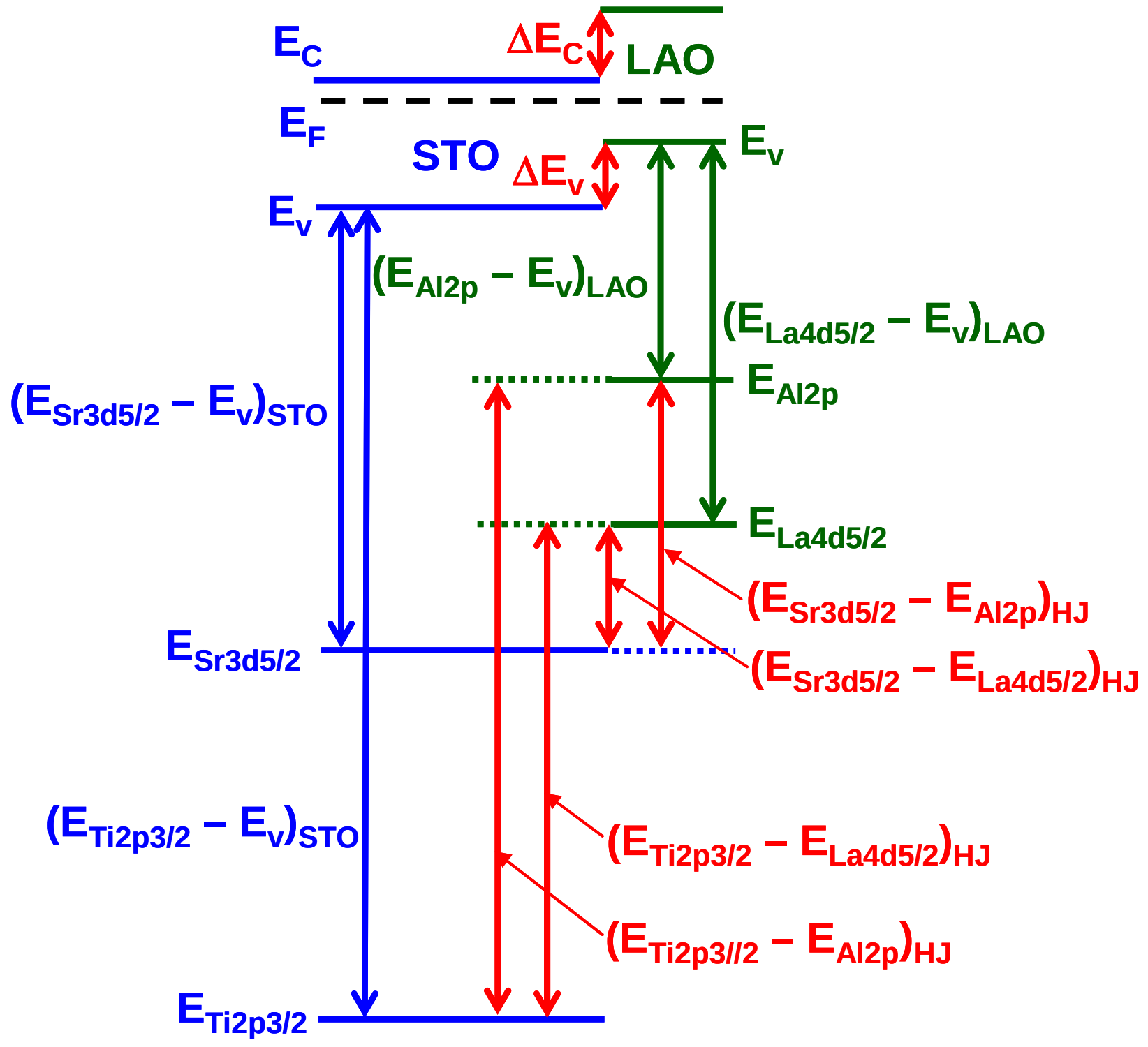

Fig. 35. Energy level diagram illustrating how band offsets and band bending are extracted from core-level and valence-band XPS spectra for LAO/STO.

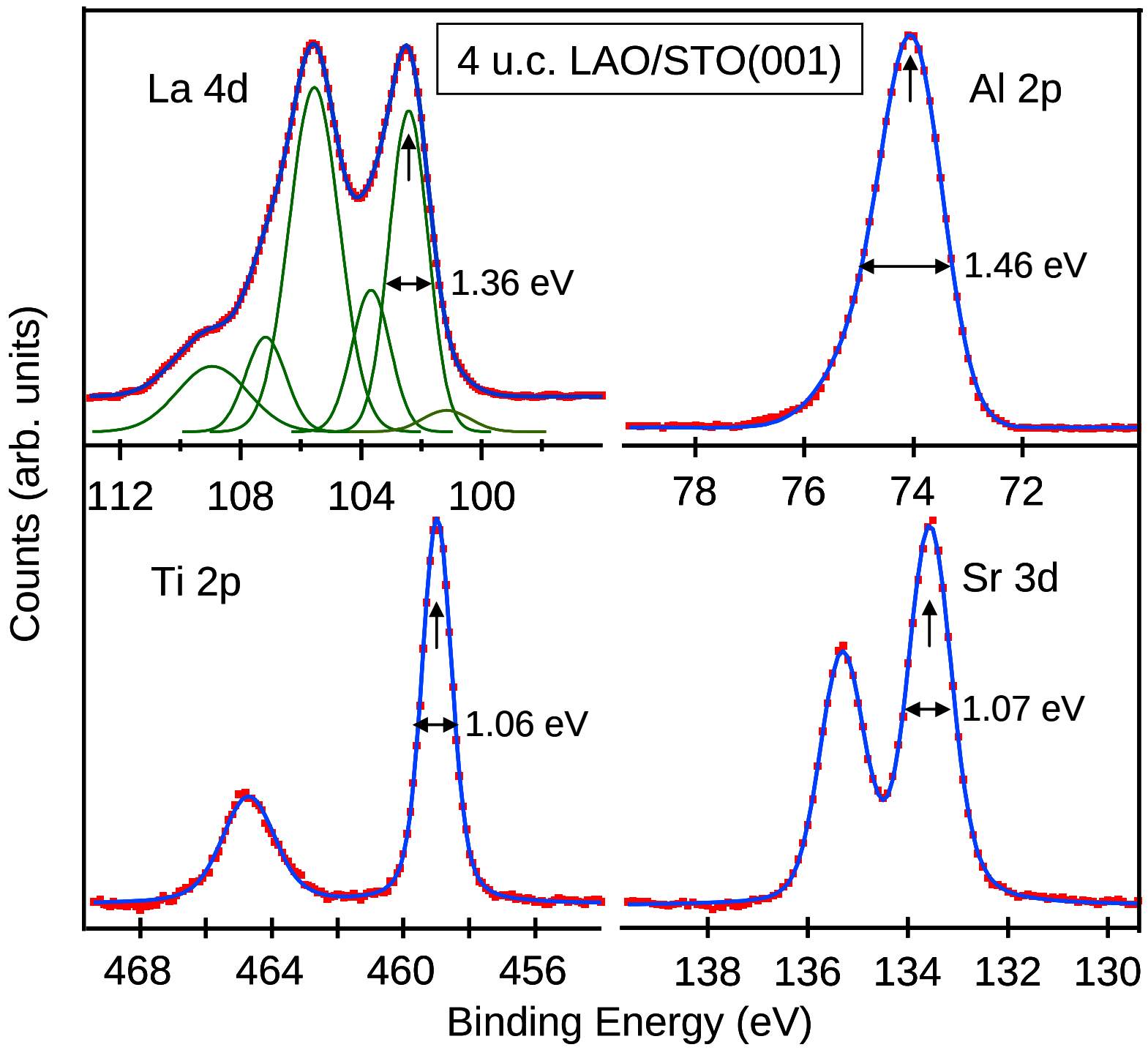

Fig. 36. Core-level spectra for 4 u.c. LAO on undoped STO(001) (Tokyo). Vertical arrows mark the spectral features used to determine the band offsets and band bending.

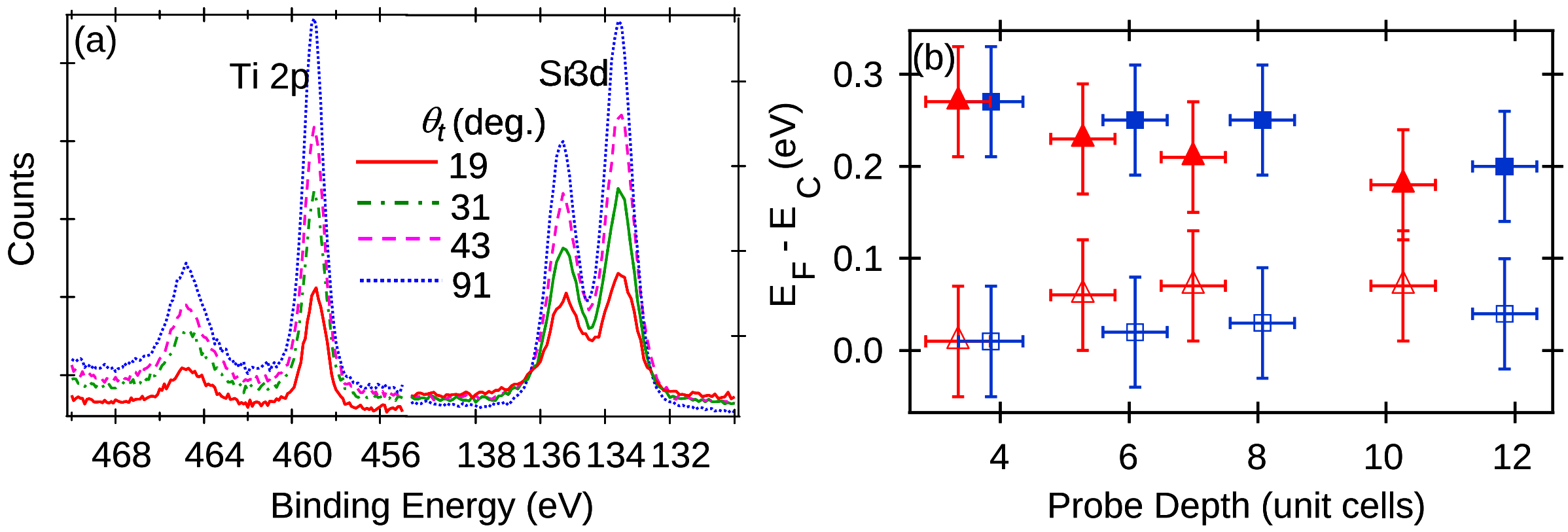

Fig. 37. (a) Typical core-level spectra (Tokyo), and (b) energy difference between the Fermi level and the conduction band minimum as a function of depth derived from Sr $3d_{5/2}$ (squares) and Ti $2p_{3/2}$ (triangles) binding energies (right) for 4 u.c. LAO/STO(001) from Tokyo (solid symbols) and Augsburg (open symbols).

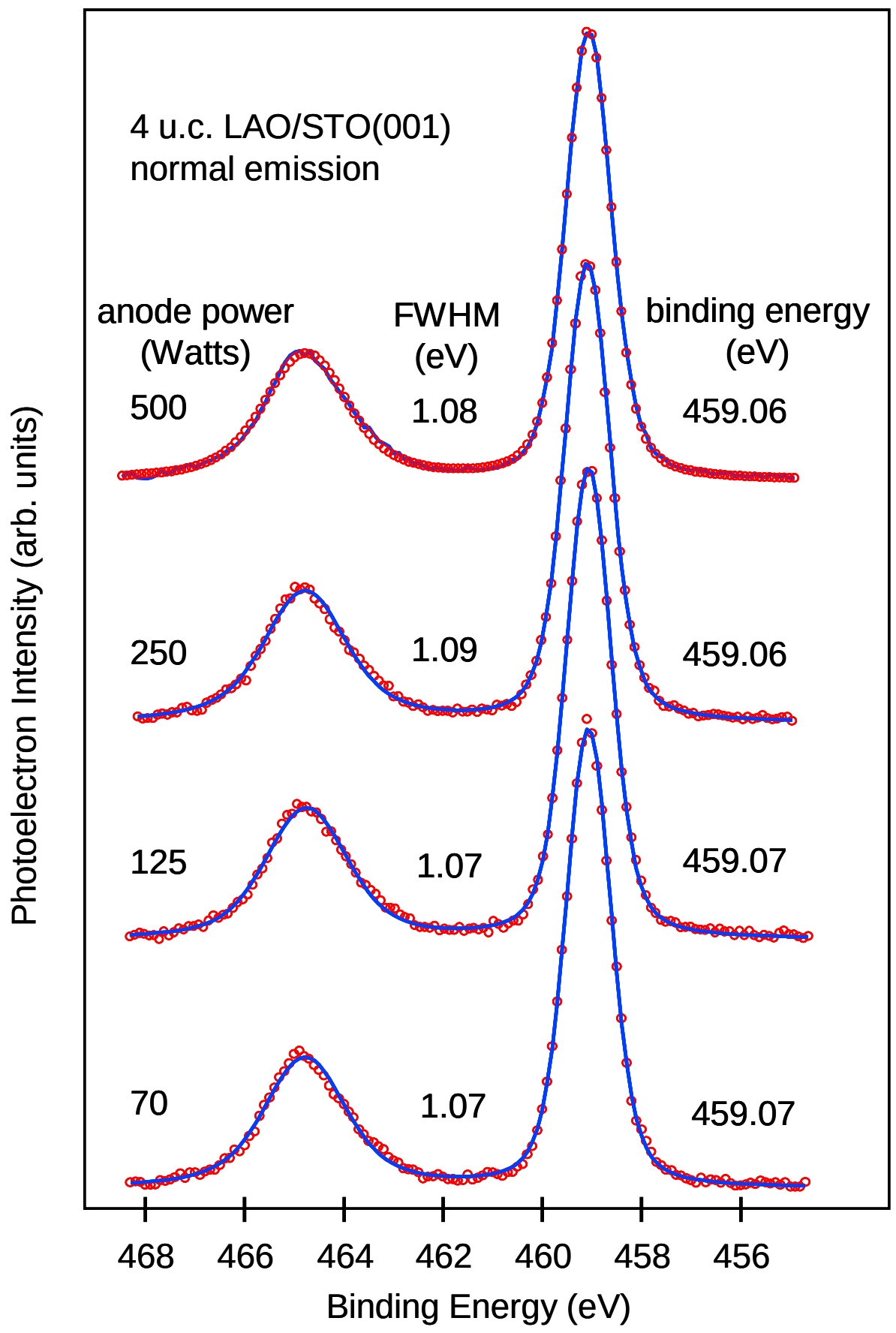

Fig. 38. Ti 2p spectra for 4 u.c. LAO/STO(001) as a function of x-ray anode power. The invariance of both binding energy and peak width with x-ray flux reveals that x-ray induced persistent photoconductivity, if it occurs, does not perturb the spectra from which the magnitude of the band bending and band offsets are extracted.

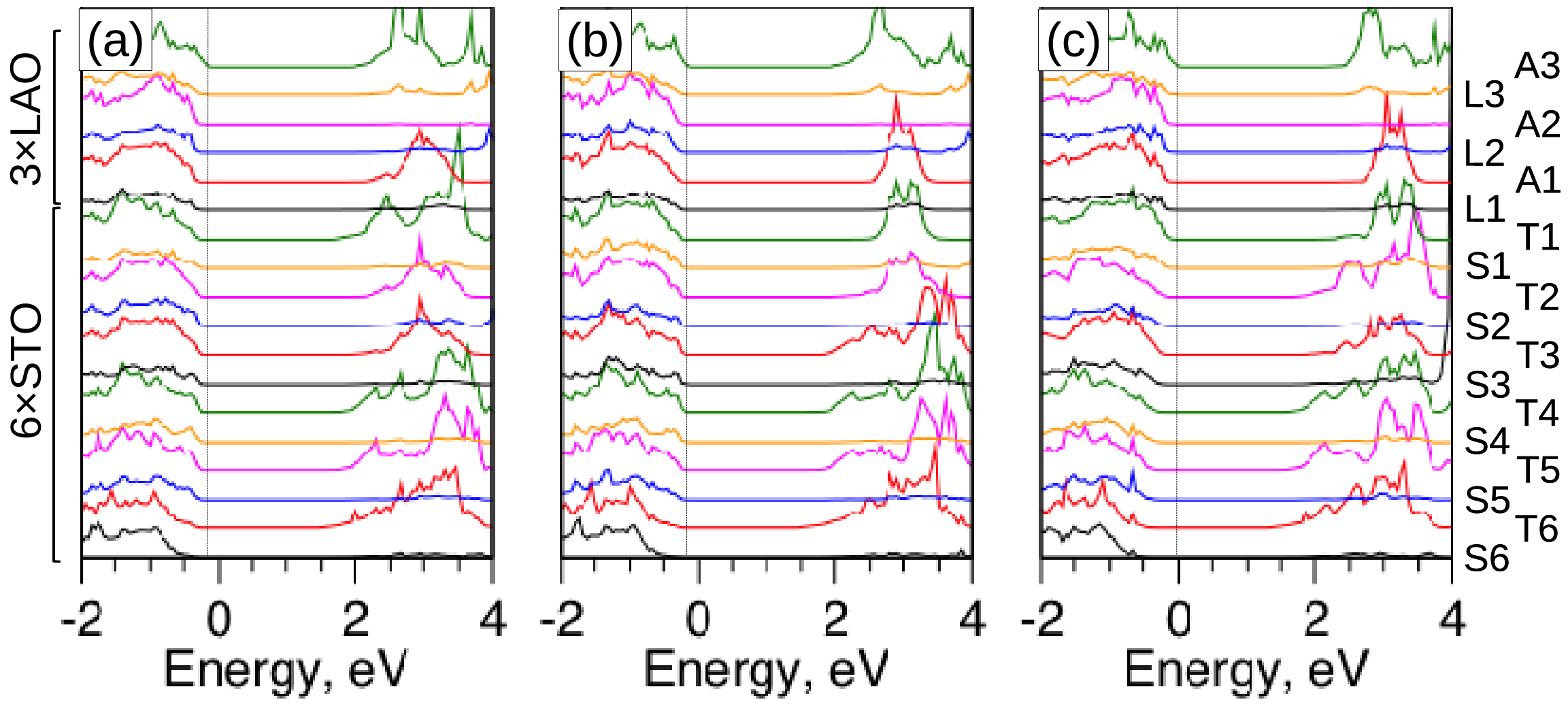

Fig. 39. Layer-projected density of states calculated for several of the lowest energy intermixed configurations of LAO(3 u.c.)/STO(6 u.c.) slabs. The atomistic structure and relative energies of these configurations are shown in Table 11.